\def\includesupplement{1}
\documentclass[11pt]{article}
\usepackage[margin=1in]{geometry}
\usepackage[T1]{fontenc}
\usepackage[utf8]{inputenc}
\usepackage{lmodern}
\usepackage{amsmath,amssymb,amsthm,booktabs,array,longtable,graphicx,caption}
\usepackage[numbers,sort&compress]{natbib}
\usepackage[hidelinks]{hyperref}
\usepackage{microtype}
\usepackage{pdfpages}
\DeclareMathSizes{10.95}{10.95}{8}{8}
\DeclareMathSizes{10}{10}{8}{8}
\usepackage{tikz}
\usetikzlibrary{arrows.meta,positioning}
\usepackage{enumitem}
\setlist{nosep,leftmargin=*}
\graphicspath{{figures/}}
\newtheorem{proposition}{Proposition}

\newcommand{\PP}{\mathbb P}
\newcommand{\E}{\mathbb E}
\newcommand{\ind}{\mathbf 1}
\newcommand{\Tk}{T_K}
\newcommand{\Healthy}{\mathsf H}
\newcommand{\Sick}{\mathsf S}
\newcommand{\Sicker}{\mathsf V}
\newcommand{\Dead}{\mathsf D}
\newcommand{\Cens}{\mathsf C}
\usepackage{authblk}

\title{\bfseries ATLAS: Autoregressive Trajectories for\\Longitudinal Analysis of Survival}
\author[1]{Morgan Nissen\textsuperscript{*}}
\author[2,3]{Chunyang Li}
\author[4]{Seth Glazier}
\author[2,3]{Julia Bohman}
\author[2,3]{Ahmad Halwani}
\author[1]{Jeffrey Humpherys}
\affil[1]{Department of Mathematics and Statistics, Missouri S\&T, Rolla, MO, USA}
\affil[2]{George E. Wahlen Department of Veterans Affairs Medical Center, Salt Lake City, UT, USA}
\affil[3]{Division of Hematology and Hematologic Malignancies, University of Utah School of Medicine, Salt Lake City, UT, USA}
\affil[4]{Independent researcher}
\date{}
\newcommand{\runningtitle}{Autoregressive Trajectories for Longitudinal Analysis of Survival}
\usepackage{fancyhdr}
\begin{document}
\maketitle
\begin{center}\small
\textsuperscript{*}Corresponding and submitting author: Morgan Nissen, \href{mailto:mkngqx@umsystem.edu}{mkngqx@umsystem.edu}.
\end{center}
\begin{abstract}
Dynamic survival prediction requires estimates that reflect a patient's observed clinical history. We introduce ATLAS (Autoregressive Trajectories for Longitudinal Analysis of Survival), a framework that trains a sequence model on longitudinal data and autoregressively samples possible futures conditional on an individual's covariates and observed history. These trajectories form a patient-specific synthetic cohort from which Kaplan--Meier survival curves and Aalen--Johansen cumulative incidence curves are estimated. We evaluate ATLAS using two stochastic finite-state automaton benchmarks with exact conditional reference curves: a single-covariate process with duration dependence and a late rule change, and a three-covariate Markov process with recovery and competing causes of death. Comparisons include eight direct survival models, including validation-optimized DRSA and Dynamic-DeepHit and history-aware SurvTRACE adaptations, with each neural model trained nine times in the first experiment and five times in the second. After baseline, every ATLAS configuration had lower mean survival-curve error than every fitted direct comparator in both experiments. With a common Sobol sampler and 10,000 trajectories per patient, the best ATLAS model did so in every individual training run, and its mean error was 62\% to 84\% lower than that of the strongest direct comparator. Baseline advantages were smaller and less consistent in the duration-dependent experiment. ATLAS provides a generative framework for history-conditioned survival estimation, while the accompanying benchmarks enable direct assessment of individual survival and competing-risk predictions against known reference curves.
\end{abstract}


\section{Introduction}
A prognosis is not a single number. When a clinician tells a patient how things are likely to go, the honest answer is a distribution over several possible futures, and that distribution should shift as new information arrives. A patient who has been stable for a year is not the same patient, prognostically, as one with the same baseline covariates who has spent the last three months getting worse. Classical survival models capture the covariates well, and dynamic methods such as Dynamic-DeepHit can fold in longitudinal observations~\citep{lee2019dynamic}. These methods predict event probabilities directly from the information supplied to them.

In this paper, we take a different route. If a model can predict what happens to a patient next, given everything that has happened so far, then we can run that model forward autoregressively and generate multiple random walks of different clinical journeys. Starting from the patient's own record, we sample a next state, append it, sample again, and keep going until the walk ends in death, dropout, or the end of follow-up. That is one possible future. Ten thousand such autoregressive random walks give ten thousand possible futures for the same person, and we can treat them as a cohort. The Kaplan--Meier estimator applied to that cohort is a survival curve for this one patient~\citep{kaplan1958}. When there are competing causes of death, the Aalen--Johansen estimator gives cause-specific cumulative incidence in the same way~\citep{aalen1978}. We call the framework ATLAS, for Autoregressive Trajectories for Longitudinal Analysis of Survival, and the generated futures a patient-specific synthetic cohort. The sequence model is trained on a full cohort; the synthetic trajectories are generated conditional on an individual patient's supplied covariates and observed history. These trajectories represent possible futures under the fitted model, rather than additional clinical observations. Increasing their number reduces Monte Carlo variability but does not remove uncertainty or error in the fitted model. The idea of estimating survival by sampling futures from a trained sequence model was developed in Glazier's thesis~\citep{glazier2019}, and this paper builds on it.

The appeal of this approach is that the survival curve can reflect what the model has learned about duration dependence, recovery, or abrupt changes in the disease process. It does not require proportional hazards or a fixed parametric survival distribution. There are still two sources of error: the model may learn the transition probabilities imperfectly, and a finite set of sampled futures may not fully represent the learned distribution. More samples can reduce the second source of error, but cannot correct the first. Part of what we want to understand is how much each contributes to the accuracy of the resulting curve.

Sampling futures from a learned model of health trajectories is an established technique, and it is worth saying where ATLAS sits. ETHOS trains a Transformer on tokenized patient timelines and estimates inpatient mortality and readmission by generating many future timelines for each patient~\citep{renc2024}. Delphi-2M, a GPT-style model trained on UK Biobank records, samples synthetic future disease trajectories and uses them to project disease burden over decades~\citep{shmatko2025}. Both show that the generative route works at scale on real records. Our contribution sits on the other side of the problem. We say exactly how a cohort of sampled futures, with censoring and competing causes, becomes a proper survival or cumulative-incidence curve, and we build benchmarks on which the right curve is known, so that the whole pipeline can be scored rather than only its downstream proxies. The approach also has older relatives. Multi-state and landmark models update a prognosis from the current state and the current time~\citep{vanhouwelingen2007,putter2007}, joint models tie a longitudinal process to the event time~\citep{rizopoulos2012}, and individual-level state-transition microsimulation in health economics draws patient trajectories from a known transition law and reads outcomes off the simulated cohort~\citep{siebert2012}. ATLAS keeps the simulation step and replaces the known transition law with a learned one, which is what lets it condition on an arbitrary history.

Related work on optimal dynamic treatment regimes provides a complementary perspective on history-dependent survival outcomes. Goldberg and Kosorok (2012) developed a censoring-adjusted Q-learning algorithm for multistage treatment decisions~\citep{goldberg2012}, while Cho et al. (2023) proposed a reinforcement-learning approach using generalized random survival forests to optimize mean survival or survival probability at a specified time~\citep{cho2023survival}. These approaches share ATLAS's emphasis on information contained in evolving histories, but address a different objective: identifying treatment policies that improve outcomes rather than predicting survival under the process represented by the training data. ATLAS instead estimates survival distributions from generated trajectories, conditional on baseline covariates or observed partial histories; its current formulation neither identifies causal treatment effects nor optimizes treatment decisions. The two research directions are therefore complementary rather than directly competing. Extending ATLAS to treatment-policy evaluation would require an explicit causal framework, with assumptions governing treatment assignment and censoring, and could not be justified by predictive accuracy alone.

There is a second problem that shows up as soon as we try to evaluate any of this. With real data we never see the true survival curve for an individual. We see one outcome per patient, and measures such as concordance and the Brier score assess predictions using those observed outcomes. They are useful, but do not give us a known individual reference curve against which to compare each prediction. Here we do our testing where the truth is computable. We build synthetic patients from stochastic finite-state automata (SFSA). A patient has fixed covariates and moves among a handful of named health states according to known transition rules. Because the rules are known, we can compute the exact conditional survival curve for any patient at any landmark by a short recursion, with no Monte Carlo error at all. That reference curve is what we score against, and it lets us separate learned-model error from the numerical error introduced by finite sampling.

This paper makes three contributions. We formalize ATLAS, in which autoregressive walks from a trained sequence model form a patient-specific synthetic cohort and the Kaplan--Meier or Aalen--Johansen estimator turns that cohort into an individualized curve. We introduce stochastic finite-state benchmarks whose exact conditional reference curves make curve-level error measurable. And we compare ATLAS with eight direct survival models, three of which read the same observed history, with ablations over seven samplers and the sampling budget.

The clinical motivation is to update prognosis as a patient's health changes, accounting for how long they have remained in a condition, their recovery, and competing causes of death. The synthetic benchmarks isolate these features in settings where the correct conditional survival and cumulative incidence curves are known, allowing prediction errors to be measured directly.

We report two experiments. The first is a single-covariate process in which the chance of getting worse grows with the time already spent in a state, the clock governing that growth changes at step 20, and at step 24 a new rule turns a second consecutive visit to the sickest state into death. It is hostile to any model that only looks at the current state. The second is a Markov process with three covariates, seven live states, drug-dependent recovery, and two mutually exclusive causes of death. It tests whether a good all-cause survival curve also gets the allocation between causes right. In both experiments we compare ATLAS, run with small recurrent and Transformer models and seven different sampling schemes, against eight direct survival models. Five of the direct models see only covariates. Three are adaptations of Dynamic-DeepHit and SurvTRACE that read the same observed history as ATLAS. These comparisons match the available patient information, although the models differ in architecture, training objective, and checkpoint selection.

Section~\ref{sec:benchmark} describes the two benchmarks, Section~\ref{sec:targets} describes ATLAS and the direct comparators, and Section~\ref{sec:truth} explains how we compute exact reference curves and score predictions. Section~\ref{sec:sampling} covers the sampling designs and the numerical error they introduce. Sections~\ref{sec:hardresults} and~\ref{sec:crresults} give the results, and Section~\ref{sec:discussion} discusses what they do and do not show. The models are small and the data synthetic. Our goal is to apply ATLAS to longitudinal healthcare records, possibly with a large language model as the sequence model, and Section~\ref{sec:discussion} says what that would require. The printed appendices give the generator specifications, the reference recursions, and the oracle argument. A separate supplement holds the model settings, per-run tables, full comparison figures, and the sampler and budget studies.

\section{Two benchmarks where the truth is known}\label{sec:benchmark}
Both experiments follow synthetic patients for 30 integer steps. Each patient has fixed covariates and moves among named health states until death, dropout, or the end of follow-up. The coefficients below are design choices for a benchmark, not clinical estimates. We named the covariates calcium, age, and drug level to make the processes easier to talk about; the numbers do not come from any clinical dose--response relationship. Appendix~\ref{app:generator} gives the complete transition matrices, initial distributions, and observation rules.

\subsection{A single-covariate process with duration dependence}
This benchmark is built so that the current state alone does not determine what happens next. The health states are Healthy ($\Healthy$), Sick ($\Sick$), and Sicker ($\Sicker$), with absorbing outcomes Death ($\Dead$) and dropout ($\Cens$). A calcium-like scalar covariate $X$ controls the base transition probabilities through
\[
F(x)=0.2+1.5(x/10)^4\exp[-(x/10)^5],\qquad
X\sim\operatorname{TruncNormal}(10,1.5^2;[7,13]),
\]
where TruncNormal denotes a normal distribution truncated to $[7,13]$. The map is hump-shaped in $x$; its functional form is a design choice and nothing else in the paper depends on it. At step 1 the probabilities of Healthy, Sick, and Sicker are $(2/3,1/4,1/12)$, with no death or dropout during initialization.

What takes the process beyond a Markov chain is that the next transition depends on how many consecutive steps the patient has already spent in the current state. We call this number the run length $u$, and in the usual terminology the process is semi-Markov. Staying in the same live state increases $u$ by one, and entering a different live state resets it to one. Here $u$ is the length of the current visit, not the total time ever spent in that state.

Longer Sick runs make progression to Sicker more likely, and longer Sicker runs make death more likely. For a transition ending at step $t$, the added probability is
\begin{equation}\label{eq:ramp}
\Delta_s(u,t)=\min\{\lambda_s(u/\rho(t))^{\nu_s},\kappa_s,0.95\Omega_s\},
\qquad \rho(t)=\begin{cases}5,&t<20,\\3,&t\ge20,\end{cases}
\end{equation}
where $s$ is the source state, $\lambda_s$ sets the size of the increase, $\nu_s$ its dependence on duration, and $\kappa_s$ its cap. The triples $(\lambda_s,\nu_s,\kappa_s)$ are $(0.35,2.5,0.90)$ for Sick-to-Sicker and $(0.97,3,0.99)$ for Sicker-to-Death. The quantity $\Omega_s$ is the probability available on the donor live destinations. We remove $\Delta_s$ proportionally from those destinations and add it to the designated worsening transition, which keeps every probability nonnegative and every row summing to one. The construction is a capped discrete power law rather than an exact continuous hazard.

The change in $\rho(t)$ from five to three means that from step 20 onward a given duration effect is reached after a shorter stay. A second rule begins at step 24. After drawing the proposed next state $\widetilde s_t$, we set
\begin{equation}\label{eq:hard24}
s_t=\begin{cases}
\Dead,&t\ge24,\ s_{t-1}=\Sicker,\ \widetilde s_t=\Sicker,\\
\widetilde s_t,&\text{otherwise}.
\end{cases}
\end{equation}
In words, a proposed second consecutive Sicker state becomes death. A first arrival in Sicker is still allowed, and a proposed recovery or dropout is not overridden. The rule applies to transitions ending at steps 24 through 30, so a Sicker state first reached at step 23 is exposed to it immediately. These rules are part of data generation and of the reference calculation. They are not imposed on any fitted model's predictions.

\begin{figure}[htbp]
\centering\includegraphics[width=\textwidth]{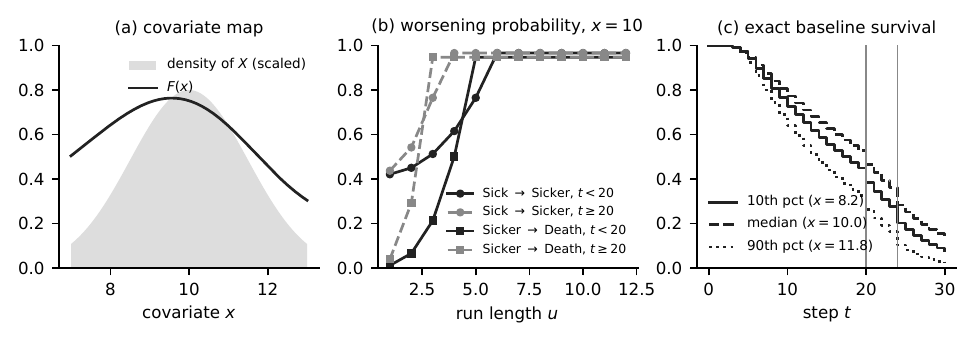}
\caption{The single-covariate process at a glance. (a) The covariate map $F(x)$, with the truncated-normal density of $X$ shaded for reference. (b) One-step probabilities of Sick-to-Sicker and Sicker-to-Death at $x=10$ as a function of run length, before and after the step-20 clock change, from Eq.~\eqref{eq:ramp} and the base matrix in Appendix~\ref{app:generator}. (c) Exact baseline reference survival for patients at the 10th, 50th, and 90th percentiles of $X$, computed by the recursion of Section~\ref{sec:truth}. Vertical lines mark steps 20 and 24.}\label{fig:sc_intuition}
\end{figure}

Figure~\ref{fig:sc_intuition} puts the pieces together. The covariate map is hump-shaped, so patients near the center of the covariate distribution stay healthy longest and risk rises toward either tail. The worsening probabilities climb steeply with run length and then saturate at their caps, and the clock change at step 20 pulls the climb forward by a couple of steps. The resulting exact baseline survival curves separate cleanly across the covariate range, and every one of them has a visible kink at step 24 where the hard rule begins.

\subsection{A multi-covariate process with competing risks}
The second experiment has seven live states, namely Healthy, Sick\_A, Sick\_B, Sick\_AB, Sicker\_A, Sicker\_B, and Sicker\_AB, where the suffix AB means both conditions are present. Death\_A and Death\_B are mutually exclusive absorbing outcomes, and dropout is a separate absorbing observation label.

Calcium, age, and drug level are sampled independently from uniform distributions on $[8,13]$, $[30,90]$, and $[0,1]$. Calcium acts through a bounded power response, age through a threshold response above 50, and drug level increases recovery from Sicker\_B to Sick\_B and from Sicker\_AB to Sick\_AB. From Sicker\_AB, the one-step probability of Death\_A is strictly greater than that of Death\_B throughout the supported covariate range.

\begin{figure}[htbp]
\centering\resizebox{\textwidth}{!}{
\begin{tikzpicture}[
  >={Stealth[length=2mm]},
  st/.style={draw,rounded corners=2pt,inner sep=3pt,font=\small,fill=white},
  dead/.style={st,draw,thick},
  x=1.05cm,y=1.05cm,
  arr/.style={->,line width=0.5pt,color=black!70},
  rec/.style={->,line width=2.2pt,color=black},
  die/.style={->,line width=0.5pt,dashed,color=black!70}]
\node[st] (H)  at (0,0)      {Healthy};
\node[st] (A)  at (2.2,1.7)  {Sick A};
\node[st] (B)  at (2.2,-1.7) {Sick B};
\node[st] (AB) at (4.6,0)    {Sick AB};
\node[st] (VA) at (7.0,1.7)  {Sicker A};
\node[st] (VB) at (7.0,-1.7) {Sicker B};
\node[st] (VAB) at (9.2,0)   {Sicker AB};
\node[dead] (DA) at (11.8,1.7)  {Death A};
\node[dead] (DB) at (11.8,-1.7) {Death B};
\draw[arr] (H) to[bend left=12] (A);   \draw[arr] (A) to[bend left=12] (H);
\draw[arr] (H) to[bend right=12] (B);  \draw[arr] (B) to[bend right=12] (H);
\draw[arr] (H) -- (AB);
\draw[arr] (A) to[bend left=12] (AB);  \draw[arr] (AB) to[bend left=12] (A);
\draw[arr] (B) to[bend right=12] (AB); \draw[arr] (AB) to[bend right=12] (B);
\draw[arr] (A) to[bend left=10] (VA);  \draw[arr] (VA) to[bend left=10] (A);
\draw[arr] (B) to[bend right=10] (VB); \draw[rec] (VB) to[bend right=10] (B);
\draw[arr] (AB) to[bend left=10] (VAB); \draw[rec] (VAB) to[bend left=10] (AB);
\draw[arr] (VA) -- (VAB); \draw[arr] (VB) -- (VAB);
\draw[die] (VA) -- (DA); \draw[die] (VB) -- (DB);
\draw[die] (VAB) -- (DA); \draw[die] (VAB) -- (DB);
\end{tikzpicture}}
\caption{State graph for the multi-covariate competing-risks experiment. The two heavy arrows are the recovery transitions that drug level increases, and dashed arrows are death exits. Each live state also has a self-transition, and dropout is omitted for readability. Death A is reachable only from Sicker A or Sicker AB, and Death B only from Sicker B or Sicker AB.}\label{fig:crgraph}
\end{figure}
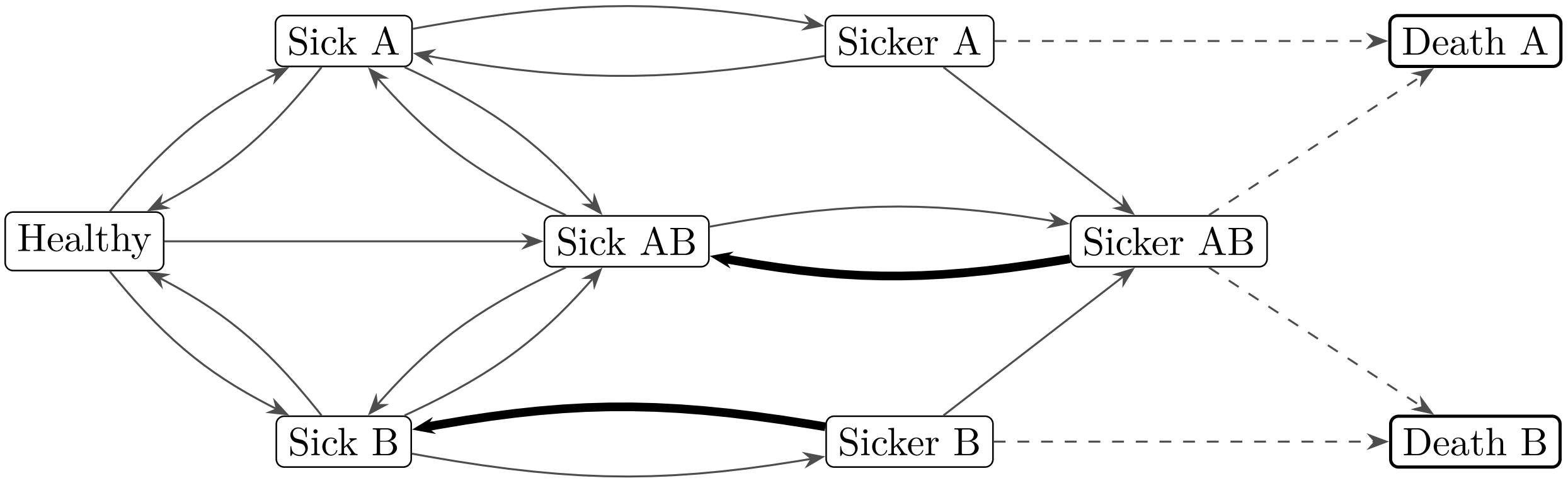

\begin{figure}[htbp]
\centering\includegraphics[width=\textwidth]{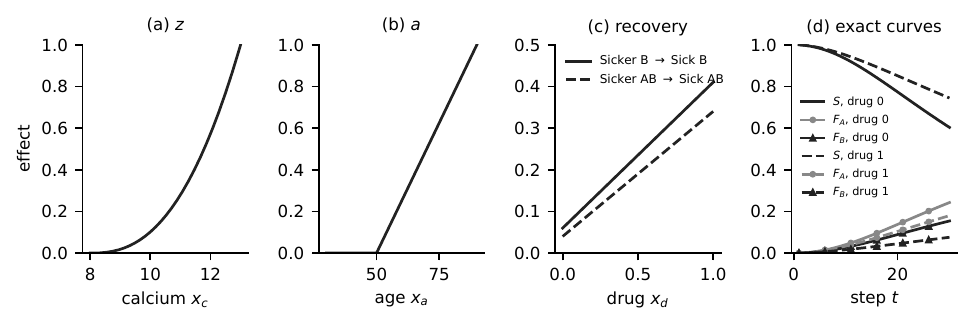}
\caption{The competing-risks process at a glance. (a) The calcium effect $z$ and (b) the age effect $a$ from Appendix~\ref{app:generator}. (c) The two recovery probabilities as functions of drug level. (d) Exact baseline survival and the two cumulative incidence functions for a patient with calcium 10.5 and age 60, at drug level 0 (solid) and 1 (dashed). At every step the three curves sum to one.}\label{fig:cr_intuition}
\end{figure}

Unlike the first experiment, this process has no duration ramp, no step-20 change, and no step-24 rule. Given the covariates, its biological transition probabilities depend only on the current state, so it is a Markov competing-risks benchmark. Independent dropout is applied with probability 0.008 after a surviving biological transition, and a same-step death is recorded first. Figure~\ref{fig:crgraph} shows the state graph, and Figure~\ref{fig:cr_intuition} shows the covariate responses, the drug effect, and the exact curves for a representative patient with and without the drug. Appendix~\ref{app:generator} gives every probability.

\section{From next-state predictions to a survival curve}\label{sec:targets}
ATLAS answers one question for one patient at one point in time. Given everything we know now, what could happen next? This section shows how a trained next-state model becomes a survival curve.

A \emph{landmark} $L$ is a time at which we update a prediction using the observations available so far~\citep{vanhouwelingen2007}. Let $A_L$ denote being alive and observed at that time, $X$ the covariates, and $H_L=(s_1,\ldots,s_L)$ the observed state history. Baseline prediction has $L=0$ and no observed history.

For a biological death time $T$, the conditional survival target is
\begin{equation}\label{eq:landmark}
S_L(t\mid x,h)=\PP(T>t\mid X=x,H_L=h,A_L),\qquad t>L.
\end{equation}
The two experiments treat this target slightly differently. The single-covariate experiment scores against a precisely defined observed-generator Kaplan--Meier functional, without assuming that it equals latent biological survival. The competing-risks experiment specifies independent censoring of a biological process, so its population Kaplan--Meier and Aalen--Johansen references coincide with the biological survival and incidence curves under the stated event-first convention. Section~\ref{sec:truth} defines both.

\subsection{Generating a patient-specific synthetic cohort}\label{sec:walks}
The sequence model is deliberately small. ATLAS-GRU and ATLAS-RNN are recurrent networks with two layers of width 32, and ATLAS-Transformer is a causal Transformer with two layers, four attention heads, and width 32; the second experiment uses the GRU and the Transformer only. At each step the model takes the patient's covariates and the states so far and returns a probability for each possible next state. Supplementary Section~S1 gives the full specification and training settings.

For model parameters $\theta$ and horizon $\Tk=30$, the learned trajectory distribution factorizes as
\begin{equation}\label{eq:factor}
p_\theta(s_{1:\Tk}\mid x)=\prod_{j=1}^{\Tk}p_\theta(s_j\mid s_{<j},x),
\end{equation}
so each next-state prediction uses the covariates and the preceding states of that particular walk. At baseline, every walk begins with a start token and draws its own initial health state. At a landmark, every walk begins with the same patient's actual observed prefix and only the future is sampled. Training uses next-state cross-entropy on observed transitions, including the first terminal outcome and excluding later repetitions of an absorbing label.

Let $D_m$ be the observed endpoint time of walk $m$, and let $J_m$ identify its endpoint, zero for censoring and a positive value for a death cause. At future step $t$ the risk count is $R_t=\sum_{m=1}^M\ind\{D_m\ge t\}$ and the death count is $E_t=\sum_{m=1}^M\ind\{D_m=t,J_m>0\}$. The Kaplan--Meier curve is
\begin{equation}\label{eq:synthetickm}
\widehat S_{\theta,M,L}(t)=\prod_{\substack{L<j\le t\\R_j>0}}\left(1-\frac{E_j}{R_j}\right).
\end{equation}
The product begins at one at the landmark. Walks censored at a time remain in that time's denominator, and deaths are counted before censoring removals. If the risk set becomes empty, the curve keeps its last value.

With competing causes, the cumulative incidence function (CIF) for cause $k$ is the probability that death from cause $k$ occurs by step $t$, with the other cause acting as a competing risk, so that a patient who dies of the other cause first can no longer die of cause $k$. We use the Aalen--Johansen update, with $E_{k,t}$ the count of cause-$k$ deaths at step $t$,
\begin{equation}\label{eq:aj}
\widehat F_{k,L}(t)=\widehat F_{k,L}(t-1)+\widehat S_L(t-1)\frac{E_{k,t}}{R_t}.
\end{equation}
Each CIF starts at zero at $L$, and the update applies when $R_t>0$. If no walks remain at risk, survival and both CIFs keep their last values. Survival uses the sum of the death counts across causes, which guarantees $\widehat S_L+\widehat F_{A,L}+\widehat F_{B,L}=1$. Estimating each cause as a separate one-minus-Kaplan--Meier curve would not produce these competing-risk probabilities. Supplementary Section~S2 gives the weighted form used by the importance and particle samplers.

\begin{figure}[htbp]
\centering\includegraphics[width=\textwidth]{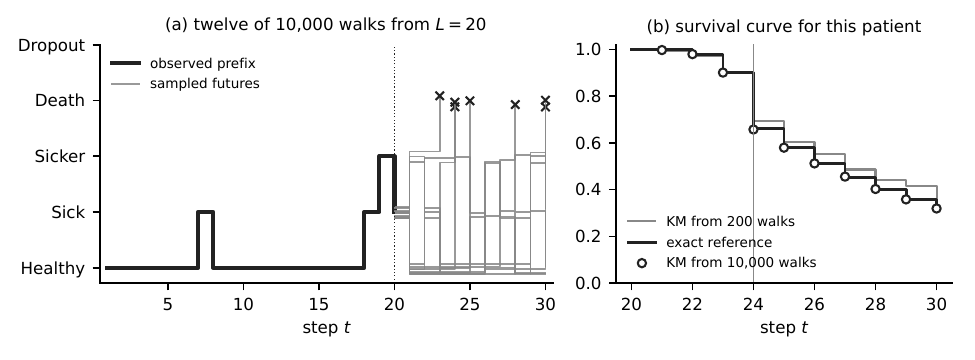}
\caption{Autoregressive random walks for one patient in the single-covariate experiment, with $x=10$ and landmark $L=20$. (a) The observed prefix (heavy line) and twelve of the 10,000 sampled futures (light lines, vertically jittered), with crosses marking deaths. (b) The Kaplan--Meier curve from 200 walks, the curve from all 10,000 walks (circles), and the exact reference from the landmark. The walks here are drawn from the true generator to illustrate the construction; the vertical line marks step 24.}\label{fig:walks}
\end{figure}

Figure~\ref{fig:walks} shows the whole construction for one patient. Every walk shares the observed prefix up to the landmark and then goes its own way, and the Kaplan--Meier curve built from the full cohort of walks sits on top of the exact reference. To keep the illustration honest we drew these walks from the true generator rather than from a trained model, so the picture shows what the sampling step does with a correct transition law. The experiments replace the generator with a learned one.

\subsection{Direct models and what they are told}
The comparison includes five covariate-only direct models, namely Cox proportional hazards (CPH), a Weibull multilayer perceptron (Weibull-MLP), deep recurrent survival analysis (DRSA; optimized)~\citep{ren2019}, DynamicDeepHit-Lite, and SurvTransformer-Lite. The two Lite models are DeepHit-style joint event-time classifiers~\citep{lee2018deephit} that read only the baseline covariates through a recurrent or Transformer time grid. At a nonzero landmark all five condition their baseline predictions on survival to $L$ but never read the patient's state sequence. The Lite label distinguishes these restricted implementations from the published methods. We write Dynamic-DeepHit and SurvTRACE for the published methods, Dynamic-DeepHit (optimized) for the selected history-aware Dynamic-DeepHit adaptations, and SurvTransformer for our history-aware SurvTRACE adaptations.

Three further direct models receive the full observed prefix along with the covariates. Dynamic-DeepHit (optimized) uses recurrent history encoding, attention conditioned on the latest observation, a joint event-time output, and an auxiliary next-measurement objective; its selected competing-risks configuration disables ranking loss. The two SurvTransformer variants adapt the feature attention of SurvTRACE to state and time tokens. They share the architecture and auxiliary tasks but differ in likelihood, one piecewise-constant hazard and one discrete-step~\citep{lee2019dynamic,wang2022survtrace}.

These are documented adaptations to SFSA sequences, not reproductions of every setting in the original papers. The SurvTRACE history representation in particular extends what is, in the original, a baseline-feature model, and its competing-risk implementation uses the cause-specific likelihood from the released code rather than the article's inverse-propensity objective. Supplementary Section~S1 records the actual architectures, losses, and training settings. Every model sees only information available at the prediction time, and no true transition probabilities are supplied during fitting.

For the history-aware Dynamic-DeepHit comparator, we selected a benchmark-specific configuration through a staged validation-based search over architecture, optimization, regularization, and loss settings. Selection minimized conditional event/censoring negative log-likelihood, averaged equally across the evaluated landmarks. The selected configuration was then fitted using the original nine training seeds for the duration-dependent experiment and five for the competing-risks experiment. We refer to these configurations as Dynamic-DeepHit (optimized). Their specifications and selection procedure are detailed in Supplementary Section~S1, and the original adaptation's results are retained in Supplementary Section~S10.

DRSA was also optimized separately for each benchmark using a staged validation-based search over recurrent width and depth, learned covariate projection, composite-loss weight, learning rate, regularization and batch size. The search used 24 screening candidates and 12 refinement entries per experiment, followed by three-seed confirmation of two finalists. Configurations and checkpoints were selected using observed-outcome conditional validation negative log-likelihood averaged equally across the evaluated landmarks, without using analytical reference curves or test-set MAE. The selected configurations were fitted using the same nine and five training seeds and are labelled DRSA (optimized). This follow-up was motivated by the initial comparison results; Supplementary Section~S1.2 gives the search and selected settings.

The two SurvTransformer variants select their checkpoints using their combined validation losses. These selection criteria differ from the reference-curve MAE used to compare predictions in this study. Supplementary Section~S1 gives the training objectives and checkpoint-selection settings.

\section{Reference curves, information, and scoring}\label{sec:truth}
\subsection{The reference curve}
Reference curves are calculated from the known generator probabilities without drawing any reference trajectories. In the duration-dependent experiment the calculation has to track the health state and its current run length together. For example, $(\Sicker,3)$ means three consecutive Sicker observations including the current step. Three live states and run lengths 1 through 30 give $m=90$ bookkeeping combinations, and combinations that are not yet reachable simply carry zero mass.

For a fixed patient's covariates, let $b_j\in\mathbb R_{\ge0}^{1\times m}$ hold the probability of remaining alive and observed in each augmented state at step $j$. Let $Q_j\in[0,1]^{m\times m}$ be the observed live-to-live matrix, and let $d_j,c_j\in[0,1]^{m\times1}$ hold observed-death and dropout probabilities for the transition to $j+1$. With $\ind$ a column vector of ones,
\begin{equation}\label{eq:massbalance}
Q_j\ind+d_j+c_j=\ind.
\end{equation}
The matrix $Q_j$ is sub-stochastic, since its rows can sum to less than one once death and dropout remove mass. The reference updates are
\begin{equation}\label{eq:kmreference}
b_{j+1}=b_jQ_j,\qquad S^*_{\rm KM}(j+1)=S^*_{\rm KM}(j)\left(1-\frac{b_jd_j}{b_j\ind}\right).
\end{equation}
The denominator is the probability still at risk before the transition and the numerator is the probability of an observed death on that transition. Death and dropout are mutually exclusive outcomes of the generator's categorical draw, so $d_j$ must not receive an additional dropout adjustment. Same-time dropouts remain in the denominator, matching Eq.~\eqref{eq:synthetickm}.

At baseline, the three initial-state probabilities initialize $b_1$, with $b_1\ind=1$ and $S^*_{\rm KM}(0)=S^*_{\rm KM}(1)=1$. At a landmark $L\ge1$ we put probability one on the patient's observed augmented state, keep the covariates fixed, and set $S^*_{\rm KM}(L)=1$. The updated curve describes future outcomes from a known live starting point. This patient-specific reference is a different object from a pooled Kaplan--Meier curve across patients with different covariates and histories.

The kernels include both clock changes. The hard rule redirects Sicker-to-Sicker mass to death for $j\in\{23,\ldots,29\}$, corresponding to transitions ending at 24 through 30, while conserving each row's total. Empty risk sets keep the last curve value. The recursion has no Monte Carlo error, though floating-point rounding remains.

Observed live mass $b_j\ind$ decreases through both death and dropout, whereas the Kaplan--Meier curve decreases only through the observed death fraction. That is why we call the single-covariate reference the observed-generator Kaplan--Meier functional. In the competing-risks generator, the explicitly independent, event-first observation mechanism makes censoring cancel from the population hazard, so the exact reference can be computed by propagating biological live mass and accumulating each death cause. Appendix~\ref{app:truth} gives both recursions.

\subsection{An oracle with restricted information}\label{sec:oracle}
Before the formal statement, here is the intuition. Suppose an oracle knows the true process exactly but is handed only a patient's covariates. At a later landmark, two patients with the same covariates can have different histories and therefore different true curves, yet the oracle must give them the same prediction, so it will be wrong for at least one of them. Handing it the history removes that error. We now make this precise. Fix $L$ and a future time $t>L$, and let $Y_t=S^*_{\rm KM,L}(t\mid X,H_L)$ among eligible patients $A_L$, where $\PP(A_L)>0$. Then $Y_t$ is a personalized reference curve value that varies across patients through their covariates and histories. It is not an indicator of an observed death.

Let $\mathcal I$ denote the information generated by the allowed inputs, so that $\mathcal I=\sigma(X)$ permits covariates only, while $\sigma(X,s_L,u_L)$ also permits the current state and run length. A prediction $g$ is measurable in $\mathcal I$ if it can be computed from that information alone. The smallest achievable expected absolute error is
\begin{equation}\label{eq:floor}
\inf_{\substack{g\text{ measurable in }\mathcal I\\0\le g\le1}}\E[|Y_t-g|\mid A_L]
=\E\!\left[|Y_t-\operatorname{med}(Y_t\mid\mathcal I,A_L)|\mid A_L\right],
\end{equation}
where $\E[\cdot\mid A_L]$ averages over eligible patients. A conditional median has at least half the conditional probability on each side, equality included, and any measurable median choice minimizes absolute error. Covariate-only predictions can have positive error even for an oracle, simply because patients with the same covariates can have different histories and hence different reference values.

\begin{proposition}\label{prop:floor}
The minimum in Eq.~\eqref{eq:floor} is zero exactly when the allowed information determines $Y_t$ with probability one among eligible patients. Allowing additional inputs cannot increase this minimum.
\end{proposition}
This is a statement about information, not about any particular estimator. At a fixed landmark, covariates and augmented state determine the duration-dependent reference, and covariates and current state suffice for the Markov competing-risks reference. So in both experiments the informed direct predictors and ATLAS share the same ideal floor of zero. At baseline, no patient history has yet been observed, and covariates alone determine the reference curve in both benchmarks. Covariate-only predictors therefore also have an ideal information floor of zero at baseline. Fitted models still make errors from learning and from numerical approximation. Appendix~\ref{app:oracle} gives the median argument and the population weighting. The finite-sample true-generator sampler that appears in our tables is a different kind of oracle, a sampling control, and it does not by itself provide an information-theoretic lower bound.

\subsection{Evaluation protocol}\label{sec:evaluation}
Each experiment has 50,000 patients, split into 40,000 training, 5,000 validation, and 5,000 test patients before histories are expanded into training prefixes. Each neural model is fitted nine times in the single-covariate experiment and five times in the competing-risks experiment, each time from a different random initialization and data order. We call each fit a training run; the seeds are listed in Supplementary Section~S1. The training limit is 200 epochs with validation patience 15. For Dynamic-DeepHit (optimized), this limit applies to joint training following five auxiliary-only burn-in epochs.

Each evaluation begins with 300 saved test patients, and nonzero landmarks keep only those still alive and observed. The single-covariate eligible counts are 300, 102, 78, and 57 at $L=0,20,23,24$. The competing-risks counts are 300, 251, and 179 at $L=0,10,20$. The 5,000-patient competing-risks test split contains 1,206 observed Death A outcomes and 693 Death B outcomes, along with 847 dropouts and 2,254 administrative survivors, so both causes are well represented. Curve errors use the 300-patient evaluation subset and its landmark survivors, not the full split. The 300-patient evaluation subset was used to keep the computational cost of trajectory generation manageable while evaluating all methods on the same patients. At each landmark, evaluation was restricted to eligible patients within this subset. The smaller eligible samples at later landmarks limit the precision of the landmark-specific error estimates.

For a target $Z$ (survival or one of the two CIFs) the primary metric is
\begin{equation}\label{eq:mae}
\mathrm{MAE}_{L,Z}=\frac{1}{n_L(30-L)}\sum_{i:A_L}\sum_{t=L+1}^{30}
|\widehat Z_{i,L}(t)-Z^*_{i,L}(t)|,
\end{equation}
where $n_L$ is the number of eligible scored patients, $Z^*$ the exact reference, and $\widehat Z$ the fitted prediction. An MAE of 0.01 means an average absolute difference of one percentage point across patients and scored times. Errors are computed for individual curves before averaging, so errors of opposite signs do not cancel. Post-rule MAE in the single-covariate experiment restricts the time sum to $t\ge24$.

Later landmarks involve different eligible patients and shorter prediction windows. Changes in MAE across landmarks therefore do not, by themselves, measure the benefit of providing a model with more history.

For each training run, stochastic configurations (the sampling designs of Section~\ref{sec:sampling}) are scored separately at five independent sampling seeds, and their MAEs are averaged. The per-run means are then averaged with equal weight, nine in the single-covariate analysis and five in the competing-risks analysis.

Their sample standard deviation describes between-run variation. Separately, pooled within-run sampling variation is the square root of the mean within-run variance, calculated over the nine or five training runs, respectively. Each within-run variance uses five sampling repetitions. The between-run statistic retains finite-sampling error; neither standard deviation is a confidence interval.

Direct predictions are deterministic for a fixed training run and need no rollout budget; each fitted neural direct model contributes one error per training run. The CPH and true-generator oracle controls are reported separately from the independently trained neural models. CPH predictions are unchanged to numerical precision, and the oracle predictions are reused across runs.

For consistency, every main table and headline comparison uses one common sampler, scrambled Sobol, for every ATLAS model, landmark, and target. The other six samplers appear only in the ablation. For comparisons across training runs, the same ATLAS architecture and sampler are used in every run. Exact sign tests summarize these paired comparisons descriptively.

To assess patient-level uncertainty, we used 5,000 paired bootstrap resamples. In each resample, 300 patient IDs were drawn with replacement from the original test evaluation subset. The same draw was used for all models, landmarks, and targets within an experiment, retaining only patients eligible at each landmark. Before resampling, patient-specific full-curve MAE was averaged over sampling repetitions, where applicable, within each training run, and then equally across training runs. We then calculated pointwise 95\% percentile confidence intervals for the difference in mean MAE (ATLAS minus comparator), with negative values favoring ATLAS. Relative error reductions were also recalculated within each resample. Throughout this procedure, models and comparison pairs were held fixed, including the strongest direct comparator identified from the unresampled mean errors. This approach quantified patient-sampling uncertainty for the saved fitted models while preserving comparisons on the same patients. Further details and results are provided in Supplementary Section~S8.

\paragraph{Use of generative AI.}\label{sec:aiuse}
The original manuscript was drafted by the authors. During subsequent research and manuscript development, ChatGPT (OpenAI) and Claude (Anthropic) assisted with code development and review, developing and refining evaluation and hyperparameter-search procedures, consistency checks between reported values and saved results, figure preparation, and manuscript revision. AI-drafted passages were produced under specific author instructions and independently reviewed and revised by the authors before inclusion in subsequent drafts. Code-related assistance included comparator implementations, hyperparameter-search workflows, and the patient-bootstrap analysis. The authors reviewed and checked the code used in the current experiments and analyzed the resulting outputs. The authors retained responsibility for methodological decisions, interpretation of results, and the final manuscript.

\section{Sampling design and computational error}\label{sec:sampling}
Once a model is trained, we are free to vary how the cohort of walks is constructed. We compare IID sampling, stratification, scrambled Sobol sequences, antithetic pairing, importance sampling, weighted branching, and conditional Monte Carlo. Appendix~\ref{app:samplers} summarizes the constructions and Supplementary Section~S5 spells them out.

The first four use different designs for the uniform random numbers that drive each walk, but all draw complete walks from the same learned conditional probabilities. Importance sampling changes the proposal to sample deaths more often and then weights each path to undo the change. Weighted branching and conditional Monte Carlo integrate the death and dropout probabilities analytically at each step and track only live particles, so for them the budget $M$ counts particles rather than complete observed walks.

All of these methods are trying to estimate the same thing, the infinite-sample learned-model functional $S_\theta$. With finitely many walks or particles, they can differ in systematic error, in variation between repetitions, and in cost. Writing $S^*$ for the true-generator reference, the triangle inequality gives
\begin{equation}\label{eq:triangle}
|\widehat S_{\theta,M}-S^*|\le |S_\theta-S^*|+|\widehat S_{\theta,M}-S_\theta|.
\end{equation}
The first term on the right is learned-model error and the second is numerical estimation error. They bound the total but do not decompose it exactly. Increasing $M$ can reduce numerical estimation error, but it does not reduce the error introduced by an inaccurately learned transition law.

The experiments reported here fix $M=10{,}000$ and repeat each stochastic prediction with five sampling seeds. Several limitations apply to the sampling designs. The standard Sobol balance guarantee applies at powers of two, so any advantage at 10,000 points is empirical~\citep{owen1998,owen2013}. Stratification and antithetic pairing need not reduce variance for every history-dependent outcome~\citep{glasserman2004}. Branching may duplicate ancestors, and equal particle weights after resampling do not make the resulting paths independent. Ordinary Kaplan--Meier confidence formulas assume independent observations and are therefore not justified for the correlated or particle designs.

Model forward passes use batches of 512. The alternative samplers recompute growing sequence contexts rather than fully reusing recurrent or Transformer caches, so measured runtimes describe these implementations and equal $M$ does not mean equal work. An earlier experiment that varied the IID budget from 500 to 10,000 is kept separately in Supplementary Section~S6.

\section{Experiment one, the single-covariate duration-dependent process}\label{sec:hardresults}
With the common Sobol sampler, ATLAS-Transformer has the lowest training-run mean error among the learned models at each post-baseline landmark. Its mean curve error is 78\% to 84\% lower than that of the strongest direct comparator. It also outperforms every fitted direct comparator at those landmarks in each of the nine training runs. The comparison has eleven learned models, eight direct models and three ATLAS architectures (GRU, RNN, and Transformer), each paired with each of the seven samplers for 21 architecture--sampler configurations. Table~\ref{tab:hardmain} reports every direct model, the three ATLAS architectures under the common Sobol sampler, and the true-generator oracle sampler. The full sampler ablation is in Supplementary Table~S7.

\begin{center}\begin{minipage}{\textwidth}\centering\small
\begin{tabular}{lrrrr}
\toprule
Model / method & $L=0$ & $L=20$ & $L=23$ & $L=24$ \\\midrule
\multicolumn{5}{l}{\textit{Direct models, covariates only}}\\
CPH & 0.038846 & 0.157917 & 0.211900 & 0.171741 \\
Weibull-MLP & 0.024253 & 0.137024 & 0.185766 & 0.150698 \\
DRSA (optimized) & 0.118561 & 0.146562 & 0.208210 & 0.157879 \\
DynamicDeepHit-Lite & 0.008556 & 0.142591 & 0.204751 & 0.155251 \\
SurvTransformer-Lite & 0.009877 & 0.142260 & 0.204242 & 0.156199 \\
\multicolumn{5}{l}{\textit{Direct models, covariates and observed history}}\\
Dynamic-DeepHit (optimized) & 0.028605 & 0.053743 & 0.059841 & 0.064791 \\
SurvTransformer & 0.018603 & 0.050899 & 0.085288 & 0.071506 \\
SurvTransformer-Discrete & 0.013296 & 0.046541 & 0.057495 & 0.066959 \\
\multicolumn{5}{l}{\textit{ATLAS, covariates and observed history, common Sobol sampler}}\\
ATLAS-GRU / Sobol & 0.007371 & 0.009763 & 0.013647 & 0.013617 \\
ATLAS-RNN / Sobol & \textbf{0.007196} & 0.013113 & 0.015888 & 0.011492 \\
ATLAS-Transformer / Sobol & 0.008818 & \textbf{0.009040} & \textbf{0.012525} & \textbf{0.010494} \\
\multicolumn{5}{l}{\textit{Control}}\\
Oracle sampler / IID & 0.002971 & 0.002955 & 0.003200 & 0.002973 \\
\bottomrule\end{tabular}
\captionof{table}{Single-covariate survival-curve MAE (lower is better). Neural-model entries are means across nine training runs. ATLAS rows use Sobol at $M=10{,}000$, first averaging five sampling seeds within each training run. Neural direct models contribute one deterministic prediction per training run. Between-run standard deviations, ranges and pooled sampling standard deviations are given in Supplementary Table~S2. CPH and the true-generator oracle are unchanged controls; the oracle row averages its five sampling seeds. Eligible counts are 300, 102, 78 and 57. Sampler-specific scores appear in Supplementary Table~S7.}\label{tab:hardmain}
\end{minipage}\end{center}

The gain is largest once the patient has history. At each of the three landmarks after baseline, all 21 ATLAS configurations have lower mean survival MAE than every fitted direct comparator in each of the nine training runs. The Transformer with Sobol has MAEs of 0.009040, 0.012525, and 0.010494 at $L=20,23,24$. The strongest direct comparator is SurvTransformer-Discrete at landmarks 20 and 23, with MAEs of 0.046541 and 0.057495, and Dynamic-DeepHit (optimized) at landmark 24, with MAE 0.064791. The corresponding reductions are 80.6\%, 78.2\%, and 83.8\%. ATLAS-Transformer with Sobol had lower MAE than the strongest mean-error direct comparator in all nine training runs at each post-baseline landmark. The corresponding one-sided exact sign test gives $p=0.002$ at each landmark, reported descriptively across training runs. The covariate-only direct models do much worse at these landmarks, as one would expect for a process whose late behavior is driven by run length and by the step-24 rule.

At baseline the gap is smaller and less uniform. ATLAS-RNN with Sobol has the lowest mean MAE, 0.007196, against 0.008556 for DynamicDeepHit-Lite, the strongest fitted direct comparator there, a 15.9\% reduction, with ATLAS-GRU close behind at 0.007371. Averaged across training runs, 14 of the 21 configurations have lower MAE than every direct comparator, namely all seven samplers for each of ATLAS-GRU and ATLAS-RNN. The baseline ranking is not stable run by run, though. With the sampler fixed at Sobol, ATLAS-RNN and ATLAS-Transformer each have lower error than the strongest direct comparator of that run in six of the nine runs, and ATLAS-GRU in five. We read this as a modest baseline advantage on average and no reliable per-run advantage. SurvTransformer-Discrete has mean MAE 0.013296 and SurvTransformer 0.018603 at baseline. Neither history-aware family receives an observed state prefix there, though their training objectives use longitudinal supervision in different ways.

Following validation-based optimization, Dynamic-DeepHit has mean MAEs of 0.028605, 0.053743, 0.059841, and 0.064791 at $L=0,20,23,24$, respectively. Its baseline error is lower than that of CPH (0.038846), while its post-baseline errors remain above those of the ATLAS configurations in Table~\ref{tab:hardmain}. DRSA (optimized), which uses covariates only, has baseline MAE 0.118561. The original history-aware Dynamic-DeepHit results are retained in Supplementary Section~S10.

SurvTransformer-Discrete has lower mean MAE than its continuous-likelihood sibling at all four landmarks and lower MAE in 28 of the 36 matched training-run--landmark comparisons. This is consistent with the integer-time setting, but repeated training alone does not isolate the effect of the likelihood.

The oracle IID sampler has mean MAEs of approximately 0.0030 to 0.0032, below the mean errors of all fitted configurations. It samples from the true generator and still has finite-sampling error at $M=10{,}000$. It is therefore a sampling control, showing the accuracy ATLAS would reach at this budget if it had learned the transition law exactly.

Per-run scores, error distributions and comparisons of between-run and sampling variation are reported in Supplementary Section~S3, Tables~S2 and~S3. Supplementary Figures~S1 and~S2 show the complete model--sampler comparisons at baseline and landmark 20, including the individual training-run scores.

\section{Experiment two, multi-covariate competing risks}\label{sec:crresults}
The pattern repeats here (Table~\ref{tab:cr_survival}). ATLAS-GRU with the common Sobol sampler had lower MAE than every fitted direct comparator for survival and both cause-specific incidence curves at all three landmarks in each of the five training runs. In the five-run means, its survival error was 62\% and 71\% lower than that of the strongest direct comparator at landmarks 10 and 20, respectively. The experiment scores survival and the two CIFs separately. It has ten learned models, the same eight direct models and two ATLAS architectures (GRU and Transformer), each paired with all seven samplers for 14 configurations. The generator has no hard rule and no duration dependence, so this experiment tests a different capability rather than a controlled modification of the first.

\begin{center}\begin{minipage}{\textwidth}\centering\small
\begin{tabular}{lrrr}
\toprule
Model / method & $L=0$ & $L=10$ & $L=20$ \\\midrule
\multicolumn{4}{l}{\textit{Direct models, covariates only}}\\
CPH & 0.054183 & 0.115799 & 0.093079 \\
Weibull-MLP & 0.015510 & 0.089976 & 0.085369 \\
DRSA (optimized) & 0.083310 & 0.095651 & 0.077448 \\
DynamicDeepHit-Lite & 0.022706 & 0.091415 & 0.077370 \\
SurvTransformer-Lite & 0.023853 & 0.090584 & 0.076661 \\
\multicolumn{4}{l}{\textit{Direct models, covariates and observed history}}\\
Dynamic-DeepHit (optimized) & 0.016242 & 0.023061 & 0.022753 \\
SurvTransformer & 0.025686 & 0.040604 & 0.050470 \\
SurvTransformer-Discrete & 0.025778 & 0.040464 & 0.048941 \\
\multicolumn{4}{l}{\textit{ATLAS, covariates and observed history, common Sobol sampler}}\\
ATLAS-GRU / Sobol & \textbf{0.007902} & \textbf{0.008825} & \textbf{0.006704} \\
ATLAS-Transformer / Sobol & 0.010378 & 0.011727 & 0.008094 \\
\multicolumn{4}{l}{\textit{Control}}\\
Oracle sampler / IID & 0.002556 & 0.002407 & 0.001681 \\
\bottomrule\end{tabular}
\captionof{table}{Overall survival MAE in the multi-covariate competing-risks experiment. Neural-model entries are means across five training runs. ATLAS rows use Sobol at $M=10{,}000$, first averaging five sampling seeds within each training run. Training-run standard deviations, ranges and individual-run scores appear in Supplementary Tables~S4 and~S5. CPH predictions agree across runs to numerical precision; the oracle is a reused true-generator sampling control. The eligible counts are 300, 251 and 179.}\label{tab:cr_survival}
\end{minipage}\end{center}

Averaged across five training runs, all 14 ATLAS configurations have lower survival MAE than every fitted direct comparator at $L=0,10,20$. ATLAS-GRU with Sobol has mean survival MAE 0.007902, 0.008825 and 0.006704 at the three landmarks. The strongest fitted direct comparator is Weibull-MLP at baseline (0.015510) and Dynamic-DeepHit (optimized) at landmarks 10 and 20 (0.023061 and 0.022753). The corresponding reductions are 49.1\%, 61.7\% and 70.5\%.

\begin{center}\begin{minipage}{\textwidth}\centering\small
\begin{tabular}{lrrr}
\toprule
Model / method & $L=0$ & $L=10$ & $L=20$ \\\midrule
\multicolumn{4}{l}{\textit{Direct models, covariates only}}\\
CPH & 0.032316 & 0.101754 & 0.074283 \\
Weibull-MLP & 0.011940 & 0.091698 & 0.075029 \\
DRSA (optimized) & 0.051295 & 0.095362 & 0.068731 \\
DynamicDeepHit-Lite & 0.018637 & 0.092973 & 0.069686 \\
SurvTransformer-Lite & 0.019176 & 0.092291 & 0.069398 \\
\multicolumn{4}{l}{\textit{Direct models, covariates and observed history}}\\
Dynamic-DeepHit (optimized) & 0.011544 & 0.017813 & 0.016223 \\
SurvTransformer & 0.021575 & 0.032789 & 0.036171 \\
SurvTransformer-Discrete & 0.021151 & 0.031637 & 0.036103 \\
\multicolumn{4}{l}{\textit{ATLAS, covariates and observed history, common Sobol sampler}}\\
ATLAS-GRU / Sobol & \textbf{0.005878} & \textbf{0.007051} & \textbf{0.004830} \\
ATLAS-Transformer / Sobol & 0.007090 & 0.008418 & 0.005454 \\
\multicolumn{4}{l}{\textit{Control}}\\
Oracle sampler / IID & 0.002225 & 0.002071 & 0.001366 \\
\bottomrule\end{tabular}
\captionof{table}{Death A cumulative incidence MAE in the multi-covariate competing-risks experiment. Neural-model entries are means across five training runs. ATLAS rows use Sobol at $M=10{,}000$, first averaging five sampling seeds within each training run. Training-run standard deviations, ranges and individual-run scores appear in Supplementary Tables~S4 and~S5. CPH predictions agree across runs to numerical precision; the oracle is a reused true-generator sampling control. The eligible counts are 300, 251 and 179. Each CIF is the probability of that cause occurring before the competing cause, after the prediction time.}\label{tab:cr_cause_a}
\end{minipage}\end{center}
\begin{center}\begin{minipage}{\textwidth}\centering\small
\begin{tabular}{lrrr}
\toprule
Model / method & $L=0$ & $L=10$ & $L=20$ \\\midrule
\multicolumn{4}{l}{\textit{Direct models, covariates only}}\\
CPH & 0.027520 & 0.056275 & 0.040708 \\
Weibull-MLP & 0.006542 & 0.045594 & 0.033452 \\
DRSA (optimized) & 0.032559 & 0.046644 & 0.032548 \\
DynamicDeepHit-Lite & 0.008993 & 0.044257 & 0.032283 \\
SurvTransformer-Lite & 0.008929 & 0.044021 & 0.032284 \\
\multicolumn{4}{l}{\textit{Direct models, covariates and observed history}}\\
Dynamic-DeepHit (optimized) & 0.009911 & 0.012385 & 0.010620 \\
SurvTransformer & 0.012421 & 0.020794 & 0.021891 \\
SurvTransformer-Discrete & 0.012653 & 0.020366 & 0.020970 \\
\multicolumn{4}{l}{\textit{ATLAS, covariates and observed history, common Sobol sampler}}\\
ATLAS-GRU / Sobol & \textbf{0.004940} & \textbf{0.004299} & \textbf{0.003265} \\
ATLAS-Transformer / Sobol & 0.006211 & 0.006218 & 0.004700 \\
\multicolumn{4}{l}{\textit{Control}}\\
Oracle sampler / IID & 0.001713 & 0.001528 & 0.001024 \\
\bottomrule\end{tabular}
\captionof{table}{Death B cumulative incidence MAE in the multi-covariate competing-risks experiment. Neural-model entries are means across five training runs. ATLAS rows use Sobol at $M=10{,}000$, first averaging five sampling seeds within each training run. Training-run standard deviations, ranges and individual-run scores appear in Supplementary Tables~S4 and~S5. CPH predictions agree across runs to numerical precision; the oracle is a reused true-generator sampling control. The eligible counts are 300, 251 and 179. Each CIF is the probability of that cause occurring before the competing cause, after the prediction time.}\label{tab:cr_cause_b}
\end{minipage}\end{center}

The same ordering holds for each cause-specific target (Tables~\ref{tab:cr_cause_a} and~\ref{tab:cr_cause_b}). Averaged across five training runs, every ATLAS configuration has lower CIF MAE than every fitted direct comparator at every tested landmark. At baseline, ATLAS-GRU with Sobol has mean MAE 0.005878 for Death A and 0.004940 for Death B. The strongest direct comparators are Dynamic-DeepHit (optimized) for Death A (0.011544) and Weibull-MLP for Death B (0.006542), giving reductions of 49.1\% and 24.5\%, respectively. At landmarks 10 and 20, the Death A errors are 0.007051 and 0.004830 and the Death B errors are 0.004299 and 0.003265. Relative to Dynamic-DeepHit (optimized), the strongest direct comparator for all four settings, the reductions are 60.4\% and 70.2\% for Death A and 65.3\% and 69.3\% for Death B. A good all-cause survival curve does not by itself guarantee a correct split between causes, and the separate CIF scores check that split directly.

At the individual-run level, ATLAS-GRU with Sobol has lower error than the strongest fitted direct comparator in all five runs for every landmark and target, 45 of 45 settings. The 45 settings comprise nine landmark--target comparisons, each evaluated across five training runs. For each landmark--target comparison, ATLAS-GRU with Sobol had lower MAE than the strongest fitted direct comparator in all five training runs, giving a one-sided exact sign-test $p=0.031$. Because the comparisons share patients and fitted models, the 45 settings are not independent replicates; these comparison counts and tests are descriptive summaries. All 14 configurations had lower MAE than every fitted direct comparator in 43 of the 45 run--landmark--target settings. The two exceptions are training seed 45 at baseline for survival and Death B, where Weibull-MLP outperforms the Transformer configurations while every GRU configuration retains the advantage.

The true-generator sampler remains more accurate than the fitted models for all nine landmark--target combinations.

Training-run standard deviations, ranges and individual-run comparisons are reported in Supplementary Section~S3, Tables~S4 and~S5.

Across the 18 architecture--landmark--target comparisons, Sobol had lower five-run mean MAE than IID in every case. Conditional Monte Carlo has the lowest mean MAE in 17 of the 18 comparisons. For ATLAS-GRU, its largest improvement over Sobol is approximately 0.000112 MAE, for Death B at baseline. These GRU differences are much smaller than its gaps to the strongest direct comparators. Sobol remains the common sampler for the main comparisons, while rankings among all seven samplers are reported as exploratory. Full numerical scores are supplied with the source files, and the survival sampler tables are in Supplementary Table~S8.

Supplementary Figures~S3 and~S4 show the full model--sampler comparisons at baseline and landmark 20, including individual training-run scores for all three targets.

\begin{figure}[htbp]
\centering\includegraphics[width=\textwidth]{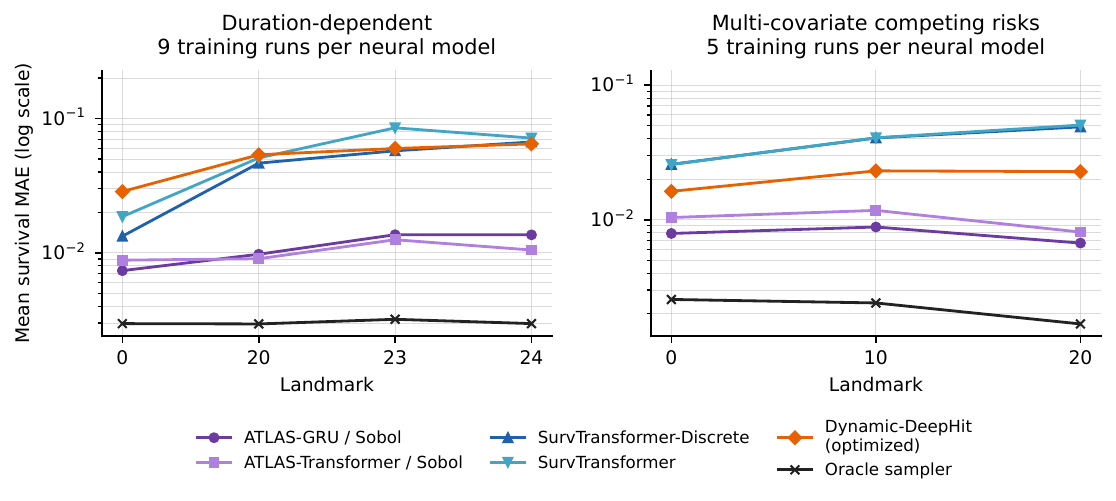}
\caption{Mean survival-curve MAE for ATLAS-GRU and ATLAS-Transformer with the common Sobol sampler and the three history-aware direct comparators. Dynamic-DeepHit (optimized) uses the validation-selected configurations described in Supplementary Section~S1. Neural-model values average nine single-covariate training runs in the left panel and five competing-risks training runs in the right panel; each ATLAS training run first averages five sampling seeds at $M=10{,}000$. The oracle curves are reused sampling controls, not independently fitted models. The vertical axes are logarithmic and landmarks are equally spaced for readability. The lines connect landmark-specific scores, not survival probabilities; both the eligible population and the prediction window change with the landmark.}\label{fig:overview}
\end{figure}

\subsection{Paired patient-bootstrap comparisons}\label{sec:bootstrapresults}
The paired patient-bootstrap analysis supported the post-baseline comparisons. Specifically, at every evaluated post-baseline landmark and target, the pointwise 95\% intervals for the MAE difference favored each ATLAS architecture using the common Sobol sampler over all eight fitted direct comparators. At baseline, the intervals also favored ATLAS, with two exceptions that included zero: ATLAS-Transformer versus DynamicDeepHit-Lite in the duration-dependent experiment, and ATLAS-Transformer versus Weibull-MLP for Death B cumulative incidence in the competing-risks experiment. For the headline ATLAS-GRU comparisons in the competing-risks experiment, intervals favored ATLAS for survival and both cumulative incidence functions at all three landmarks (Supplementary Section~S8).

\section{Discussion}
\label{sec:discussion}
The two experiments point the same way (Figure~\ref{fig:overview}). ATLAS had lower MAE than the fitted direct comparators at post-baseline landmarks in both benchmarks, including comparisons with direct models that received the same observed history. This supports autoregressive trajectory generation as a way to estimate individual survival and cause-specific incidence curves. The duration-dependent experiment tests how predictions account for time spent in a state and changes in transition rules. The competing-risks experiment tests whether the model can estimate both overall survival and the allocation of deaths between causes. The smaller and less consistent baseline advantage in the duration-dependent experiment shows that the benefit depends on the prediction setting.

Matching the inputs does not match architecture, training objective, supervision, capacity, or optimization, so these experiments compare complete fitted procedures. All methods used the same cohort, patient-level split and eligible test patients, were scored against the same exact reference curves, and saw only information available at the prediction time. The three history-aware direct models received the same observed prefix as ATLAS, so the comparison holds patient information fixed, but it does not isolate the contribution of sampling itself.

An astute reader may ask whether a restricted model, given only the information relevant to these data-generating processes, would outperform larger models that must learn their own representations of patient history. It can. A multinomial logistic regression (MLR) transition model is, in effect, an autoregressive next-state model like ATLAS whose learned history encoder has been replaced by a fixed feature map. When those features are exactly the ones the generator uses, the smaller model avoids the cost of learning them. Its training problem is also far simpler: with roughly a thousand coefficients and a convex, ridge-penalized objective that has a unique optimum, it can be fitted reliably in a single run and would be expected to need less data, whereas the neural models must solve a much larger non-convex optimization problem.

In exploratory analyses (Supplementary Section~S9), a history-aware MLR model (HAMMLR) supplied with baseline covariates, current state, run length and calendar time outperformed ATLAS on both synthetic experiments. A Markov MLR model (MMLR) supplied with baseline covariates and current state did so in four of nine competing-risks settings, where those inputs are sufficient. These comparisons benefit from analyst-supplied history summaries and direct probability propagation. When duration and time were omitted on the duration-dependent experiment, MMLR's post-baseline error was roughly 21 to 25 times that of the best ATLAS configuration, although differences in feature bases and regularization prevent attributing the full gap to history omission alone. We report these structurally informed models separately because the relevant history summary is generally unknown in clinical sequences, and because the explicit origin-specific implementations examined here can become computationally and statistically demanding as the state space and historical representation expand.

Sampling budget and learning capacity answer different questions. Increasing $M$ can reduce numerical estimation error, but it does not reduce the error introduced by an inaccurately learned transition law. The oracle sampler illustrates the potential accuracy of ATLAS if the transition distributions were learned perfectly. Sampler rankings vary with model, landmark, and target, so no single sampler is best in every setting. Because sampling is separate from sequence-model training, its design and budget can be changed without retraining. The fixed-checkpoint resource study in Supplementary Section~S6 examines this numerical component.

There are limits to the evidence. The later-landmark advantage persisted across repeated training runs, while baseline rankings in the duration-dependent experiment varied. The one-sided sign tests describe paired training-run comparisons on the fixed datasets; they do not measure variation across new datasets. Within each experiment, the runs share one generated cohort, split, and evaluation set, and hardware and software versions differed across some runs (Supplementary Section~S1). Dynamic-DeepHit (optimized) and DRSA (optimized) select checkpoints using landmark-averaged validation negative log-likelihood, while the two SurvTransformer variants use their combined validation losses. Selection did not use the known-generator reference-curve MAE. Dynamic-DeepHit and DRSA received the additional validation-based searches described in Supplementary Section~S1; the other model configurations were retained. The comparisons therefore concern these complete fitted procedures.

Validation-based optimization substantially reduced Dynamic-DeepHit's curve errors, providing a stronger history-aware comparator. Each ATLAS architecture using the common Sobol sampler retained lower mean MAE than Dynamic-DeepHit (optimized) at baseline and every evaluated landmark, for survival and both competing-risk CIFs. The search was motivated by the initial comparison results and used observed validation outcomes for selection; the original results are retained in Supplementary Section~S10. Because architecture, regularization, loss weights and checkpoint selection changed together, the improvement describes a revised fitted procedure rather than an isolated effect of checkpoint selection. Baseline prediction does not exercise Dynamic-DeepHit's principal longitudinal advantage, and these synthetic experiments do not establish a general limitation of the published method.

For DRSA (optimized), baseline curve MAEs of 0.119 and 0.083 (Tables~\ref{tab:hardmain} and~\ref{tab:cr_survival}) may partly reflect its composite training objective: the event-status term, weighted by 0.5 in our implementation (Supplementary Section~S1.2), emphasizes status prediction at observed follow-up times, whereas curve MAE assesses probability accuracy across the full evaluation grid. Good concordance therefore need not imply low curve MAE.

The paired patient bootstrap complements the repeated-training summaries by quantifying patient-sampling uncertainty around the mean performance differences. In particular, the pointwise 95\% intervals consistently favored ATLAS at post-baseline landmarks, supporting its average performance advantage under resampling of the evaluated patients. Together, these analyses describe both variation across training runs and the precision of the average performance comparisons across patients.

The two experiments also carry different observation interpretations. The single-covariate study scores the observed-generator Kaplan--Meier functional (Section~\ref{sec:truth}), whereas the competing-risks study specifies independent censoring and so has biological reference curves under that synthetic construction. A Kaplan--Meier estimator applied to simulated paths does not, on its own, license conclusions about latent survival, and reference-curve MAE is not an outcome-based calibration, concordance, or Brier-score result.

Each patient receives a separate generated cohort and curve. Patients with the same supplied information share the same modeled distribution, and increasing the sampling budget does not recover distinctions the model was never given. Simulated walks are not independent clinical observations, and a larger $M$ improves numerical resolution without reducing the training uncertainty of the underlying model or establishing clinical calibration.

The healthcare application we have in mind is dynamic prognosis. Update the possible futures as observations arrive, then summarize event timing and competing outcomes. That would represent longitudinal records as time-indexed clinical states or tokens and train an autoregressive generator, possibly a large language model, on a patient's available record, as ETHOS and Delphi-2M already do at scale~\citep{renc2024,shmatko2025}. It would require explicit mappings from tokens to times and events, event definitions, visit and dropout mechanisms, and a way of handling incomplete records, before generated records can be reduced to valid event and censoring times. Assessing clinical usefulness would then take external and temporal validation, calibration under the intended observation process, evaluation across patient groups, and comparison with direct models given the same clinical history. The small models used here are not clinical language models, and a drug-associated recovery transition in a benchmark is not evidence of treatment efficacy. Changing a treatment token in an observationally trained model does not by itself identify an intervention effect, and a causal claim would require a separate estimand and supporting design.

The practical promise of ATLAS is a prognosis rebuilt from each patient's own record. Every new observation changes the futures the model generates, and the curve follows. In these benchmarks the advantage was largest where the record was longest. Whether it holds on clinical records is the question for the next study.

\section{Conclusion}
We have presented ATLAS, a framework that trains a sequence model on a cohort, runs autoregressive random walks forward from one patient's covariates and observed history, and estimates survival or competing-event incidence from the resulting synthetic cohort of futures. The framework does not require a specified parametric form for the hazard, since survival is estimated directly from the generated walks. Its predictions still depend on the sequence model's assumptions and learned transition probabilities. We have also proposed stochastic finite-state automata as a flexible benchmark for survival models, with the useful property that the true conditional survival functional can be computed exactly for any patient at any landmark.

On two such benchmarks, one duration-dependent with a late rule change and one Markov with competing risks and drug-dependent recovery, ATLAS-Transformer in the first experiment and ATLAS-GRU in the second, both using Sobol sampling, had lower full-curve MAE than every fitted direct comparator at every evaluated post-baseline landmark in every training run. At later landmarks, reductions in training-run mean survival-curve MAE relative to the strongest fitted direct comparator ranged from 62\% to 84\%. These comparisons included validation-optimized Dynamic-DeepHit and history-aware SurvTRACE adaptations given the same observed history as ATLAS. At baseline the advantage was smaller and, in the first experiment, not consistent across training runs. Nine training runs per neural model support the first comparison and five support the second, and all data remain synthetic. The gap to the true-generator oracle leaves room for better predictive models of the next-state distribution to lower the error further. Validation on clinical records remains necessary. Together, ATLAS and the SFSA benchmarks provide a way to generate possible futures from observed patient histories and evaluate the resulting survival and competing-risk curves against exact references in controlled experiments.

\section*{Declarations}
\paragraph{Human participants and patient data.}
This study used only simulated data and did not involve human participants or identifiable patient records.

\paragraph{Data and code availability.}
All data in this paper are synthetic and are generated by the code in the accompanying package, which also contains the archived predictions, per-run metrics, checkpoint hashes, and the scripts that produce every table and figure. The ATLAS repository contains the package and instructions for reproducing the results: \url{https://github.com/mkngqx-ctrl/ATLAS}.

\paragraph{Author contributions.}
\textbf{Morgan Nissen:} Conceptualization, methodology, software, validation, formal analysis, investigation, data curation, writing---original draft, and writing---review and editing. Contributions included designing the current SFSA experiments to examine different model behaviours; evaluating ATLAS architectures and sampling methods; adapting contemporary comparators; developing and checking the current codebase with AI assistance; investigating Dynamic-DeepHit and DRSA performance and conducting their hyperparameter searches; and curating the training, validation, test and prediction outputs. 
\textbf{Chunyang Li:} Conceptualization, methodology, software, formal analysis, investigation, writing---original draft, and writing---review and editing, including development of the original ATLAS implementation and its initial experiments. 
\textbf{Seth Glazier:} Conceptualization, methodology, and writing---review and editing, including the foundational trajectory-sampling approach developed in his thesis. 
\textbf{Julia Bohman:} Writing---review and editing. All authors contributed to manuscript revision.
\textbf{Ahmad Halwani:} Supervision and writing---review and editing, including clinical consultation and oversight of the current iteration. 
\textbf{Jeffrey Humpherys:} Conceptualization, methodology, supervision, writing---original draft, and writing---review and editing, including supervision of the foundational thesis, the initial ATLAS work and the current research and manuscript development. 
\paragraph{Funding.}
No specific institutional funding was received for this research.
\paragraph{Competing interests.}
The authors declare that they have no competing interests.
\paragraph{Disclaimer.}
The views expressed in this article are those of the authors and do not necessarily reflect the position or policy of the Department of Veterans Affairs or the United States government.
\paragraph{Use of generative AI.}
AI-assisted research and manuscript-preparation activities are described in Section~\ref{sec:evaluation}. The authors take responsibility for the final manuscript.

\clearpage
\appendix
\section{Generator specifications for the two experiments}\label{app:generator}
This appendix gives the generative laws used by the result archives. The single-covariate generator matches the source hashes stored with its run, and the competing-risks generator is taken from the source snapshot inside its archive. These are two distinct data-generating processes. The second is not obtained by relabeling the first one's deaths.

\subsection{Single-covariate initialization and base matrix}
The state order is $(\Healthy,\Sick,\Sicker,\Dead,\Cens)$, meaning Healthy, Sick, Sicker, Death and dropout. For a patient with fixed covariate $x$, set
\[
f=F(x)=0.2+1.5(x/10)^4e^{-(x/10)^5},\qquad p=0.003,
\]
with $X\sim\operatorname{TruncNormal}(10,1.5^2;[7,13])$. The unmodified transition matrix is
\begin{equation}\label{eq:matrix}
T(f,p)=\begin{pmatrix}
f-p/3-0.001 &(1-f)/2-p/3 &(1-f)/2-p/3 &0.001&p\\
2f/3-p/3 &(1-f)/3-p/3-0.002 &(2-f)/3-p/3 &0.002&p\\
0 &f/8-2p/3+0.306 &0.69-f/8-p/3 &0.004&p\\
0&0&0&1&0\\0&0&0&0&1
\end{pmatrix}.
\end{equation}
Every row sums to one. All entries are nonnegative over the declared covariate domain. At step 1 the state is sampled from the live distribution $(2/3,1/4,1/12)$, independently of $x$, and the run length is one. This initial draw is not a transition from Eq.~\eqref{eq:matrix}; no death or dropout occurs at step 1.

\subsection{Duration-dependent modification and the two clock changes}
Before a transition ending at absolute step $t\in\{2,\ldots,30\}$, let $s$ be the source state and $u$ its current run length. The source is Sick or Sicker for the adjustments in the following table.
\begin{center}\small
\begin{tabular}{llllrrr}\toprule
Source & Boosted destination & Donor destinations & Donor mass & $\lambda$ & $\nu$ & $\kappa$\\\midrule
Sick & Sicker & Healthy, Sick & $\Omega_{\Sick}$ & 0.35 & 2.5 & 0.90\\
Sicker & Death & Sick, Sicker & $\Omega_{\Sicker}$ & 0.97 & 3 & 0.99\\\bottomrule
\end{tabular}
\end{center}
The donor mass $\Omega_s$ is the sum of the corresponding entries in the unmodified row $T_{s,\cdot}$. Define $\rho(t)=5$ for $t<20$ and $\rho(t)=3$ for $t\ge20$, and calculate $\Delta_s$ from Eq.~\eqref{eq:ramp}. For each donor destination $r$ and boosted destination $\tau(s)$, the adjusted row is
\begin{align}
T^{\rm sm}_{s,r}(u,t)&=T_{s,r}(1-\Delta_s/\Omega_s),\label{eq:donors}\\
T^{\rm sm}_{s,\tau(s)}(u,t)&=T_{s,\tau(s)}+\Delta_s.
\end{align}
All other entries remain unchanged. If $\Omega_s=0$, set $\Delta_s=0$ and leave the row unchanged. The amount removed from donors equals the amount added to the worsening transition. Because $\Delta_s\le.95\Omega_s$, donor entries remain nonnegative. Healthy has no duration adjustment, and no adjustment changes the dropout entry.

For example, longer consecutive Sick runs redirect probability away from recovery and continued sickness toward Sicker. Longer Sicker runs redirect probability away from recovery and continued Sicker status toward death. The cap can make two different run lengths yield the same transition row. The step-20 change strengthens the uncapped duration effect without creating an additional deterministic death threshold.

Draw a proposed outcome from the adjusted row. Beginning at step 24, apply Eq.~\eqref{eq:hard24}, so that a proposal to remain Sicker when already Sicker is replaced by Death. Equivalently, the final row entries satisfy
\[
T^{\rm final}_{\Sicker,\Dead}=T^{\rm sm}_{\Sicker,\Dead}+T^{\rm sm}_{\Sicker,\Sicker},
\qquad T^{\rm final}_{\Sicker,\Sicker}=0
\quad(t\ge24).
\]
The rule is applied after the proposal, and only to that proposal. A proposed dropout remains dropout, and an already absorbed patient stays absorbed. Among live outcomes, an unchanged state increases the run length by one and a changed state resets it to one. There is no additional hard run-length threshold of three or five in this experiment; those values enter the duration ramp through $\rho(t)$.

The observed endpoint is the first Death or dropout step. Live patients remaining after step 30 are administratively censored at 30. Death and dropout are mutually exclusive categorical outcomes of the same draw. Setting $p=0$ also changes the live entries of Eq.~\eqref{eq:matrix}, so it is not the same thing as removing an independent observation mechanism. This is why the reference target for this experiment is explicitly the observed-generator Kaplan--Meier functional.

\subsection{Multi-covariate competing-risks initialization}
Use the order
\[
(H,A,B,AB,V_A,V_B,V_{AB},D_A,D_B,C),
\]
where $H$ denotes Healthy; $A,B,AB$ denote the three Sick states; $V_A,V_B,V_{AB}$ denote their Sicker counterparts; $D_A,D_B$ are the death causes; and $C$ is dropout. These symbols are state labels rather than covariates.

The covariates $x_c$ (calcium), $x_a$ (age) and $x_d$ (drug) are independent uniform draws on $[8,13]$, $[30,90]$ and $[0,1]$. They remain fixed within a patient. Define three dimensionless effects
\begin{equation}\label{eq:crcovariates}
z=\left(\frac{x_c-8}{5}\right)^{2.5},\qquad
a=\frac{\max(x_a-50,0)}{40},\qquad d=x_d.
\end{equation}
All three lie in $[0,1]$. Calcium uses a bounded power response. Age has a ReLU-like threshold at 50. Drug changes the two specified recovery probabilities and is not a time-varying treatment assignment.

The initial biological probabilities at step 1 are
\begin{align*}
\pi_A&=0.08+0.10z,&\pi_B&=0.08+0.10a,&\pi_{AB}&=0.025+0.05za,\\
\pi_{V_A}&=0.01+0.02z,&\pi_{V_B}&=0.01+0.02a,&\pi_{V_{AB}}&=0.005+0.015za.
\end{align*}
Healthy receives the remainder $\pi_H=1-\sum_{s\ne H}\pi_s$. Both death probabilities are zero at initialization. Unlike the first experiment, independent observation dropout can occur at step 1 after the live state is drawn.

\subsection{Complete competing-risks transition matrix}
Let $P(x)$ be the biological transition matrix on the seven live states and two death states, excluding dropout. Its block form is
\begin{equation}\label{eq:crmatrix}
P(x)=\begin{pmatrix}
Q(x)&r_A(x)&r_B(x)\\
0_{1\times7}&1&0\\
0_{1\times7}&0&1
\end{pmatrix},
\end{equation}
where $Q$ is $7\times7$ and $r_A,r_B$ are $7\times1$ death-exit vectors. The following table specifies every nonzero off-diagonal entry. All unlisted off-diagonal entries are zero, and each live diagonal is one minus the sum of its row's listed entries, so the table and the absorbing rows completely specify the matrix.

\begin{center}\begin{minipage}{\textwidth}\centering\small
\begin{tabular}{lll}\toprule
From & To & Biological transition probability\\\midrule
$H$ & $A$ & $0.02+0.10z+0.025a$\\
$H$ & $B$ & $0.02+0.08a+0.01z$\\
$H$ & $AB$ & $0.005+0.02za$\\
$A$ & $H$ & $0.08/(1+z+0.25a)$\\
$A$ & $V_A$ & $0.04+0.20z+0.05a$\\
$A$ & $AB$ & $0.025+0.065a$\\
$B$ & $H$ & $0.08/(1+a+0.25z)$\\
$B$ & $V_B$ & $0.04+0.18a+0.04z$\\
$B$ & $AB$ & $0.025+0.06z$\\
$AB$ & $A$ & $0.025$\\
$AB$ & $B$ & $0.025$\\
$AB$ & $V_{AB}$ & $0.05+0.15z+0.12a$\\
$V_A$ & $A$ & $0.09$\\
$V_A$ & $V_{AB}$ & $0.025+0.055a$\\
$V_A$ & $D_A$ & $0.025+0.14z+0.08a$\\
$V_B$ & $B$ & $0.06+0.35d$\\
$V_B$ & $V_{AB}$ & $0.025+0.055z$\\
$V_B$ & $D_B$ & $0.02+0.10a+0.03z$\\
$V_{AB}$ & $AB$ & $0.04+0.30d$\\
$V_{AB}$ & $D_B$ & $b=0.02+0.06a+0.025z$\\
$V_{AB}$ & $D_A$ & $1.8b+0.03z+0.015$\\\bottomrule
\end{tabular}
\captionof{table}{Complete off-diagonal specification of the competing-risks biological matrix. The scalar $b$ in the last two rows is a transition probability, distinct from the observed-mass vector $b_j$ used in the KM reference.}
\end{minipage}\end{center}

Increasing drug level transfers mass from the self-transition to recovery in the $V_B$ and $V_{AB}$ rows, leaving their immediate death entries unchanged. From $V_{AB}$, the Death A minus Death B probability is $0.8b+0.03z+0.015>0$. This local ordering does not force Death A's cumulative incidence to exceed Death B's for every patient, because patients can visit different states.

There is no dependence on run length or absolute time in this matrix. The matrices remain valid throughout the specified covariate domain. In particular, every live row includes a nonnegative residual self-transition, and $Q\ind+r_A+r_B=\ind$.

\subsection{Competing-risks observation matrix and endpoint extraction}
Set $r=0.008$. After a biological transition, record a death immediately if one occurs. For a live outcome, apply an independent dropout draw with probability $r$. The complete observed matrix on all ten states is
\begin{equation}\label{eq:observedcr}
P_{\rm obs}(x)=\begin{pmatrix}
(1-r)Q&r_A&r_B&rQ\ind\\
0_{1\times7}&1&0&0\\
0_{1\times7}&0&1&0\\
0_{1\times7}&0&0&1
\end{pmatrix}.
\end{equation}
Each live destination probability is thus scaled by $1-r$, death probabilities are unchanged, and the dropout probability is $r$ times the probability of a live biological outcome. Every row sums to one. The observed initialization is $((1-r)\pi,0,0,r)$, with $\pi$ the seven-state live distribution. There is still no death at step 1.

Death A, Death B and dropout are absorbing. The first terminal step supplies the observed duration and cause; live patients at step 30 are administratively censored. Cause zero denotes censoring, not a third death cause. This construction specifies independent censoring with death recorded before same-time censor removal, so the biological recursion and the observed-process population Kaplan--Meier and Aalen--Johansen curves agree. That identification is specific to this experiment and is not asserted for the first generator.

\subsection{Cohort generation and archived provenance}
Both cohorts have 50,000 independently generated patients. Cohort seeds are 24001 for the single-covariate process and 34001 for the competing-risks process; split seeds are 24002 and 34002. Both splits contain 40,000 training, 5,000 validation and 5,000 test patients. The first archive reuses the original cohort and older model predictions while adding three newly fitted history-aware comparators. The original competing-risks archive supplies its own cohort, model checkpoints and generator snapshot. The expanded analyses retain the same cohort and split within each experiment while adding independently trained neural checkpoints. The original \texttt{results\_20260919} records are preserved; separately named multi-seed supplements contain per-run scores, training and rollout variability, source hashes and runtime provenance. An accompanying \texttt{supplementary/MULTISEED\_MANIFEST.json} maps manuscript tables and figures to their supporting files.

\section{Conditional reference algorithms}\label{app:truth}
\subsection{Observed-generator KM for the single-covariate experiment}
The augmented state is $(s,u)$ with health state $s\in\{\Healthy,\Sick,\Sicker\}$ and current run length $u\in\{1,\ldots,30\}$. We construct one kernel for every destination time 2 through 30, including the duration ramp, the step-20 change, and the step-24 override. For a live destination $r$, the new run length is $u+1$ if $r=s$, otherwise one. The finite 30-step horizon makes larger durations unreachable; representing them is unnecessary.

From each kernel we extract the live matrix $Q_j$, death column $d_j$, and dropout column $c_j$ for the transition from $j$ to $j+1$, checking nonnegativity and $Q_j\ind+d_j+c_j=\ind$ for every row. We start from the normalized initial distribution or from the unit vector at the observed landmark augmented state, propagate $b_{j+1}=b_jQ_j$, and calculate the Kaplan--Meier update in Eq.~\eqref{eq:kmreference}. For a landmark $L$ the first future kernel ends at $L+1$, not at $L$, so the hard rule first changes $Q_{23}$ and $d_{23}$.

For positive risk mass, define the death and dropout fractions
\[
h_j=\frac{b_jd_j}{b_j\ind},\qquad g_j=\frac{b_jc_j}{b_j\ind}.
\]
Writing $A_j=b_j\ind$ yields
\[
A_{j+1}=A_j(1-h_j-g_j),\qquad S^*_{\rm KM}(j+1)=S^*_{\rm KM}(j)(1-h_j).
\]
With the same normalized starting mass, $A_j\le S^*_{\rm KM}(j)$. Without dropout they coincide. The hazard ratios are invariant to a common positive rescaling of $b_j$, which can be useful for numerical stability, but rescaled mass must not be interpreted as the original probability of remaining observed. An exactly empty risk set has no further defined conditional death fraction; the implementation retains the final product-limit value.

\subsection{Biological reference for the competing-risks experiment}
For fixed covariates, let $v_j$ be a $1\times7$ row vector of biological live-state probabilities, and use $Q,r_A,r_B$ from Eq.~\eqref{eq:crmatrix}. Accumulate the death masses using
\begin{equation}\label{eq:dp}
v_{j+1}=v_jQ,\qquad
F_k(j+1)=F_k(j)+v_jr_k,\qquad S(j)=v_j\ind.
\end{equation}
At baseline, initialize $v_1=\pi$ and both CIFs to zero, with $S(0)=S(1)=1$. At a landmark, initialize $v_L$ as the unit vector at the observed live state and both future CIFs to zero. All future times use the same covariate-specific matrix because there are no clock changes or duration effects.

The observed live-state distribution is a common scalar multiple of $v_j$, because the independent dropout factor is identical for every surviving biological path of the same length. It therefore cancels from the observed death-to-risk ratios. Since biological deaths are recorded before dropout on each transition, the observed-process population Kaplan--Meier and Aalen--Johansen updates equal Eq.~\eqref{eq:dp}. This argument relies on the specific observation matrix in Appendix~\ref{app:generator} and should not be carried over to the single-covariate experiment without an equivalent observation construction.

Every saved reference curve and every reported metric was recomputed from the archived predictions and agreed to within $10^{-15}$; details are in Supplementary Section~S1.

\section{Oracle weighting, attainment and uncertainty}\label{app:oracle}
The oracle floor is a population statement for a specified input restriction. For each allowed-information condition, a median minimizes the expected absolute distance to the reference value. Taking the conditional expectation over eligible patients proves Eq.~\eqref{eq:floor}. A zero expected nonnegative error requires equality with the reference almost surely, meaning with probability one in that eligible population. Conversely, if the inputs determine the reference, predicting it gives zero error. When information is added, the previous predictor is still admissible, so the infimum cannot increase.

The zero-floor statement concerns reference probabilities, not the prediction of an individual's random future event with certainty. Knowing the process and sufficient current information permits exact calculation of the distribution even though realized futures remain stochastic. It also does not imply that a finite fitted network or a finite rollout sample attains that distribution.

For a finite set of survival horizons, choosing the lower conditional median consistently gives a nonincreasing median curve because the underlying reference curves are ordered across time. Cause-specific CIFs are nondecreasing, but independently selecting marginal medians for survival and two CIFs need not preserve their joint sum of one. The componentwise absolute-error bound remains valid; simultaneous attainment under joint probability constraints requires additional care.

If the calculation groups patients into augmented-state classes $z$, use weights $w_z(x,L)=\PP(Z_L=z\mid X=x,A_L)$, not an unweighted average over classes. To average covariate-specific errors $r_{L,t}(x)$ over eligible patients, write $a_L(x)=\PP(A_L\mid X=x)$ and let $f_X$ be the baseline covariate density. Then
\begin{equation}\label{eq:weight}
R_{L,t}=\frac{\int r_{L,t}(x)a_L(x)f_X(x)\,dx}{\int a_L(x)f_X(x)\,dx}.
\end{equation}
The eligibility weights account for the fact that the landmark cohort is a selected subset of the baseline cohort. Standardizing instead to the baseline distribution is a different target and must be labeled accordingly.

Removing rare classes and renormalizing need not lower the computed floor. If retained classes have mass $p$ and normalized minimum absolute error $R_{\rm ret}$, bounded reference values imply $pR_{\rm ret}\le R_{\rm full}\le pR_{\rm ret}+1-p$. The reference calculations for the present experiments retain every reachable state rather than truncate small-probability classes.

\section{Sampling methods and individual curves}\label{app:samplers}
Within each ATLAS architecture and training run, all seven samplers share the same checkpoint. They change how the walks or particles are constructed numerically, not the learned next-state probabilities. The true-generator IID sampler substitutes the known SFSA law for the learned law and serves as a finite-sampling control. Supplementary Section S5 gives the constructions in detail, including the importance weights and the particle updates.

\begin{center}\begin{minipage}{\textwidth}\centering\small
\begin{tabular}{p{0.21\textwidth}p{0.69\textwidth}}\toprule
Method & Construction\\\midrule
IID & Independent uniforms mapped through each walk's next-state categorical distribution.\\
Stratified & Divide $[0,1]$ into $M$ strata, sample once per stratum, and randomly assign the samples to walks at each step.\\
Scrambled Sobol & Assign one randomized low-discrepancy point to each walk; its coordinates supply the successive categorical draws.\\
Antithetic & Pair a uniform vector $U$ with $1-U$, using each path's own conditional probabilities.\\
Importance & Multiply proposed death probabilities by two before normalizing, then correct event and risk counts with likelihood-ratio weights.\\
Weighted branching & Integrate death and dropout exits, expand live successors, and resample $M$ live histories by their probability masses.\\
Conditional MC & Integrate death and dropout exits and sample a conditional live successor for each particle, retaining evolving weights.\\\bottomrule
\end{tabular}
\captionof{table}{The seven numerical methods compared in both experiments. The final two rows use live particles rather than complete observed walks. Branching has three live successors in the single-covariate experiment and seven in the competing-risks experiment.}
\end{minipage}\end{center}

Figure~\ref{fig:individual} shows how individual predicted curves differ across models for three baseline patients.

\begin{center}\begin{minipage}{\textwidth}\centering
\includegraphics[width=\textwidth]{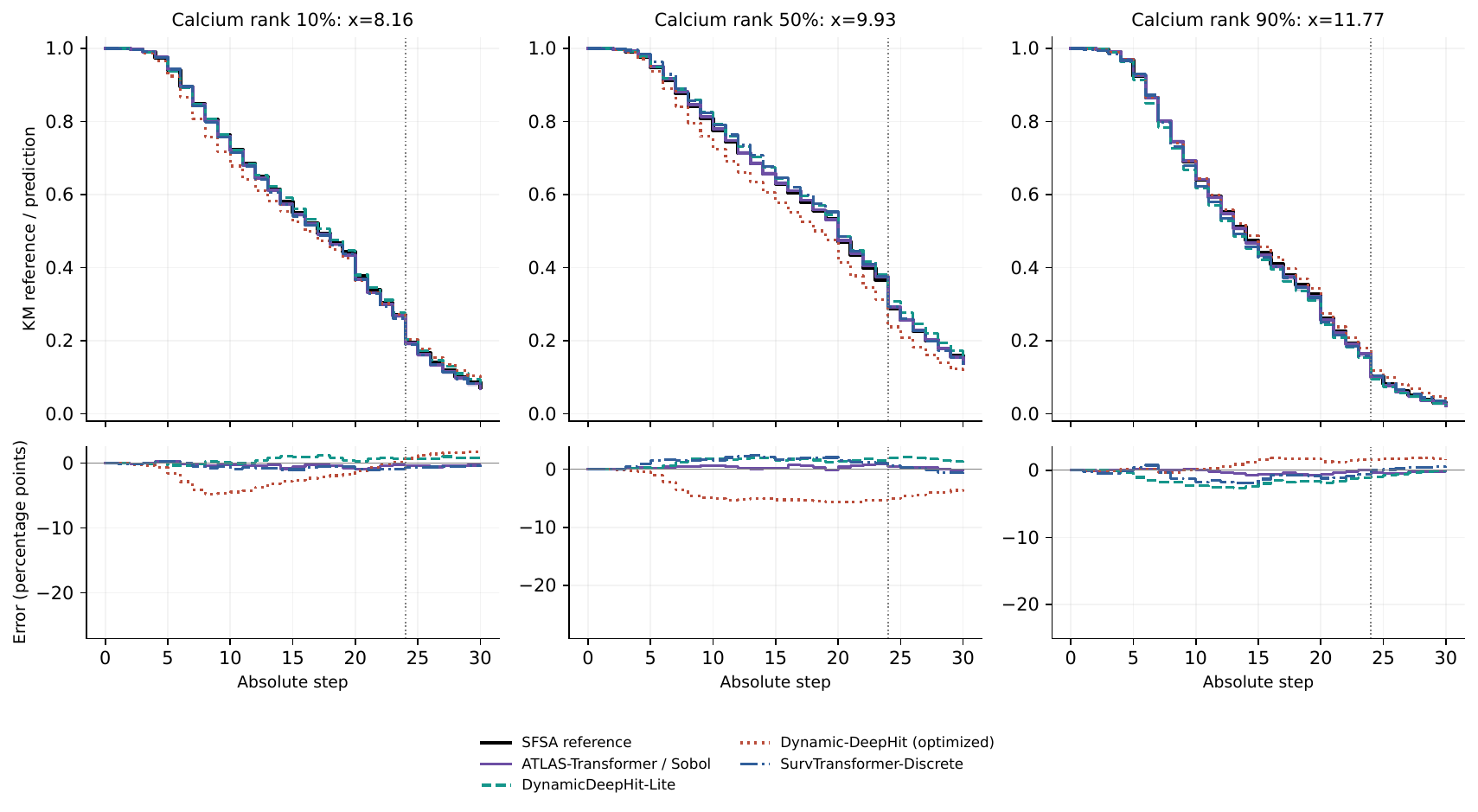}
\captionof{figure}{Individual baseline predictions in the single-covariate duration-dependent experiment. Fitted-model curves use training seed 42; Dynamic-DeepHit (optimized) uses its validation-selected checkpoint. Patients are selected nearest the 10th, 50th and 90th covariate ranks within the 300 scored patients, not by model error. Black is the deterministic reference, and purple is ATLAS-Transformer with Sobol with $M=10{,}000$ and sampling seed 41001. Purple shading is the range across five sampling seeds, not a confidence interval. The other direct-model curves retain their original saved predictions. Lower panels show signed errors in percentage points, and the vertical line marks step 24.}\label{fig:individual}
\end{minipage}\end{center}

Every upper-panel curve is a prediction for one specified covariate value. At baseline there is no observed state prefix. The ATLAS curve comes from 10,000 separately generated futures for that patient, not from walks leading up to the first state and not from pooled walks across different patients. The lower panels show where each curve differs from the reference. A cohort mean prediction curve, by contrast, averages across patients, so overprediction for some and underprediction for others can cancel. The primary MAE in the paper averages individual absolute errors before averaging patients and therefore retains those discrepancies.

\ifdefined\includesupplement
\ifdefined\includesupplement\clearpage\fi
\setcounter{section}{0}
\setcounter{subsection}{0}
\setcounter{table}{0}
\setcounter{figure}{0}
\setcounter{equation}{0}
\renewcommand{\thesection}{S\arabic{section}}
\renewcommand{\thetable}{S\arabic{table}}
\renewcommand{\thefigure}{S\arabic{figure}}
\renewcommand{\theequation}{S\arabic{equation}}
\ifdefined\includesupplement
\begin{center}
{\Large\bfseries Supplementary Material}\\[4pt]
{\large ATLAS: Autoregressive Trajectories for Longitudinal Analysis of Survival}
\end{center}
\fi
\noindent This supplement contains the model specifications and training settings, the competing-risk estimator in detail, the training-run dispersion and per-run tables, the full model--sampler comparison figures, the sampling estimator details, the earlier sampling-resource study, the sampler ablation tables, the paired patient-bootstrap analysis, the exploratory MMLR and HAMMLR transition-model comparisons, and the original history-aware Dynamic-DeepHit results. Section, table, figure and equation numbers carry the prefix S. References to Sections, Tables, Figures and Equations without that prefix point to the main text.

\noindent The synthetic data, analysis code and instructions for reproducing the supplementary results are in the ATLAS repository: \url{https://github.com/mkngqx-ctrl/ATLAS}.

\section{Model formulations and actual run settings}\label{app:models}
\subsection{Prediction inputs and common training protocol}
The covariate-only models are CPH, Weibull-MLP, DRSA (optimized), DynamicDeepHit-Lite and SurvTransformer-Lite. The three newer direct models receive all live state observations through the landmark, the covariates and the prediction time. Future state fields are masked. ATLAS uses the same observed information as the starting point for its walks but optimizes a next-state objective. Scalar covariate inputs in the first experiment become three-dimensional inputs in the second. Patients are split into training, validation and test sets before their histories are expanded into prefixes. Numerical scaling parameters are fitted on training patients only and then applied unchanged to the validation and test sets.

The single-covariate analysis uses training seeds 42, 43, 44, 45, 50, 51, 52, 62, and 63; the competing-risks analysis uses training seeds 42, 43, 45, 46 and 47. Stochastic predictions use sampling seeds 41001, 52001, 63001, 74001, and 85001 in both experiments. The maximum is 200 epochs with early-stopping patience 15. The original single-covariate seed-42 archive reused eight existing checkpoints and 460 existing predictions, and then added the three history-aware checkpoints and their 12 deterministic predictions. The new direct-model settings are taken from checkpoint metadata rather than the notebook's earlier defaults. In the competing-risks runs, the SurvTransformer models use validation-index seed 145 throughout, selecting the same 2,048 validation prefixes from the fixed cohort. The optimized Dynamic-DeepHit selection uses fixed eligible validation patients at each evaluated landmark, as described below.

\begin{center}\begin{minipage}{\textwidth}\centering\small
\begin{tabular}{p{0.42\textwidth}rr}\toprule
Setting & Single covariate & Competing risks\\\midrule
Optimized DDH GRU width / layers & 64 / 4 & 64 / 4\\
Attention-network hidden width & 64 & 64\\
Optimized DDH cause-head width / layers & 32 / 2 & 100 / 1\\
SurvTransformer token width / layers & 16 / 6 & 16 / 6\\
SurvTransformer attention heads & 2 & 2\\
Feed-forward width / task-head layers & 64 / 4 & 64 / 4\\
History-model dropout & 0.1 & 0.1\\
SurvTransformer learning rate & 0.001 & 0.001\\
History-model training batch size & 256 & 256\\
Optimized DDH auxiliary burn-in epochs & 5 & 5\\
Optimized DDH learning rate (rounded) & 0.000963 & 0.000604\\
Optimized DDH ranking weight & 0.001 & 0\\
Optimized DDH ranking scale & 0.2 & 0.2\\
Optimized DDH auxiliary weight & 0.01 & 1\\
Optimized DDH head/output L1 & 0.00001 & 0.00001\\
Optimized DDH validation patients / landmark & up to 512 & up to 512\\
Training prefixes per history model & 672,886 & 864,247\\
SurvTransformer validation prefixes & 2,048 & 2,048\\
Selected optimized DDH epoch & 8--12 & 11--14\\
Selected SurvTransformer epoch & 3--7 & 2--3\\
Selected SurvTransformer-Discrete epoch & 3--6 & 1--3\\\bottomrule
\end{tabular}
\captionof{table}{Actual history-comparator settings and selected checkpoint epochs. Epoch entries give the minimum and maximum across nine single-covariate training runs and five competing-risks training runs. These are the epochs of the retained checkpoints, not the epochs when training stopped. Optimized Dynamic-DeepHit epoch numbers include the five auxiliary-only burn-in epochs. The SurvTransformer variants have no burn-in phase. Per-seed SurvTransformer values are supplied in \texttt{supplementary/}, in \texttt{hard24\_selected\_epochs.csv} and \texttt{cr3\_selected\_epochs.csv}; optimized Dynamic-DeepHit values are in \texttt{optimized\_DDH\_selected\_epochs.csv}. The two SurvTransformer likelihood variants share architecture settings. Early selected epochs are reported as observed; their causes have not been isolated by additional training experiments.}
\end{minipage}\end{center}

Adam is used for these adapted neural models. Dynamic-DeepHit (optimized) selects its checkpoint by landmark-averaged conditional validation negative log-likelihood, with the selected losses and weights specified below. Its weight decay is zero. Both SurvTransformer variants select by their combined validation loss, with mortality and observed-duration auxiliary weights one and weight decay $10^{-4}$. Observed duration is scaled by the 30-step horizon. These are not the authors' original tuning results. The competing-risks archive contains only four rows in its top-level training summary, but all ten fitted models and the per-model training records are present in the checkpoints. The model inventory rests on that metadata rather than on the incomplete summary.

\subsection{Covariate-only direct models}
CPH uses proportional hazards with the supplied covariates. The competing-risks version fits one cause-specific Cox model per cause and combines hazard increments with Eq.~\eqref{eq:crconversion}; its penalizer is 0.1. Weibull-MLP predicts positive scales and shapes from a small covariate network. The single-cause model uses discrete event-bin likelihood, while the competing-risks model uses the continuous additive cause-specific Weibull likelihood and numerical integration for CIFs.

The Lite models use recurrent or Transformer time-grid representations of the baseline covariates, not the observed state sequence. Their output is a joint softmax over event-time bins and one shared beyond-horizon probability. With $K$ causes and horizon $B=30$, this gives $KB+1$ entries. For one cause,
\[
\widehat S(l)=p_{>B}+\sum_{j=l+1}^{B}p_j.
\]
An uncensored event contributes its cause-time log probability, while a censored endpoint contributes the log probability of survival beyond that time. The shared tail permits administrative survivors. A pairwise ranking term compares patients at the same event time. It is an additional objective rather than a guarantee of probability accuracy.

DRSA (optimized) represents interval probabilities using an LSTM over the 30-step prediction grid. A learned ReLU projection transforms the training-standardized baseline covariates before normalized prediction time is appended. The model does not receive the observed health-state sequence. The duration-dependent head predicts a sigmoid death hazard, while the competing-risks head uses a softmax over conditional survival and the two death causes. Survival and CIFs are obtained by probability propagation; survival beyond step 30 remains possible. These are adaptations of the published model families rather than exact ports of all released settings~\citep{ren2019,martinsson2017,lee2018deephit}.

The DRSA objective is $\alpha L_z+(1-\alpha)L_c$, where $L_z$ is the uncensored event-time negative log-likelihood and $L_c$ combines the event-status and censoring terms, averaged over the minibatch. The competing-risks event-time term uses the observed cause. The search retained this composite objective and considered $\alpha\in\{0.05,0.10,0.25,0.50,0.75,0.90\}$. Both selected configurations use $\alpha=0.50$, giving equal weights to the two objective components. Observed deaths use the recorded terminal time for both event-time and event-status terms; this differs from an observation scheme that follows individuals beyond their event time.

The search was frozen before optimization and used the original 40,000 training and 5,000 validation patients in each experiment. Screening crossed the six alpha values with four architectures: 32 units/one layer/no projection; 128 units/one layer/no projection; 128 units/one layer/ReLU-64; and 256 units/two layers/ReLU-128. These 24 candidates used seed 42 and cumulative budgets of 5, 15 and 30 epochs. Two candidates per alpha were retained after the first budget and one per alpha after the second, so all six alpha values remained represented through screening. The two leading alpha values then entered refinement. Each was crossed with five fixed new recipes and its screening winner, giving 12 refinement entries. The new recipes explored widths 64, 128 and 256, one or two layers, ReLU projections of width 64 or 128, learning rates from 0.0001 to 0.003, dropout 0 to 0.2, L2 coefficients 0 to 0.0001, and batch sizes 128 or 256. Refinement used cumulative budgets of 10, 30 and 60 epochs, retaining six then four entries. Two finalists were confirmed using training seeds 42, 43 and 45; the selected configuration minimized their mean best validation score.

Every checkpoint and promotion decision used conditional observed-outcome validation negative log-likelihood averaged equally over the evaluated landmarks, using all eligible validation patients. Eligibility required the recorded event or censoring time to exceed the landmark. The score did not include the training penalty. Analytical reference curves and test-set MAE were not used for selection. Confirmation and final training used Adam with $(\beta_1,\beta_2)=(0.9,0.99)$, gradient clipping at 5, a maximum of 200 epochs and patience 15, with a minimum validation improvement of $10^{-6}$. The L2 penalty was $\lambda\|\theta\|_2^2/(2\cdot128)$, with the fixed denominator preserving its strength across batch sizes. Dropout was applied after the covariate projection and recurrent output, and between recurrent layers when applicable. Final fits used all training patients, one record per patient; selected confirmation fits were reused unchanged for seeds 42, 43 and 45.

\begin{center}\begin{minipage}{\textwidth}\centering\small
\begin{tabular}{lrr}\toprule
Selected DRSA setting & Duration-dependent & Competing risks\\\midrule
LSTM width / layers & 256 / 1 & 256 / 2\\
ReLU projection width & 128 & 128\\
Alpha & 0.50 & 0.50\\
Learning rate & 0.003 & 0.0015\\
Dropout & 0.20 & 0.10\\
L2 coefficient & 0.0001 & 0.00001\\
Batch size & 128 & 256\\
Selected epoch across final fits & 12--66 & 11--15\\\bottomrule
\end{tabular}
\captionof*{table}{Selected DRSA configurations. The nine duration-dependent and five competing-risk fits use the original training seeds and evaluation patients. Selected epochs refer to retained checkpoints, not stopping epochs. Per-fit settings and logs are supplied in the accompanying data/code package.}
\end{minipage}\end{center}

For all covariate-only models, survival at a landmark is obtained by conditioning on survival to $L$. In a competing-risk joint distribution, the future CIF and survival are
\[
\frac{F_k(t)-F_k(L)}{S(L)}\quad\text{and}\quad\frac{S(t)}{S(L)},
\]
respectively. This changes the prediction origin without supplying the observed state history.

\subsection{History-aware Dynamic-DeepHit (optimized)}
The optimized Dynamic-DeepHit adaptations retain a recurrent history encoder, attention conditioned on the latest measurement, cause-specific subnetworks, and a joint event-time probability distribution with a shared beyond-horizon category. In both selected configurations, hidden states from all recurrent layers are concatenated for attention and context construction. The cause-specific representations are then combined through a common output projection. The auxiliary task predicts observed next-step live-state indicators using masked binary cross-entropy, while L1 regularization applies to the cause-specific and output weights.

The search evaluated 12 configurations per experiment using successive training budgets of 5, 15, and 30 joint epochs, preceded by five auxiliary-only burn-in epochs. Six configurations advanced to the second stage and three to the third. The two strongest validation configurations were evaluated at two additional training seeds, and the final configuration was selected by mean validation performance across the three search and confirmation seeds. The selection criterion equally weighted conditional negative log-likelihood across the experiment's evaluated landmarks, using up to 512 fixed eligible validation patients at each landmark. Analytical reference curves and test-set MAE were not used for selection.

Final fits used all training prefixes, Adam optimization, batch size 256, and a maximum of 200 joint epochs with early-stopping patience 15. Each fit retained the checkpoint with the lowest landmark-averaged validation negative log-likelihood. The duration-dependent configuration retained a nonzero ranking term. In contrast, the competing-risks configuration selected a ranking weight of zero and therefore represents an optimized adaptation with ranking loss disabled.

The history representation contains one-hot live-state indicators, covariates and time information. The latest measurement bypasses the recurrent past, and a baseline context with no observed state is handled explicitly. The cause-time probabilities and shared tail are conditioned on survival through the landmark. The retained duration-dependent ranking term compares conditional incidence at common elapsed horizons and excludes prefixes of the same patient. The SFSA encoding, finite-horizon tail and binary auxiliary target are adaptations of the longitudinal framework~\citep{lee2019dynamic}; the optimized configurations are distinct from both DynamicDeepHit-Lite and an exact port of the released PBC2 example.

\subsection{SurvTRACE-based history models}
Each observed state/time position is a categorical feature token. Covariates and the landmark are additional feature tokens. Future positions are masked from attention and readout, so the model receives the complete observed prefix without future leakage. The adapted encoder follows the article's SELU residual construction and combines contextual representations with the original embeddings before the task heads. This differs from the inspected public BERT-style implementation's GELU/layer-normalization architecture~\citep{wang2022survtrace}.

Let $z_{k,t}$ be a cause-hazard logit and $a_{k,t}=\operatorname{softplus}(z_{k,t})$. Set $a_t=\sum_k a_{k,t}$. For observed endpoint $D>L$ and cause $J$, the primary piecewise-constant-hazard negative log likelihood is
\[
\mathcal L_{\rm PCH}=\sum_{t=L+1}^{D}a_t-\ind\{J>0\}\log a_{J,D}.
\]
Integer endpoints occupy a full final interval. This is the operational cause-specific likelihood from the released trainer, not the article's inverse-propensity-weighted objective~\citep{survtracecode}. Both variants convert hazards to coherent interval probabilities using Eq.~\eqref{eq:crconversion}.

SurvTransformer-Discrete instead uses $\sum_{t=L+1}^{D-1}a_t-\log p_{J,D}$ for an event, and $\sum_{t=L+1}^{D}a_t$ for censoring. It therefore scores the observed integer event bin rather than a continuous event density. Both variants retain observed-event-status and observed-duration auxiliary tasks, where a censored endpoint supplies an observed duration rather than proof of eventual survival. The history representation and retained article components make these explicitly documented SFSA adaptations.

\subsection{ATLAS generative implementations}
The ATLAS recurrent models~\citep{cho2014,hochreiter1997} use state embeddings of dimension eight and two recurrent layers of width 32, concatenating standardized covariates. The causal Transformer~\citep{vaswani2017} has width 32, two layers, four attention heads, and feed-forward width 64. The first experiment includes GRU, RNN and Transformer; the second includes GRU and Transformer. Their observed state vocabularies have five and ten outcomes, respectively, plus a distinct input start token.

For patient $i$ and step $j$, let $m_{ij}$ equal one if the preceding token is live or the start token, and zero if it is absorbed. The teacher-forced objective is
\begin{equation}\label{eq:atlasloss}
\mathcal L=-\frac{\sum_{i,j}m_{ij}\log p_\theta(s_{ij}\mid s_{i,<j},x_i)}{\sum_{i,j}m_{ij}}.
\end{equation}
Inputs are shifted relative to targets. The loss includes the first terminal transition but masks subsequent repeated absorbing states. Its normalization weights observed transitions rather than giving every patient the same total contribution. At inference, each walk conditions on its own complete history. The generator's transition graph and hard rules are not enforced on the learned logits.

\paragraph{Runtime and hardware provenance.}\label{app:runtime}
GPU hardware and software versions differ across runs. The optimized Dynamic-DeepHit models were fitted on an NVIDIA RTX PRO 6000 Blackwell Server Edition GPU, whereas the optimized DRSA models were fitted locally on an NVIDIA GeForce RTX 3060 using PyTorch 2.11.0 with CUDA 13.0. Hardware and software details are retained in the corresponding run records. For the original Dynamic-DeepHit results retained in Section~S10, training seeds 50, 51 and 52 used native PyTorch CUDA kernels with cuDNN disabled for compatibility. In the original competing-risks comparison, seed 42 used an NVIDIA L4, whereas the other four runs used an NVIDIA RTX PRO 6000 Blackwell. These differences should be considered when interpreting the recorded timings, which do not constitute a controlled hardware comparison.

\paragraph{Archived prediction and metric audits.}
For the original seed-42 archives, we regenerated every saved conditional reference curve from its recorded covariates, prefix, and landmark. Maximum absolute differences were below $6\times10^{-16}$. Recomputing MAE, post-rule MAE where applicable, and RMSE from the 721 saved prediction files reproduced 1,219 metric rows to below $2\times10^{-16}$. Separately, the nine-training-run single-covariate audit recomputed MAE, post-rule MAE and RMSE from 4,248 saved prediction files, reproducing all 4,248 metric rows to within $1.01\times10^{-16}$. The saved reference arrays were identical across the nine runs; this comparison reused the archived reference arrays rather than regenerating them for each training run. For the five-run competing-risks analysis, recomputing MAE and RMSE from 1,245 saved prediction files reproduced 3,735 metric rows to within $1.01\times10^{-16}$. Aggregate means and standard deviations agreed with the rounded supplied tables to within $5\times10^{-13}$. Saved reference arrays were identical across runs, and checkpoint tensors confirmed five distinct fitted weight sets for each neural model. The original archive audit is included in the later multi-run checks, so these file counts are not disjoint. These checks verify the consistency of the archived calculations. They do not establish that any trained model is optimal.

\section{Competing-risk estimation and probability conservation}\label{app:cr}
Let $T$ be the biological first-death time and $J\in\{A,B\}$ its cause. The cause-$k$ cumulative incidence at a landmark is
\[
F_{k,L}(t\mid x,h)=\PP(L<T\le t,J=k\mid X=x,H_L=h,A_L).
\]
It counts cause $k$ only when that cause occurs first. Together with the corresponding survival probability, $S_L(t)+F_{A,L}(t)+F_{B,L}(t)=1$. This is an observable competing-event target rather than the hypothetical probability of cause $k$ after eliminating the other cause~\citep{putter2007}.

For a generated path $m$, let $D_m$ be its observed endpoint, $J_m$ its cause (zero for censoring), and $w_m\ge0$ its weight. Define
\[
R_t=\sum_mw_m\ind\{D_m\ge t\},\qquad
E_{k,t}=\sum_mw_m\ind\{D_m=t,J_m=k\}.
\]
Initialize $\widehat S_L(L)=1$ and $\widehat F_{k,L}(L)=0$. At each future integer step with $R_t>0$, update
\begin{align}
\widehat F_{k,L}(t)&=\widehat F_{k,L}(t-1)+\widehat S_L(t-1)E_{k,t}/R_t,\\
\widehat S_L(t)&=\widehat S_L(t-1)\{1-(E_{A,t}+E_{B,t})/R_t\}.
\end{align}
The same pre-update survival is used in both cause increments. Empty risk sets carry all previous values forward. These updates preserve total probability and reduce to ordinary Kaplan--Meier for all-cause survival~\citep{aalen1978}. Every $w_m=1$ for unweighted complete walks. Importance sampling uses likelihood-ratio weights, and the particle methods use evolving weighted transition counts instead of complete-path endpoints.

For cause-specific hazard predictions, the separate quantities $1-S_A$ and $1-S_B$ must not be read as the competing-risk CIFs. The implemented conversion combines both hazards through a shared probability of surviving each interval. If $a_{k,t}\ge0$ is the cumulative cause-$k$ hazard within the unit interval and $a_t=a_{A,t}+a_{B,t}$, then
\begin{equation}\label{eq:crconversion}
p_{0,t}=e^{-a_t},\qquad
p_{k,t}=(1-e^{-a_t})\frac{a_{k,t}}{a_t}.
\end{equation}
The zero-total-hazard limit is $p_{0,t}=1,p_{k,t}=0$. Accumulating $S_L(t-1)p_{k,t}$ gives coherent CIFs. CPH and the SurvTransformer variants use this constant-rate interval conversion; the Weibull-MLP comparator integrates its continuous cause-specific density over each interval. Neither conversion assumes that the SFSA's induced event times are Weibull.

\section{Metrics, sampling uncertainty, and computational cost}\label{supp:metrics}
The primary metrics compare individual predicted curves to deterministic references on the strictly future grid $L+1,\ldots,30$. MAE averages absolute differences. RMSE takes the square root of the corresponding mean squared difference within each prediction replicate, and the archived summary averages replicate RMSE values rather than their squares. In the competing-risks run, survival and each CIF are scored separately rather than combined into some unreported aggregate target.

For a stochastic method, the sample standard deviation of its five replicate MAEs describes sensitivity to the sampling seed with the checkpoint fixed. It does not include variation from retraining. Each checkpoint first contributes its mean over the five sampling seeds. The nine single-covariate training-run means, or the five competing-risks training-run means, are then averaged equally. Their sample standard deviation describes between-run variation and still contains finite-sampling error. Pooled sampling standard deviation is the square root of the average within-run variance across the corresponding nine or five training runs. The 45 single-covariate sampling measurements represent nine training runs; the 25 competing-risks measurements represent five. They are not 45 or 25 independently trained models. Matched sampling seeds may correlate numerical errors across training runs. These standard deviations are descriptive and are not confidence intervals. Neural direct models contribute one deterministic prediction per training run and have no within-run sampling standard deviation. Averaging the prediction curves first and then scoring would answer a different question and would generally shrink the numerical variation, so we instead average already-computed losses. The original seed-42 source records include descriptive paired-patient intervals calculated from patient losses averaged over the saved sampling seeds. These intervals are conditional on the original seed-42 checkpoints and sampled prediction sets; they are not intervals for the nine-run or five-run mean results. They also do not account for selecting a sampler or comparator after looking at the results.

Outcome-based Brier scores~\citep{graf1999} and concordance answer different questions from reference-curve error. A censoring-weighted Brier calculation requires a justified censoring model, positivity at scored times, and a tie convention that matches the data. Concordance measures ordering rather than absolute probability accuracy. Neither metric can be inferred from the MAE rankings, and the distinction matters most when a checkpoint is selected by concordance but evaluated by curve MAE.

Prediction latency depends on architecture, sampled path length, batching, sampler bookkeeping, and the estimator calculation. The recorded batch size of 512 is an inference setting, while the history-model training batch size is 256. Equal particle and trajectory counts do not imply equal computational work. The reused single-covariate prediction timings and the newer model timings were not necessarily measured in the same session, and the training runs span different GPU hardware and software versions. The original Dynamic-DeepHit backend differences for seeds 50, 51 and 52 pertain to the results retained in Section~S10. The optimized Dynamic-DeepHit fits used an NVIDIA RTX PRO 6000 Blackwell Server Edition GPU; the earlier DDH timing summaries describe the original adaptation. For competing risks, the supplied pooled timing summaries use only seeds 43, 45, 46 and 47 on the same recorded GPU model; seed-42 L4 timings are retained separately. All five runs contribute to accuracy summaries. DRSA (optimized) was fitted locally on an NVIDIA GeForce RTX 3060 using PyTorch 2.11.0 with CUDA 13.0; its per-fit elapsed times and hardware metadata are supplied with the optimized results. The recorded timings therefore do not constitute a controlled speed comparison.

For a Transformer of width $d$, recomputing a full prefix of length $t$ costs order $td^2+t^2d$ at fixed depth. Summing these passes through length $T$ gives order $T^2d^2+T^3d$ per walk, and caching can bring this down to order $Td^2+T^2d$. Recurrent caching similarly avoids repeated prefix work. Actual wall-clock comparisons have to be measured with the implemented kernels and hardware rather than read off asymptotic formulas.

\begingroup\small
\setlength{\tabcolsep}{4pt}
\begin{longtable}{lrrrr}
\caption{Variation across the nine single-covariate training runs for the neural configurations in Table~1 of the main text. Each ATLAS training run contributes its mean MAE over five sampling seeds at $M=10{,}000$; ATLAS rows use Sobol. Between-run SD is the sample standard deviation of the nine training-run means. Minimum and maximum refer to those means. Sampling SD is the square root of the average within-run sampling variance; it is undefined for direct predictions. CPH and the true-generator oracle are unchanged controls and are omitted here. These summaries are descriptive, not confidence intervals.}\label{tab:harddispersion}\\
\toprule
Model / method & Between SD & Minimum & Maximum & Sampling SD \\\midrule
\endfirsthead
\multicolumn{5}{l}{\tablename\ \thetable\ (continued)}\\
\toprule
Model / method & Between SD & Minimum & Maximum & Sampling SD \\\midrule
\endhead
\midrule\multicolumn{5}{r}{Continued on next page}\\\endfoot
\bottomrule\endlastfoot
\multicolumn{5}{l}{\textit{Landmark $L=0$}}\\*
Weibull-MLP & 0.000239 & 0.023829 & 0.024598 & --- \\*
DRSA (optimized) & 0.007890 & 0.110197 & 0.135464 & --- \\*
DynamicDeepHit-Lite & 0.002011 & 0.004167 & 0.010698 & --- \\*
SurvTransformer-Lite & 0.002998 & 0.006564 & 0.015891 & --- \\*
Dynamic-DeepHit (optimized) & 0.003181 & 0.024291 & 0.033932 & --- \\*
SurvTransformer & 0.004859 & 0.012103 & 0.025143 & --- \\*
SurvTransformer-Discrete & 0.002161 & 0.010735 & 0.017151 & --- \\*
ATLAS-GRU / Sobol & 0.002379 & 0.004393 & 0.010309 & 0.000055 \\*
ATLAS-RNN / Sobol & 0.002659 & 0.003993 & 0.011482 & 0.000052 \\*
ATLAS-Transformer / Sobol & 0.004954 & 0.003316 & 0.018088 & 0.000064 \\
\addlinespace
\multicolumn{5}{l}{\textit{Landmark $L=20$}}\\*
Weibull-MLP & 0.000449 & 0.136597 & 0.138008 & --- \\*
DRSA (optimized) & 0.004728 & 0.140973 & 0.152804 & --- \\*
DynamicDeepHit-Lite & 0.001936 & 0.139121 & 0.145121 & --- \\*
SurvTransformer-Lite & 0.002338 & 0.139655 & 0.146120 & --- \\*
Dynamic-DeepHit (optimized) & 0.008377 & 0.037261 & 0.065852 & --- \\*
SurvTransformer & 0.009822 & 0.037009 & 0.067081 & --- \\*
SurvTransformer-Discrete & 0.004971 & 0.039368 & 0.053463 & --- \\*
ATLAS-GRU / Sobol & 0.003520 & 0.005468 & 0.016233 & 0.000072 \\*
ATLAS-RNN / Sobol & 0.004691 & 0.008798 & 0.023873 & 0.000056 \\*
ATLAS-Transformer / Sobol & 0.001328 & 0.006616 & 0.011309 & 0.000058 \\
\addlinespace
\multicolumn{5}{l}{\textit{Landmark $L=23$}}\\*
Weibull-MLP & 0.000381 & 0.185343 & 0.186221 & --- \\*
DRSA (optimized) & 0.004006 & 0.202487 & 0.214769 & --- \\*
DynamicDeepHit-Lite & 0.002553 & 0.200399 & 0.208035 & --- \\*
SurvTransformer-Lite & 0.004553 & 0.196337 & 0.212297 & --- \\*
Dynamic-DeepHit (optimized) & 0.009616 & 0.046243 & 0.074294 & --- \\*
SurvTransformer & 0.013379 & 0.067916 & 0.102157 & --- \\*
SurvTransformer-Discrete & 0.006843 & 0.048906 & 0.073837 & --- \\*
ATLAS-GRU / Sobol & 0.003435 & 0.008747 & 0.018402 & 0.000050 \\*
ATLAS-RNN / Sobol & 0.003054 & 0.013198 & 0.022547 & 0.000051 \\*
ATLAS-Transformer / Sobol & 0.002452 & 0.009737 & 0.017807 & 0.000046 \\
\addlinespace
\multicolumn{5}{l}{\textit{Landmark $L=24$}}\\*
Weibull-MLP & 0.000833 & 0.149664 & 0.152170 & --- \\*
DRSA (optimized) & 0.004159 & 0.151963 & 0.163361 & --- \\*
DynamicDeepHit-Lite & 0.001884 & 0.151875 & 0.158396 & --- \\*
SurvTransformer-Lite & 0.003835 & 0.149574 & 0.163230 & --- \\*
Dynamic-DeepHit (optimized) & 0.008846 & 0.051325 & 0.075680 & --- \\*
SurvTransformer & 0.008680 & 0.059305 & 0.083381 & --- \\*
SurvTransformer-Discrete & 0.010842 & 0.056364 & 0.091665 & --- \\*
ATLAS-GRU / Sobol & 0.004328 & 0.008064 & 0.020148 & 0.000056 \\*
ATLAS-RNN / Sobol & 0.004282 & 0.006977 & 0.019457 & 0.000059 \\*
ATLAS-Transformer / Sobol & 0.002933 & 0.007748 & 0.015835 & 0.000049 \\
\addlinespace
\end{longtable}
\endgroup

Table~\ref{tab:hard24_checkpoint_scores} gives the individual training-run scores for the common Sobol comparison. At baseline, the median MAEs across training runs are 0.006930 for ATLAS-GRU, 0.006546 for ATLAS-RNN and 0.007273 for ATLAS-Transformer. The corresponding ranges are 0.004393--0.010309, 0.003993--0.011482 and 0.003316--0.018088. Their between-run standard deviations are about 43, 51 and 77 times their pooled within-run sampling standard deviations, respectively. At this sampling budget, differences between fitted models are therefore much larger than the numerical variation from repeating the sampler.

Transformer training seeds 43 and 51 have the largest baseline errors, 0.015513 and 0.018088, but there is no single uniformly poor training run. For example, seed 42 has the smallest Transformer baseline MAE (0.003316) and its largest MAE at $L=23$ (0.017807). All nine runs remain in the analysis; these descriptive comparisons are not used to discard checkpoints.

\begingroup\small
\setlength{\tabcolsep}{4pt}
\begin{longtable}{rrrrlr}
\caption{Individual training-run survival MAEs for the single-covariate experiment. The GRU, RNN and Transformer columns refer to ATLAS with Sobol; each value averages the same five sampling seeds at $M=10{,}000$. The last two columns identify the lowest-error fitted direct comparator within that training run and landmark. The comparator may therefore differ between rows. All nine training runs are retained.}\label{tab:hard24_checkpoint_scores}\\
\toprule
Seed & GRU & RNN & Transformer & Best direct model & MAE \\\midrule
\endfirsthead
\multicolumn{6}{l}{\tablename\ \thetable\ (continued)}\\
\toprule
Seed & GRU & RNN & Transformer & Best direct model & MAE \\\midrule
\endhead
\midrule\multicolumn{6}{r}{Continued on next page}\\\endfoot
\bottomrule\endlastfoot
\multicolumn{6}{l}{\textit{Landmark $L=0$}}\\*
42 & 0.010119 & 0.008393 & 0.003316 & DynamicDeepHit-Lite & 0.010698 \\*
43 & 0.009730 & 0.005007 & 0.015513 & SurvTransformer-Lite & 0.006564 \\*
44 & 0.010309 & 0.006546 & 0.006588 & DynamicDeepHit-Lite & 0.008745 \\*
45 & 0.006742 & 0.005418 & 0.008925 & DynamicDeepHit-Lite & 0.010660 \\*
50 & 0.005138 & 0.011482 & 0.005617 & DynamicDeepHit-Lite & 0.007758 \\*
51 & 0.004393 & 0.007287 & 0.018088 & SurvTransformer-Lite & 0.007031 \\*
52 & 0.004549 & 0.005528 & 0.007273 & SurvTransformer-Lite & 0.007881 \\*
62 & 0.006930 & 0.003993 & 0.009394 & DynamicDeepHit-Lite & 0.004167 \\*
63 & 0.008425 & 0.011111 & 0.004648 & DynamicDeepHit-Lite & 0.007353 \\
\addlinespace
\multicolumn{6}{l}{\textit{Landmark $L=20$}}\\*
42 & 0.005468 & 0.013813 & 0.009060 & SurvTransformer-Discrete & 0.042674 \\*
43 & 0.008250 & 0.008798 & 0.009060 & SurvTransformer & 0.041463 \\*
44 & 0.007018 & 0.010462 & 0.008369 & SurvTransformer-Discrete & 0.044347 \\*
45 & 0.010287 & 0.009300 & 0.009128 & SurvTransformer & 0.037009 \\*
50 & 0.016233 & 0.013494 & 0.010128 & SurvTransformer-Discrete & 0.039368 \\*
51 & 0.009268 & 0.013098 & 0.009674 & SurvTransformer-Discrete & 0.052113 \\*
52 & 0.010414 & 0.015662 & 0.006616 & SurvTransformer-Discrete & 0.042726 \\*
62 & 0.006785 & 0.009518 & 0.011309 & Dynamic-DeepHit (optimized) & 0.052988 \\*
63 & 0.014140 & 0.023873 & 0.008017 & SurvTransformer-Discrete & 0.044037 \\
\addlinespace
\multicolumn{6}{l}{\textit{Landmark $L=23$}}\\*
42 & 0.008776 & 0.022547 & 0.017807 & Dynamic-DeepHit (optimized) & 0.052758 \\*
43 & 0.014619 & 0.013229 & 0.010707 & SurvTransformer-Discrete & 0.057438 \\*
44 & 0.010653 & 0.015780 & 0.012856 & SurvTransformer-Discrete & 0.053347 \\*
45 & 0.014978 & 0.013939 & 0.012150 & Dynamic-DeepHit (optimized) & 0.055410 \\*
50 & 0.015540 & 0.014584 & 0.010352 & Dynamic-DeepHit (optimized) & 0.051255 \\*
51 & 0.018402 & 0.014542 & 0.009737 & SurvTransformer-Discrete & 0.056618 \\*
52 & 0.016420 & 0.016393 & 0.012481 & Dynamic-DeepHit (optimized) & 0.046243 \\*
62 & 0.008747 & 0.013198 & 0.014520 & Dynamic-DeepHit (optimized) & 0.056692 \\*
63 & 0.014685 & 0.018786 & 0.012113 & SurvTransformer-Discrete & 0.048906 \\
\addlinespace
\multicolumn{6}{l}{\textit{Landmark $L=24$}}\\*
42 & 0.008064 & 0.019457 & 0.014011 & Dynamic-DeepHit (optimized) & 0.054247 \\*
43 & 0.010115 & 0.006977 & 0.008518 & SurvTransformer-Discrete & 0.066654 \\*
44 & 0.010159 & 0.008110 & 0.015835 & SurvTransformer-Discrete & 0.056364 \\*
45 & 0.015913 & 0.007499 & 0.009553 & SurvTransformer & 0.059305 \\*
50 & 0.020148 & 0.008532 & 0.010934 & Dynamic-DeepHit (optimized) & 0.069270 \\*
51 & 0.017121 & 0.015504 & 0.007777 & SurvTransformer & 0.063273 \\*
52 & 0.012868 & 0.012163 & 0.007748 & Dynamic-DeepHit (optimized) & 0.051325 \\*
62 & 0.009980 & 0.010619 & 0.011982 & SurvTransformer-Discrete & 0.062187 \\*
63 & 0.018180 & 0.014563 & 0.008086 & Dynamic-DeepHit (optimized) & 0.056427 \\
\addlinespace
\end{longtable}
\endgroup

For competing risks, Table~\ref{tab:crdispersion} reports training-run dispersion and Table~\ref{tab:cr_checkpoint_scores} gives individual-run scores for the common Sobol comparison. At baseline, the survival MAEs are $0.007902\pm.000969$ for ATLAS-GRU and $0.010378\pm.004211$ for ATLAS-Transformer, where the standard deviations describe between-run variation. The corresponding ranges are 0.006763--0.009192 and 0.005860--0.016214; pooled within-run sampling standard deviations are 0.000062 and 0.000080. The numerical variation from repeating Sobol sampling is therefore much smaller than the variation among these fitted checkpoints.

Transformer training seed 45 has the largest baseline survival and Death B errors, whereas seed 46 has the largest Death A errors at all three landmarks and the largest survival error at landmark 10. These observations describe variation in the saved fits; they do not establish its cause or justify discarding a run. All five runs remain in the analysis.

\begingroup\small
\setlength{\tabcolsep}{4pt}
\begin{longtable}{lrrrr}
\caption{Variation across five competing-risks training runs for the neural configurations in Tables~2 to 4 of the main text. ATLAS uses Sobol; each training run contributes its mean MAE over five sampling seeds at $M=10{,}000$. Between SD is the sample standard deviation across the five training-run means. Minimum and maximum refer to these means. Sampling SD is the square root of the average within-run sampling variance and is undefined for direct predictions. CPH and the reused true-generator oracle are omitted. These are descriptive standard deviations, not confidence intervals.}\label{tab:crdispersion}\\
\toprule
Model / method & Between SD & Minimum & Maximum & Sampling SD \\\midrule
\endfirsthead
\multicolumn{5}{l}{\tablename\ \thetable\ (continued)}\\
\toprule
Model / method & Between SD & Minimum & Maximum & Sampling SD \\\midrule
\endhead
\midrule\multicolumn{5}{r}{Continued on next page}\\\endfoot
\bottomrule\endlastfoot
\multicolumn{5}{l}{\textit{Overall survival, landmark $L=0$}}\\*
Weibull-MLP & 0.000941 & 0.014796 & 0.017149 & --- \\*
DRSA (optimized) & 0.008051 & 0.073429 & 0.095392 & --- \\*
DynamicDeepHit-Lite & 0.002235 & 0.019689 & 0.024970 & --- \\*
SurvTransformer-Lite & 0.004247 & 0.020683 & 0.029893 & --- \\*
Dynamic-DeepHit (optimized) & 0.001566 & 0.014159 & 0.018062 & --- \\*
SurvTransformer & 0.006237 & 0.015549 & 0.030736 & --- \\*
SurvTransformer-Discrete & 0.005740 & 0.016079 & 0.030699 & --- \\*
ATLAS-GRU / Sobol & 0.000969 & 0.006763 & 0.009192 & 0.000062 \\*
ATLAS-Transformer / Sobol & 0.004211 & 0.005860 & 0.016214 & 0.000080 \\
\addlinespace
\multicolumn{5}{l}{\textit{Overall survival, landmark $L=10$}}\\*
Weibull-MLP & 0.000419 & 0.089528 & 0.090527 & --- \\*
DRSA (optimized) & 0.001908 & 0.093233 & 0.098538 & --- \\*
DynamicDeepHit-Lite & 0.000970 & 0.090057 & 0.092382 & --- \\*
SurvTransformer-Lite & 0.000527 & 0.090191 & 0.091504 & --- \\*
Dynamic-DeepHit (optimized) & 0.001064 & 0.021463 & 0.024414 & --- \\*
SurvTransformer & 0.003960 & 0.034238 & 0.044783 & --- \\*
SurvTransformer-Discrete & 0.006668 & 0.032131 & 0.048810 & --- \\*
ATLAS-GRU / Sobol & 0.000603 & 0.008022 & 0.009545 & 0.000061 \\*
ATLAS-Transformer / Sobol & 0.003193 & 0.009096 & 0.016585 & 0.000088 \\
\addlinespace
\multicolumn{5}{l}{\textit{Overall survival, landmark $L=20$}}\\*
Weibull-MLP & 0.000819 & 0.084534 & 0.086462 & --- \\*
DRSA (optimized) & 0.001269 & 0.075799 & 0.079182 & --- \\*
DynamicDeepHit-Lite & 0.001231 & 0.076062 & 0.078888 & --- \\*
SurvTransformer-Lite & 0.000742 & 0.075834 & 0.077457 & --- \\*
Dynamic-DeepHit (optimized) & 0.002024 & 0.020137 & 0.025122 & --- \\*
SurvTransformer & 0.006152 & 0.043631 & 0.059791 & --- \\*
SurvTransformer-Discrete & 0.006717 & 0.040613 & 0.057940 & --- \\*
ATLAS-GRU / Sobol & 0.000962 & 0.005345 & 0.007767 & 0.000049 \\*
ATLAS-Transformer / Sobol & 0.002568 & 0.006094 & 0.012472 & 0.000047 \\
\addlinespace
\multicolumn{5}{l}{\textit{Death A cumulative incidence, landmark $L=0$}}\\*
Weibull-MLP & 0.000326 & 0.011411 & 0.012281 & --- \\*
DRSA (optimized) & 0.008298 & 0.044096 & 0.064585 & --- \\*
DynamicDeepHit-Lite & 0.001934 & 0.015571 & 0.020233 & --- \\*
SurvTransformer-Lite & 0.005961 & 0.014263 & 0.026971 & --- \\*
Dynamic-DeepHit (optimized) & 0.001721 & 0.009296 & 0.013874 & --- \\*
SurvTransformer & 0.006619 & 0.013729 & 0.028147 & --- \\*
SurvTransformer-Discrete & 0.005366 & 0.015126 & 0.027732 & --- \\*
ATLAS-GRU / Sobol & 0.000316 & 0.005458 & 0.006254 & 0.000085 \\*
ATLAS-Transformer / Sobol & 0.002202 & 0.004788 & 0.010554 & 0.000097 \\
\addlinespace
\multicolumn{5}{l}{\textit{Death A cumulative incidence, landmark $L=10$}}\\*
Weibull-MLP & 0.000295 & 0.091250 & 0.092017 & --- \\*
DRSA (optimized) & 0.001538 & 0.094405 & 0.097997 & --- \\*
DynamicDeepHit-Lite & 0.000848 & 0.091948 & 0.093940 & --- \\*
SurvTransformer-Lite & 0.001127 & 0.091304 & 0.094065 & --- \\*
Dynamic-DeepHit (optimized) & 0.002066 & 0.015511 & 0.021124 & --- \\*
SurvTransformer & 0.005751 & 0.027236 & 0.041741 & --- \\*
SurvTransformer-Discrete & 0.004956 & 0.025591 & 0.038542 & --- \\*
ATLAS-GRU / Sobol & 0.000816 & 0.006142 & 0.008229 & 0.000072 \\*
ATLAS-Transformer / Sobol & 0.002119 & 0.006648 & 0.012090 & 0.000061 \\
\addlinespace
\multicolumn{5}{l}{\textit{Death A cumulative incidence, landmark $L=20$}}\\*
Weibull-MLP & 0.000749 & 0.073952 & 0.076010 & --- \\*
DRSA (optimized) & 0.001191 & 0.067890 & 0.070812 & --- \\*
DynamicDeepHit-Lite & 0.001343 & 0.068056 & 0.071293 & --- \\*
SurvTransformer-Lite & 0.000892 & 0.068454 & 0.070783 & --- \\*
Dynamic-DeepHit (optimized) & 0.001854 & 0.014048 & 0.018435 & --- \\*
SurvTransformer & 0.006075 & 0.030907 & 0.046494 & --- \\*
SurvTransformer-Discrete & 0.004897 & 0.030069 & 0.041021 & --- \\*
ATLAS-GRU / Sobol & 0.000581 & 0.003901 & 0.005467 & 0.000045 \\*
ATLAS-Transformer / Sobol & 0.001945 & 0.004011 & 0.008758 & 0.000040 \\
\addlinespace
\multicolumn{5}{l}{\textit{Death B cumulative incidence, landmark $L=0$}}\\*
Weibull-MLP & 0.000737 & 0.005582 & 0.007418 & --- \\*
DRSA (optimized) & 0.004795 & 0.029774 & 0.041060 & --- \\*
DynamicDeepHit-Lite & 0.000678 & 0.008407 & 0.010114 & --- \\*
SurvTransformer-Lite & 0.001600 & 0.007171 & 0.011227 & --- \\*
Dynamic-DeepHit (optimized) & 0.003015 & 0.007918 & 0.014989 & --- \\*
SurvTransformer & 0.001977 & 0.009643 & 0.015115 & --- \\*
SurvTransformer-Discrete & 0.001696 & 0.010788 & 0.014838 & --- \\*
ATLAS-GRU / Sobol & 0.000898 & 0.004206 & 0.006248 & 0.000063 \\*
ATLAS-Transformer / Sobol & 0.003685 & 0.003800 & 0.012715 & 0.000074 \\
\addlinespace
\multicolumn{5}{l}{\textit{Death B cumulative incidence, landmark $L=10$}}\\*
Weibull-MLP & 0.000204 & 0.045392 & 0.045891 & --- \\*
DRSA (optimized) & 0.001435 & 0.045525 & 0.049102 & --- \\*
DynamicDeepHit-Lite & 0.000454 & 0.043682 & 0.044718 & --- \\*
SurvTransformer-Lite & 0.000566 & 0.043382 & 0.044782 & --- \\*
Dynamic-DeepHit (optimized) & 0.001836 & 0.011190 & 0.015593 & --- \\*
SurvTransformer & 0.002759 & 0.018397 & 0.024082 & --- \\*
SurvTransformer-Discrete & 0.001679 & 0.018373 & 0.022484 & --- \\*
ATLAS-GRU / Sobol & 0.000551 & 0.003839 & 0.005024 & 0.000057 \\*
ATLAS-Transformer / Sobol & 0.001359 & 0.004914 & 0.008365 & 0.000061 \\
\addlinespace
\multicolumn{5}{l}{\textit{Death B cumulative incidence, landmark $L=20$}}\\*
Weibull-MLP & 0.000153 & 0.033250 & 0.033638 & --- \\*
DRSA (optimized) & 0.000454 & 0.031955 & 0.033077 & --- \\*
DynamicDeepHit-Lite & 0.000126 & 0.032086 & 0.032402 & --- \\*
SurvTransformer-Lite & 0.000299 & 0.031806 & 0.032552 & --- \\*
Dynamic-DeepHit (optimized) & 0.001152 & 0.008779 & 0.011758 & --- \\*
SurvTransformer & 0.002111 & 0.018771 & 0.024704 & --- \\*
SurvTransformer-Discrete & 0.001577 & 0.019173 & 0.022617 & --- \\*
ATLAS-GRU / Sobol & 0.000408 & 0.002855 & 0.003898 & 0.000037 \\*
ATLAS-Transformer / Sobol & 0.000999 & 0.003938 & 0.006411 & 0.000034 \\
\addlinespace
\end{longtable}
\endgroup

\begingroup\small
\setlength{\tabcolsep}{4pt}
\begin{longtable}{rrrlr}
\caption{Individual competing-risks training-run MAEs. GRU and Transformer refer to ATLAS with Sobol, averaging five sampling seeds at $M=10{,}000$. The final columns identify the lowest-error fitted direct comparator within each training run, landmark and target. All five runs are retained.}\label{tab:cr_checkpoint_scores}\\
\toprule
Seed & GRU & Transformer & Best direct model & MAE \\\midrule
\endfirsthead
\multicolumn{5}{l}{\tablename\ \thetable\ (continued)}\\
\toprule
Seed & GRU & Transformer & Best direct model & MAE \\\midrule
\endhead
\midrule\multicolumn{5}{r}{Continued on next page}\\\endfoot
\bottomrule\endlastfoot
\multicolumn{5}{l}{\textit{Overall survival, landmark $L=0$}}\\*
42 & 0.007311 & 0.005860 & Dynamic-DeepHit (optimized) & 0.014159 \\*
43 & 0.006763 & 0.010463 & Weibull-MLP & 0.017149 \\*
45 & 0.007703 & 0.016214 & Weibull-MLP & 0.015290 \\*
46 & 0.009192 & 0.012458 & Weibull-MLP & 0.015318 \\*
47 & 0.008540 & 0.006897 & Weibull-MLP & 0.014998 \\
\addlinespace
\multicolumn{5}{l}{\textit{Overall survival, landmark $L=10$}}\\*
42 & 0.008543 & 0.009096 & Dynamic-DeepHit (optimized) & 0.023364 \\*
43 & 0.008740 & 0.009979 & Dynamic-DeepHit (optimized) & 0.024414 \\*
45 & 0.008022 & 0.013370 & Dynamic-DeepHit (optimized) & 0.023185 \\*
46 & 0.009545 & 0.016585 & Dynamic-DeepHit (optimized) & 0.022880 \\*
47 & 0.009274 & 0.009605 & Dynamic-DeepHit (optimized) & 0.021463 \\
\addlinespace
\multicolumn{5}{l}{\textit{Overall survival, landmark $L=20$}}\\*
42 & 0.006173 & 0.006094 & Dynamic-DeepHit (optimized) & 0.022848 \\*
43 & 0.006890 & 0.007659 & Dynamic-DeepHit (optimized) & 0.025122 \\*
45 & 0.005345 & 0.007877 & Dynamic-DeepHit (optimized) & 0.020137 \\*
46 & 0.007345 & 0.012472 & Dynamic-DeepHit (optimized) & 0.024227 \\*
47 & 0.007767 & 0.006370 & Dynamic-DeepHit (optimized) & 0.021430 \\
\addlinespace
\multicolumn{5}{l}{\textit{Death A cumulative incidence, landmark $L=0$}}\\*
42 & 0.005870 & 0.004788 & Dynamic-DeepHit (optimized) & 0.009296 \\*
43 & 0.006254 & 0.007143 & Weibull-MLP & 0.012281 \\*
45 & 0.006104 & 0.007306 & Dynamic-DeepHit (optimized) & 0.011359 \\*
46 & 0.005458 & 0.010554 & Dynamic-DeepHit (optimized) & 0.010783 \\*
47 & 0.005703 & 0.005658 & Weibull-MLP & 0.011903 \\
\addlinespace
\multicolumn{5}{l}{\textit{Death A cumulative incidence, landmark $L=10$}}\\*
42 & 0.008229 & 0.007508 & Dynamic-DeepHit (optimized) & 0.017679 \\*
43 & 0.007381 & 0.006648 & Dynamic-DeepHit (optimized) & 0.021124 \\*
45 & 0.006142 & 0.008001 & Dynamic-DeepHit (optimized) & 0.017833 \\*
46 & 0.006468 & 0.012090 & Dynamic-DeepHit (optimized) & 0.016918 \\*
47 & 0.007033 & 0.007842 & Dynamic-DeepHit (optimized) & 0.015511 \\
\addlinespace
\multicolumn{5}{l}{\textit{Death A cumulative incidence, landmark $L=20$}}\\*
42 & 0.005104 & 0.004011 & Dynamic-DeepHit (optimized) & 0.014806 \\*
43 & 0.004904 & 0.005644 & Dynamic-DeepHit (optimized) & 0.017678 \\*
45 & 0.003901 & 0.004401 & Dynamic-DeepHit (optimized) & 0.016148 \\*
46 & 0.004777 & 0.008758 & Dynamic-DeepHit (optimized) & 0.018435 \\*
47 & 0.005467 & 0.004457 & Dynamic-DeepHit (optimized) & 0.014048 \\
\addlinespace
\multicolumn{5}{l}{\textit{Death B cumulative incidence, landmark $L=0$}}\\*
42 & 0.004206 & 0.003800 & Weibull-MLP & 0.005582 \\*
43 & 0.004357 & 0.005023 & SurvTransformer-Lite & 0.007171 \\*
45 & 0.004380 & 0.012715 & Weibull-MLP & 0.006350 \\*
46 & 0.006248 & 0.005295 & Weibull-MLP & 0.007133 \\*
47 & 0.005508 & 0.004221 & Weibull-MLP & 0.006228 \\
\addlinespace
\multicolumn{5}{l}{\textit{Death B cumulative incidence, landmark $L=10$}}\\*
42 & 0.003839 & 0.004914 & Dynamic-DeepHit (optimized) & 0.011498 \\*
43 & 0.003907 & 0.005570 & Dynamic-DeepHit (optimized) & 0.015593 \\*
45 & 0.003965 & 0.008365 & Dynamic-DeepHit (optimized) & 0.012237 \\*
46 & 0.004759 & 0.006684 & Dynamic-DeepHit (optimized) & 0.011190 \\*
47 & 0.005024 & 0.005556 & Dynamic-DeepHit (optimized) & 0.011406 \\
\addlinespace
\multicolumn{5}{l}{\textit{Death B cumulative incidence, landmark $L=20$}}\\*
42 & 0.002855 & 0.003938 & Dynamic-DeepHit (optimized) & 0.011758 \\*
43 & 0.003123 & 0.004382 & Dynamic-DeepHit (optimized) & 0.011256 \\*
45 & 0.003034 & 0.006411 & Dynamic-DeepHit (optimized) & 0.008779 \\*
46 & 0.003414 & 0.004685 & Dynamic-DeepHit (optimized) & 0.010325 \\*
47 & 0.003898 & 0.004083 & Dynamic-DeepHit (optimized) & 0.010982 \\
\addlinespace
\end{longtable}
\endgroup

\paragraph{Competing-risks sensitivity to including training seed 42.} The primary analysis retains all five full runs. A descriptive sensitivity analysis uses the four runs on the same recorded GPU model, seeds 43, 45, 46 and 47. An ATLAS configuration still has the lowest fitted mean MAE in all nine landmark--target settings; GRU leads all nine, with conditional Monte Carlo in eight and Sobol for survival at landmark 10. However, for baseline Death B, only eight of the fourteen ATLAS configurations beat the best direct comparator in the four-run means, compared with all fourteen in the five-run means. The best-configuration result is therefore retained, while the broader all-configurations comparison is more sensitive to which runs are included. This is a sensitivity check, not a reason to exclude seed 42 or a test of statistical significance. The complete comparison is supplied in the competing-risks multi-seed supplement.

\section{Full model--sampler comparisons}\label{supp:figures}
Figures~\ref{fig:hard24_nine_L0} to~\ref{fig:cr3_five_L20} show every model and sampler at baseline and at landmark 20 in both experiments, with individual training-run scores. In the figure labels, Weibull denotes the Weibull-MLP direct model and the oracle is the true-generator sampler.
\clearpage
\begingroup
\newgeometry{layoutwidth=11in,layoutheight=8.5in,margin=0.5in}
\paperwidth=11in\paperheight=8.5in
\ifdefined\pdfpagewidth\pdfpagewidth=\paperwidth\pdfpageheight=\paperheight
\else\ifdefined\pagewidth\pagewidth=\paperwidth\pageheight=\paperheight
\else\fi\fi
\captionsetup{font=small}
\centering
\noindent\includegraphics[width=0.95\linewidth,height=0.80\textheight,keepaspectratio]{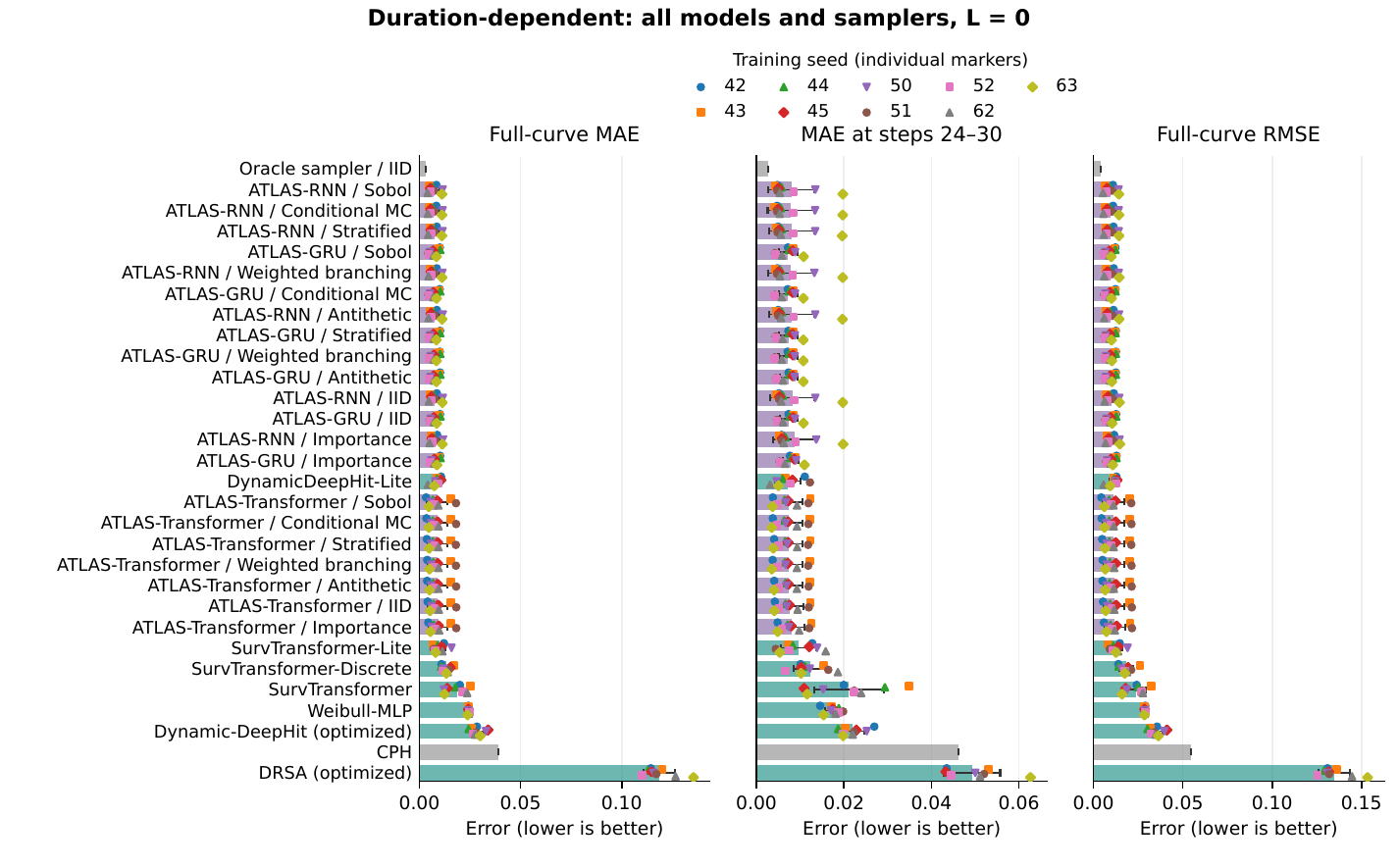}\par
\captionof{figure}{Nine-training-run comparison at $L=0$ in the single-covariate duration-dependent experiment. The panels show full-curve MAE, MAE restricted to steps 24--30, and RMSE. Bars are means over training runs; colored markers show individual runs and error bars show their standard deviation. Each stochastic training-run score averages five sampling seeds at $M=10{,}000$. CPH and the true-generator oracle are unchanged controls, shown from the seed-42 source without training-variation error bars. Dynamic-DeepHit (optimized) and DRSA (optimized) use the validation-selected fits, and row order follows the updated mean survival MAE. Lower error is better.}\label{fig:hard24_nine_L0}\par
\clearpage
\noindent\includegraphics[width=0.95\linewidth,height=0.80\textheight,keepaspectratio]{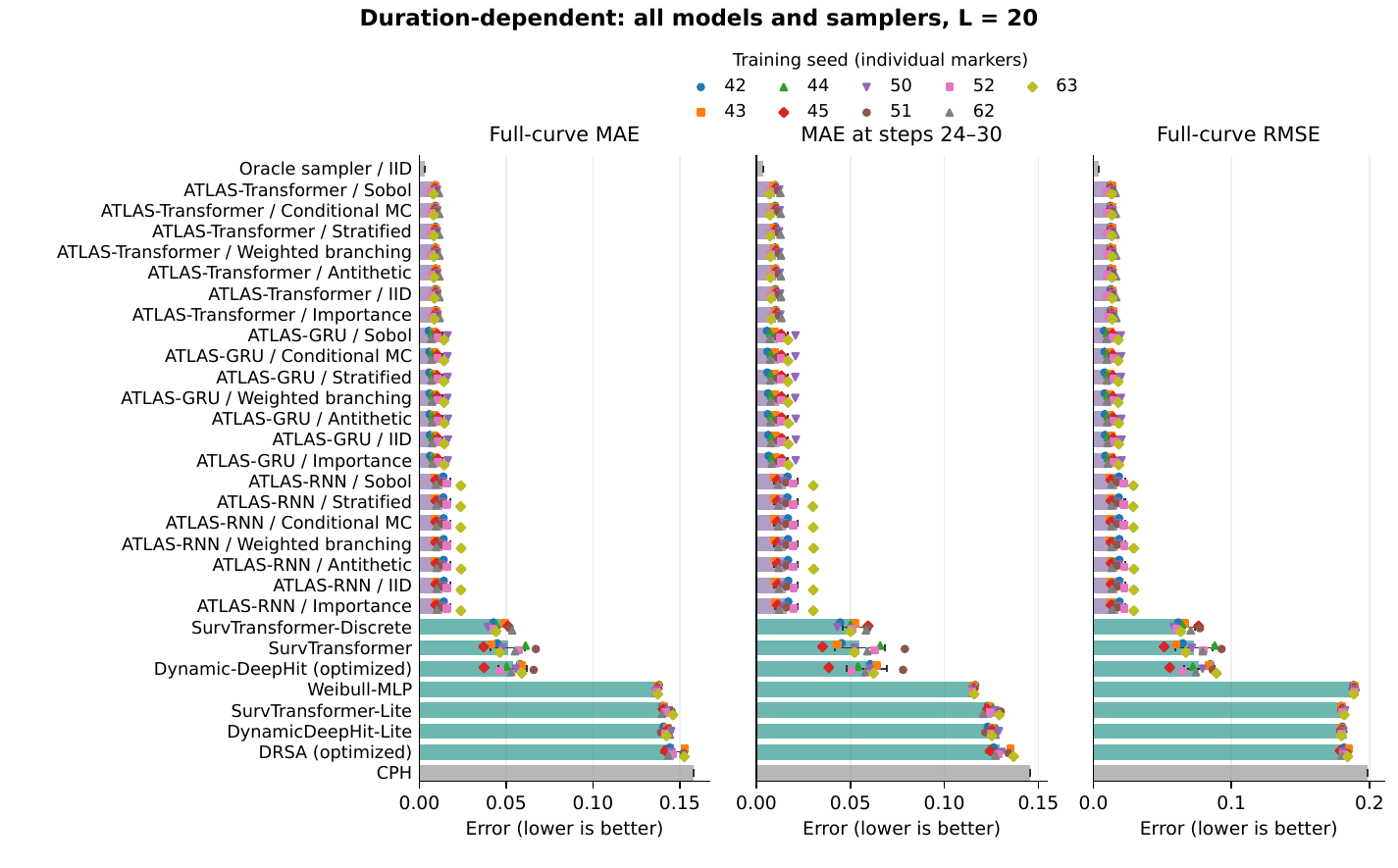}\par
\captionof{figure}{Nine-training-run comparison at $L=20$ in the single-covariate duration-dependent experiment. The panels show full-curve MAE, MAE restricted to steps 24--30, and RMSE. Bars are means over training runs; colored markers show individual runs and error bars show their standard deviation. Each stochastic training-run score averages five sampling seeds at $M=10{,}000$. CPH and the true-generator oracle are unchanged controls, shown from the seed-42 source without training-variation error bars. Dynamic-DeepHit (optimized) and DRSA (optimized) use the validation-selected fits, and row order follows the updated mean survival MAE. Lower error is better.}\label{fig:hard24_nine_L20}\par
\clearpage
\restoregeometry
\paperwidth=8.5in\paperheight=11in
\ifdefined\pdfpagewidth\pdfpagewidth=\paperwidth\pdfpageheight=\paperheight
\else\ifdefined\pagewidth\pagewidth=\paperwidth\pageheight=\paperheight
\else\fi\fi
\endgroup

\clearpage
\begingroup
\newgeometry{layoutwidth=11in,layoutheight=8.5in,margin=0.5in}
\paperwidth=11in\paperheight=8.5in
\ifdefined\pdfpagewidth\pdfpagewidth=\paperwidth\pdfpageheight=\paperheight
\else\ifdefined\pagewidth\pagewidth=\paperwidth\pageheight=\paperheight
\else\fi\fi
\captionsetup{font=small}
\centering

\noindent\includegraphics[width=0.95\linewidth,height=0.80\textheight,keepaspectratio]{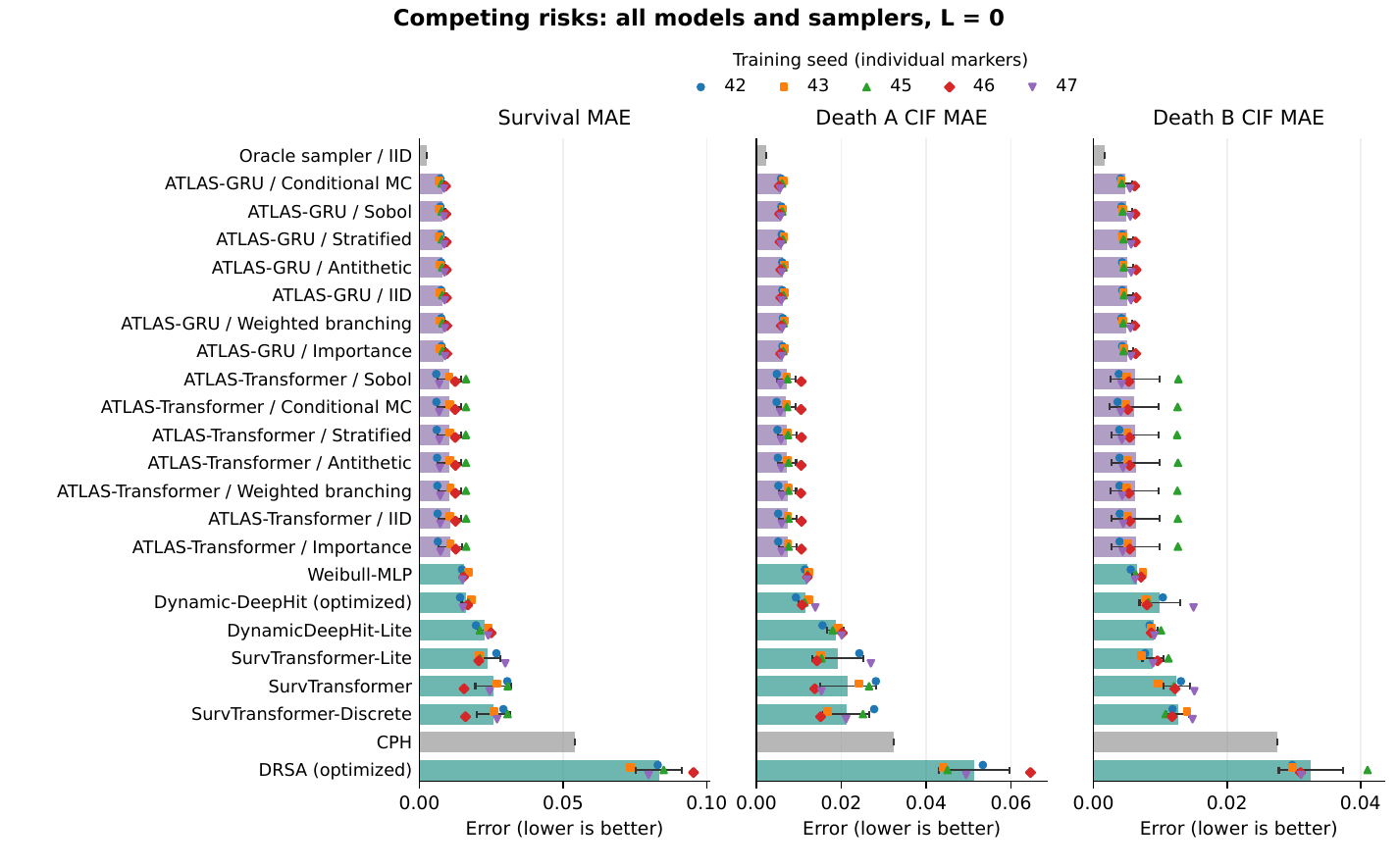}\par
\captionof{figure}{Five-training-run comparison at $L=0$ in the multi-covariate competing-risks experiment. The panels show survival, Death A cumulative incidence and Death B cumulative incidence MAE. Bars are means across training runs; markers show individual runs and error bars show one sample standard deviation across runs. Each stochastic training-run score averages five sampling seeds at $M=10{,}000$. These are training-run standard deviations, not confidence intervals or within-run sampling standard deviations. CPH predictions agree to numerical precision; the oracle is a reused sampling control. Dynamic-DeepHit (optimized) and DRSA (optimized) use the validation-selected fits, and row order follows the updated mean survival MAE. Lower error is better.}\label{fig:cr3_five_L0}\par
\clearpage
\noindent\includegraphics[width=0.95\linewidth,height=0.80\textheight,keepaspectratio]{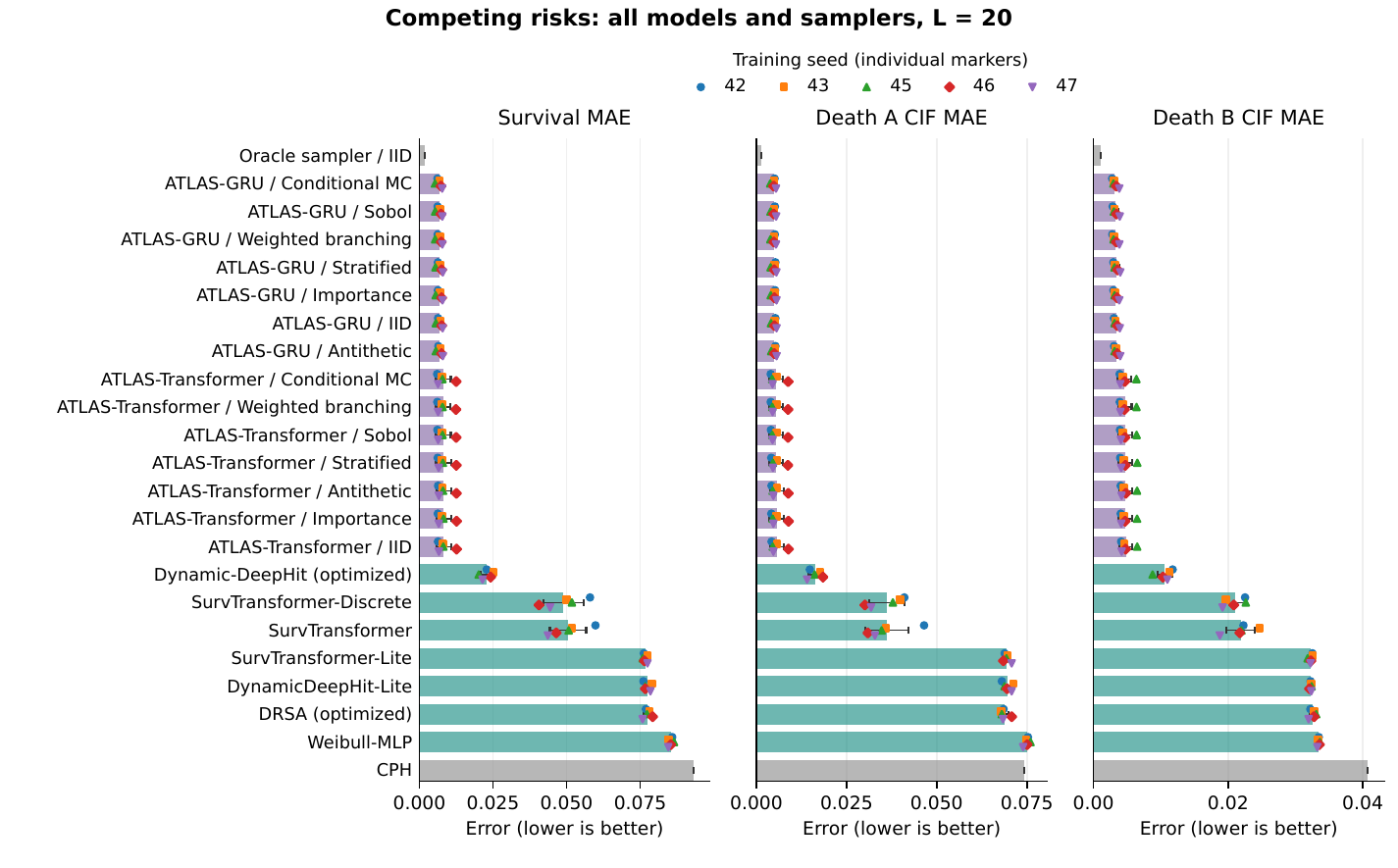}\par
\captionof{figure}{Five-training-run comparison at $L=20$ in the multi-covariate competing-risks experiment. The panels show survival, Death A cumulative incidence and Death B cumulative incidence MAE. Bars are means across training runs; markers show individual runs and error bars show one sample standard deviation across runs. Each stochastic training-run score averages five sampling seeds at $M=10{,}000$. These are training-run standard deviations, not confidence intervals or within-run sampling standard deviations. CPH predictions agree to numerical precision; the oracle is a reused sampling control. Dynamic-DeepHit (optimized) and DRSA (optimized) use the validation-selected fits, and row order follows the updated mean survival MAE. Lower error is better.}\label{fig:cr3_five_L20}\par
\clearpage
\restoregeometry
\paperwidth=8.5in\paperheight=11in
\ifdefined\pdfpagewidth\pdfpagewidth=\paperwidth\pdfpageheight=\paperheight
\else\ifdefined\pagewidth\pagewidth=\paperwidth\pageheight=\paperheight
\else\fi\fi
\endgroup

\section{Sampling estimator details}\label{supp:samplers}
This section expands the sampler table in Appendix D of the main text. Within each ATLAS architecture and training run, all seven samplers share the same checkpoint and change only how walks or particles are constructed.

Stratification uses fresh randomized assignments at each step. Antithetic and Sobol walks are deliberately dependent. Although the intended marginal path law is retained, the finite-sample product-limit estimator is nonlinear, so identical marginal draws do not guarantee unbiased or lower-variance Kaplan--Meier and Aalen--Johansen estimates.

In importance sampling, let $q$ be the altered proposal and $p_\theta$ the learned law. For a path ending at observed time $D_m$, the weight is
\[
w_m=\prod_{j=L+1}^{D_m}\frac{p_\theta(s_{mj}\mid h_{m,j-1},x)}{q(s_{mj}\mid h_{m,j-1},x)}.
\]
The product stops at the first absorbing outcome or at the horizon. Both death causes are tilted in the competing-risks experiment. A common scaling of the weights cancels from the weighted risk and event ratios. The weight-based effective sample size $(\sum_mw_m)^2/\sum_mw_m^2$ describes weight concentration and is not a measure of learned-model accuracy.

For the particle methods, normalized weights $a_m$ represent the current live observed-history distribution. The death fraction for cause $k$ is integrated as $\sum_ma_mp_{m,k}$ and enters the Kaplan--Meier and Aalen--Johansen updates. Conditional Monte Carlo samples a live successor from the normalized live probabilities and multiplies the path weight by its total live probability before renormalizing. Branching instead enumerates live successors with masses $a_mp_{m,s}$ and resamples from their combined distribution.

Resampling can copy the same ancestor more than once, and equal weights afterward do not make the resulting paths independent. The competing-risks implementation records the particle effective sample size before resampling; the earlier implementation's post-resampling value must not be read as a count of independent paths. Ordinary independent-observation Kaplan--Meier uncertainty formulas do not apply to these correlated designs without an appropriate sampling-design argument.

At exactly 10,000 points, Sobol lacks its usual power-of-two balance guarantee~\citep{owen1998,owen2013}. The reported sampler comparisons are empirical for this budget, these training runs, and these targets. The source package keeps the original records and separately named multi-seed results, with per-run scores, training-run and within-run sampling standard deviations, and timing provenance, so that point-estimate improvements are not mistaken for universal guarantees.

\section{Sampling resources on the original duration-dependent scenarios}\label{sec:budget}
This is an earlier, separate resource experiment. It holds the trained models fixed and changes only the number of generated walks, using the original duration-dependent scenarios without the later step-24 rule. The two scenarios, labeled Weibull and ReLU, differ in the covariate map and covariate distribution; the Weibull scenario is the one described in Section~2 of the main text, and the label refers to that scenario, not to the Weibull-MLP direct model. The direct comparators are the Lite implementations.
\begin{center}\begin{minipage}{\textwidth}\centering
\includegraphics[width=\textwidth]{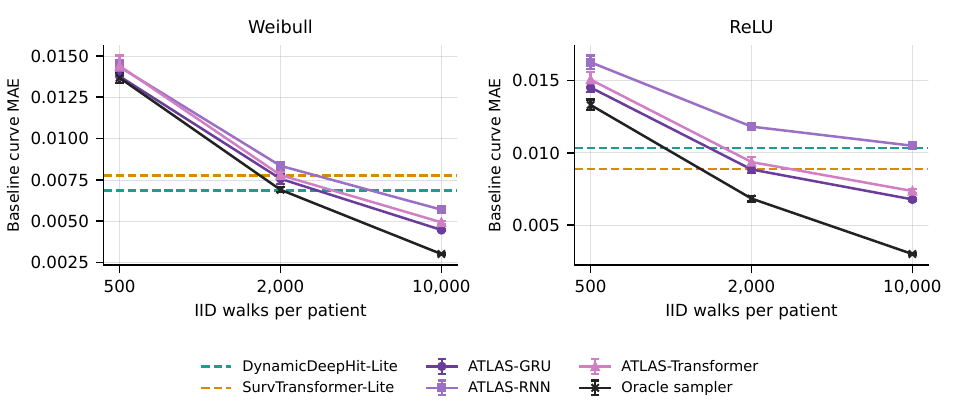}
\captionof{figure}{Resource study without the step-24 rule. Baseline MAE against deterministic references, using fixed previously trained checkpoints and five IID sampling seeds. Error bars are one seed standard deviation, and dashed lines are the fixed direct predictions.}
\end{minipage}\end{center}

Increasing the IID budget reduces mean MAE in all 24 ATLAS scenario--landmark combinations, from 500 to 2,000 and again to 10,000 samples. The oracle-sampler error falls from roughly 0.0133 to 0.0142 at 500 down to 0.00296 to 0.00311 at 10,000. So a substantial share of the low-budget error is numerical. That does not mean fitted-model error disappears as sampling increases.

In the Weibull scenario at baseline, the GRU, RNN, and Transformer errors fall from 0.01377, 0.01434, and 0.01440 to 0.00447, 0.00569, and 0.00493. All three then beat DynamicDeepHit-Lite (0.00686) and SurvTransformer-Lite (0.00777). In other words, increasing prediction resources alone reverses those baseline mean-MAE comparisons, without any retraining.

The ReLU scenario supplies a useful counterexample. At 10,000 samples the GRU (0.00677) and Transformer (0.00735) beat both direct comparators, but the RNN (0.01048) remains above SurvTransformer-Lite (0.00889) and DynamicDeepHit-Lite (0.01035). At the tested nonzero landmarks, all three ATLAS models lead the two restricted direct comparators, though as before this involves different input information. Full budgets, landmarks, and variance diagnostics appear in Section~\ref{app:budgetresults}.

The single-covariate study in the main text uses a different cohort and checkpoints fitted for that process, later augmented with the history-aware comparators. Its baseline advantage should not be read as a controlled consequence of adding the hard rule alone. The resource study also replaces the earlier Monte Carlo reference with a deterministic one, so its absolute scores need not reproduce older tables computed against 2,000-trajectory reference estimates.

\subsection{Complete sampling-resource results}\label{app:budgetresults}
This separate study uses the original duration-dependent ramp with the step-20 change and no added step-24 deterministic override. Checkpoints, patients and reference curves are fixed across budgets. Each entry is mean curve MAE across five IID sampling seeds. All 24 ATLAS rows improve monotonically across the three budgets. ``Oracle'' is the true-generator sampler, not a fitted method.
\small
\begin{longtable}{llrrrr}
\caption{Complete sampling-resource results, mean curve MAE at each IID walk budget.}\label{tab:budgetall}\\
\toprule Scenario & Model & $L$ & $M=500$ & $M=2000$ & $M=10000$\\\midrule\endfirsthead
\toprule Scenario & Model & $L$ & $M=500$ & $M=2000$ & $M=10000$\\\midrule\endhead
ReLU & ATLAS-GRU & 0 & 0.014518 & 0.008841 & 0.006769 \\
ReLU & ATLAS-RNN & 0 & 0.016263 & 0.011810 & 0.010484 \\
ReLU & ATLAS-Transformer & 0 & 0.015063 & 0.009355 & 0.007348 \\
ReLU & Oracle & 0 & 0.013307 & 0.006825 & 0.002985 \\
ReLU & ATLAS-GRU & 10 & 0.014247 & 0.008761 & 0.006089 \\
ReLU & ATLAS-RNN & 10 & 0.016386 & 0.011745 & 0.010171 \\
ReLU & ATLAS-Transformer & 10 & 0.015216 & 0.009460 & 0.007468 \\
ReLU & Oracle & 10 & 0.013598 & 0.006749 & 0.003088 \\
ReLU & ATLAS-GRU & 15 & 0.014179 & 0.007618 & 0.004774 \\
ReLU & ATLAS-RNN & 15 & 0.016224 & 0.010833 & 0.008807 \\
ReLU & ATLAS-Transformer & 15 & 0.015783 & 0.009693 & 0.007584 \\
ReLU & Oracle & 15 & 0.014118 & 0.006693 & 0.003040 \\
ReLU & ATLAS-GRU & 20 & 0.014101 & 0.007697 & 0.005037 \\
ReLU & ATLAS-RNN & 20 & 0.015506 & 0.010321 & 0.008430 \\
ReLU & ATLAS-Transformer & 20 & 0.014544 & 0.009720 & 0.007861 \\
ReLU & Oracle & 20 & 0.013414 & 0.006423 & 0.003024 \\
Weibull & ATLAS-GRU & 0 & 0.013768 & 0.007565 & 0.004467 \\
Weibull & ATLAS-RNN & 0 & 0.014339 & 0.008350 & 0.005694 \\
Weibull & ATLAS-Transformer & 0 & 0.014399 & 0.007788 & 0.004929 \\
Weibull & Oracle & 0 & 0.013677 & 0.006900 & 0.003010 \\
Weibull & ATLAS-GRU & 10 & 0.014386 & 0.008124 & 0.005239 \\
Weibull & ATLAS-RNN & 10 & 0.015153 & 0.009876 & 0.007502 \\
Weibull & ATLAS-Transformer & 10 & 0.015230 & 0.009429 & 0.006946 \\
Weibull & Oracle & 10 & 0.013939 & 0.006827 & 0.003113 \\
Weibull & ATLAS-GRU & 15 & 0.014626 & 0.009031 & 0.006469 \\
Weibull & ATLAS-RNN & 15 & 0.015845 & 0.010777 & 0.008936 \\
Weibull & ATLAS-Transformer & 15 & 0.015952 & 0.010113 & 0.008064 \\
Weibull & Oracle & 15 & 0.014236 & 0.006960 & 0.003054 \\
Weibull & ATLAS-GRU & 20 & 0.015615 & 0.010581 & 0.009266 \\
Weibull & ATLAS-RNN & 20 & 0.015003 & 0.010222 & 0.008346 \\
Weibull & ATLAS-Transformer & 20 & 0.013618 & 0.007538 & 0.004785 \\
Weibull & Oracle & 20 & 0.013597 & 0.006617 & 0.002957 \\
\bottomrule\end{longtable}\normalsize
The archive also records seed SD, ensemble-curve MAE, pointwise sampling variance, and an exact finite-replicate MSE identity.  For $R$ prediction replicates with mean curve $\bar S$, the averaged squared error satisfies
\[
 \frac1R\sum_{r=1}^R\|\widehat S_r-S^*\|^2
 =\|\bar S-S^*\|^2+\frac1R\sum_{r=1}^R\|\widehat S_r-\bar S\|^2,
\]
where the norm averages over patients and scored horizons. The mean-curve term still contains finite-replicate Monte Carlo error and is not pure model bias. The raw variance and comparison tables ship with the source. Baseline references here are deterministic, whereas the older local runs used 2,000 simulated reference trajectories, so their absolute MAEs are not directly interchangeable.

\section{Sampler ablation tables}\label{supp:ablation}
The tables below compare all seven sampling methods for each ATLAS architecture.

\begin{center}\begin{minipage}{\textwidth}\centering\small
\begin{tabular}{llrrrr}
\toprule
Architecture & Sampler & $L=0$ & $L=20$ & $L=23$ & $L=24$ \\\midrule
ATLAS-GRU & IID & 0.007687 & 0.010125 & 0.013912 & 0.013947 \\
ATLAS-GRU & Stratified & 0.007485 & 0.009886 & 0.013760 & 0.013628 \\
ATLAS-GRU & Sobol & 0.007371 & 0.009763 & 0.013647 & 0.013617 \\
ATLAS-GRU & Antithetic & 0.007582 & 0.010035 & 0.013958 & 0.013886 \\
ATLAS-GRU & Importance & 0.007840 & 0.010221 & 0.014056 & 0.013933 \\
ATLAS-GRU & Branching & 0.007556 & 0.009951 & 0.013734 & 0.013664 \\
ATLAS-GRU & Conditional MC & 0.007430 & 0.009879 & 0.013757 & 0.013597 \\
\addlinespace
ATLAS-RNN & IID & 0.007590 & 0.013357 & 0.016144 & 0.012008 \\
ATLAS-RNN & Stratified & 0.007350 & 0.013159 & 0.015964 & 0.011597 \\
ATLAS-RNN & Sobol & 0.007196 & 0.013113 & 0.015888 & 0.011492 \\
ATLAS-RNN & Antithetic & 0.007465 & 0.013332 & 0.016192 & 0.011835 \\
ATLAS-RNN & Importance & 0.007789 & 0.013399 & 0.016215 & 0.011986 \\
ATLAS-RNN & Branching & 0.007409 & 0.013232 & 0.015923 & 0.011623 \\
ATLAS-RNN & Conditional MC & 0.007262 & 0.013192 & 0.016006 & 0.011578 \\
\addlinespace
ATLAS-Transformer & IID & 0.009145 & 0.009392 & 0.012812 & 0.010920 \\
ATLAS-Transformer & Stratified & 0.008942 & 0.009178 & 0.012648 & 0.010649 \\
ATLAS-Transformer & Sobol & 0.008818 & 0.009040 & 0.012525 & 0.010494 \\
ATLAS-Transformer & Antithetic & 0.009014 & 0.009306 & 0.012866 & 0.010862 \\
ATLAS-Transformer & Importance & 0.009328 & 0.009497 & 0.012875 & 0.010937 \\
ATLAS-Transformer & Branching & 0.009008 & 0.009236 & 0.012667 & 0.010524 \\
ATLAS-Transformer & Conditional MC & 0.008864 & 0.009155 & 0.012621 & 0.010566 \\
\addlinespace
\bottomrule\end{tabular}
\captionof{table}{All seven sampling methods: mean survival-curve MAE at $M=10{,}000$ for the single-covariate experiment, averaged equally over nine training checkpoints after averaging five sampling seeds within each checkpoint. All methods within an architecture and training run share its checkpoint. Table~\ref{tab:harddispersion} reports training-run and sampling dispersion for the common Sobol configurations, and Table~\ref{tab:hard24_checkpoint_scores} gives their individual training-run scores.}
\end{minipage}\end{center}

For the single-covariate experiment, Sobol has the lowest MAE averaged across the nine training runs in eleven of the twelve architecture--landmark comparisons. The exception is ATLAS-GRU at $L=24$, where conditional Monte Carlo has mean MAE 0.013597 and Sobol has 0.013617. The difference is approximately $2.0\times10^{-5}$. These are comparisons of the recorded mean errors at this sampling budget.

Sobol improves on IID in all twelve comparisons of nine-training-run mean MAE. For Transformer at baseline, Sobol reduces mean MAE from 0.009145 to 0.008818 (3.57\%); for GRU at landmark 24, the reduction is from 0.013947 to 0.013617 (2.37\%).

\begin{center}\begin{minipage}{\textwidth}\centering\small
\begin{tabular}{llrrr}
\toprule
Architecture & Sampler & $L=0$ & $L=10$ & $L=20$ \\\midrule
ATLAS-GRU & IID & 0.008187 & 0.009099 & 0.006903 \\
ATLAS-GRU & Stratified & 0.008032 & 0.008932 & 0.006844 \\
ATLAS-GRU & Sobol & 0.007902 & 0.008825 & 0.006704 \\
ATLAS-GRU & Antithetic & 0.008131 & 0.009001 & 0.006933 \\
ATLAS-GRU & Importance & 0.008286 & 0.009170 & 0.006853 \\
ATLAS-GRU & Branching & 0.008208 & 0.008851 & 0.006728 \\
ATLAS-GRU & Conditional MC & 0.007894 & 0.008820 & 0.006674 \\
\addlinespace
ATLAS-Transformer & IID & 0.010625 & 0.011943 & 0.008316 \\
ATLAS-Transformer & Stratified & 0.010425 & 0.011799 & 0.008198 \\
ATLAS-Transformer & Sobol & 0.010378 & 0.011727 & 0.008094 \\
ATLAS-Transformer & Antithetic & 0.010518 & 0.011879 & 0.008268 \\
ATLAS-Transformer & Importance & 0.010711 & 0.012003 & 0.008274 \\
ATLAS-Transformer & Branching & 0.010584 & 0.011785 & 0.008071 \\
ATLAS-Transformer & Conditional MC & 0.010385 & 0.011676 & 0.008058 \\
\addlinespace
\bottomrule\end{tabular}
\captionof{table}{All seven sampling methods: mean survival-curve MAE at $M=10{,}000$ for the competing-risks experiment, averaged equally across five training runs after averaging five sampling seeds within each checkpoint. All methods within an architecture and training run share its checkpoint. Cause-specific sampler scores and both levels of variability are supplied in the multi-seed supplement.}
\end{minipage}\end{center}

For the competing-risks experiment, Sobol improves on IID in all eighteen comparisons of five-training-run mean MAE. Conditional Monte Carlo has the lowest mean error in seventeen comparisons; Sobol leads only for Transformer baseline survival. These rankings describe the tested models, targets and sampling budget, rather than a universal sampler ordering. Complete cause-specific sampler scores, along with training-run and within-run sampling standard deviations, are supplied in the multi-seed results accompanying the source.

The archive supports Sobol as a useful alternative to IID, especially where complete walks are wanted. It does not support calling it uniformly best. The particle methods also estimate different quantities internally and incur different work at the same nominal $M$, so their small MAE differences should be interpreted alongside their construction and cost.

\clearpage
\section{Paired patient-bootstrap uncertainty}\label{app:patientbootstrap}
\subsection{Bootstrap estimands and resampling}
\paragraph{Prediction errors} The analysis used the original 300 test evaluation patients in each experiment and their eligible landmark subsets. The duration-dependent eligible counts were 300, 102, 78 and 57 at landmarks 0, 20, 23 and 24; competing-risks counts were 300, 251 and 179 at landmarks 0, 10 and 20. Survival was evaluated in both experiments and the Death A and Death B CIFs were evaluated separately in the competing-risks experiment.

\paragraph{Averaging over fitted procedures} For each patient, target and landmark, absolute prediction error was averaged over the scored time grid. These patient-specific MAEs were averaged over the five saved ATLAS sampling seeds within a training fit and then equally over nine duration-dependent or five competing-risks fits. Direct neural predictions contributed one patient error per training fit. CPH was a numerically unchanged control. Prediction curves were not averaged before calculating errors.

\paragraph{Paired resampling} Within each experiment, each bootstrap draw sampled 300 baseline evaluation patient IDs with replacement. A patient's multiplicity was shared by every model, landmark and target; observations from different patients were never mixed into one history. The original alive-and-observed eligibility rule was then applied at each landmark. Consequently, later-landmark sample sizes varied between resamples. The two experiments were resampled separately, using random seed 20261030 and independent random-number streams derived from it.

\paragraph{Intervals and contrasts} For model m, let e(i,m,L,k) denote the patient error averaged over the saved fits and sampling repetitions, and let I(i,L) indicate landmark eligibility. The mean-error difference is defined below. In resample b, replace each patient's unit weight by its resampling multiplicity w(i,b). The 2.5th and 97.5th percentiles of the resulting 5,000 differences form the pointwise 95\% interval. Relative reductions, 100 times one minus ATLAS MAE divided by comparator MAE, were also recalculated in each resample.

\[
\widehat\Delta_{AC,Lk}=\frac{\sum_i I_{iL}(e_{iA,Lk}-e_{iC,Lk})}{\sum_i I_{iL}},\qquad
\widehat\Delta^{(b)}_{AC,Lk}=\frac{\sum_i w_i^{(b)}I_{iL}(e_{iA,Lk}-e_{iC,Lk})}{\sum_i w_i^{(b)}I_{iL}}.
\]
\paragraph{Fixed comparisons} Every ATLAS architecture with the common Sobol sampler was compared with each of the eight fitted direct comparators, including Dynamic-DeepHit (optimized) and DRSA (optimized). This gives 96 duration-dependent and 144 competing-risk comparisons. The 13 headline comparisons use ATLAS-Transformer for the duration-dependent experiment and ATLAS-GRU for competing risks. At each setting, the strongest direct comparator was identified from the unresampled updated mean MAEs and held fixed during resampling. These descriptive architecture/comparator choices were not reselected inside the bootstrap; their selection uncertainty is not included.

\paragraph{Interpretation} The resampling unit is the patient, not the trajectory, prediction time, training seed or patient-prefix row. The intervals describe patient-sampling uncertainty conditional on the saved fitted models, their realized sampling outputs and the underlying synthetic cohort. They do not incorporate new-training-cohort variation, model-selection uncertainty or clinical transport. No multiplicity adjustment was made, and the intervals should not be interpreted as a simultaneous confidence statement across comparisons.

\subsection{Bootstrap results}
\paragraph{Duration-dependent experiment} All 72 post-baseline intervals favored ATLAS, covering three architectures, eight direct comparators and three landmarks. Across all four landmarks, 95 of 96 intervals were below zero. The exception was baseline ATLAS-Transformer versus DynamicDeepHit-Lite: the MAE difference was 0.000262, with a 95\% interval from -0.000146 to 0.000685. Thus the baseline Transformer comparison does not establish a clear advantage for either method. GRU and RNN comparisons against the strongest baseline direct comparator had intervals favoring ATLAS, conditional on their averages over saved fits.

\begingroup\emergencystretch=5em
\paragraph{Headline duration-dependent comparisons} ATLAS-Transformer is compared with SurvTransformer-Discrete at landmarks 20 and 23 and Dynamic-DeepHit (optimized) at landmark 24. The MAE differences were -0.037501, -0.044970, -0.054298, respectively, with pointwise intervals [-0.041791, -0.033499], [-0.049395, -0.040651], [-0.064475, -0.044688]. The corresponding relative reductions were 80.6\%, 78.2\%, 83.8\%, with pointwise intervals [78.1\%, 82.8\%], [75.4\%, 81.0\%], [81.7\%, 85.7\%].
\par\endgroup

\paragraph{Competing-risks experiment} All 96 post-baseline intervals favored ATLAS, covering two architectures, eight comparators, three targets and two landmarks. Across baseline and later landmarks, 143 of 144 intervals were below zero. The exception was baseline ATLAS-Transformer versus Weibull-MLP for the Death B CIF: difference -0.000331, with interval [-0.000773, 0.000104]. The ATLAS-GRU headline comparisons had intervals favoring ATLAS for survival and both CIFs at all three landmarks.

\paragraph{Headline competing-risk survival} For ATLAS-GRU versus Dynamic-DeepHit (optimized), survival-MAE differences were -0.014236 at landmark 10 and -0.016049 at landmark 20, with intervals [-0.015992, -0.012561] and [-0.018552, -0.013629]. Relative reductions were 61.7\% and 70.5\%, with intervals [58.3\%, 64.8\%] and [65.7\%, 74.5\%]. Separate CIF results appear in Table~\ref{tab:patientboot_cr} and Figures~\ref{fig:patientboot_1} and~\ref{fig:patientboot_3}.

\paragraph{Reading the tables and figures} Tables~\ref{tab:patientboot_hard} and~\ref{tab:patientboot_cr} report absolute MAE differences; negative values favor ATLAS. Figure~\ref{fig:patientboot_1} instead displays relative MAE reductions, for which positive values favor ATLAS. Figures~\ref{fig:patientboot_2} and~\ref{fig:patientboot_3} show all 240 architecture-comparator settings; hatching identifies the two intervals that include zero. Counts of intervals excluding zero are descriptive summaries of pointwise intervals, not family-wise significance tests or independent replications.

\begin{table}[htbp]\centering\small
\caption{Duration-dependent survival: ATLAS-Transformer versus the strongest fitted direct comparator. Differences are ATLAS minus comparator; negative favors ATLAS. Pointwise 95\% paired patient-bootstrap intervals condition on the saved fits; comparison pairs are held fixed.}\label{tab:patientboot_hard}
\resizebox{\textwidth}{!}{%

\begin{tabular}{lllll}
\hline
L & n & Comparator & Difference & 95\% interval\\\hline
0 & 300 & DynamicDeepHit-Lite & 0.000262 & [-0.000146, 0.000685]\\
20 & 102 & SurvTransformer-Discrete & -0.037501 & [-0.041791, -0.033499]\\
23 & 78 & SurvTransformer-Discrete & -0.044970 & [-0.049395, -0.040651]\\
24 & 57 & Dynamic-DeepHit (optimized) & -0.054298 & [-0.064475, -0.044688]\\
\hline\end{tabular}}
\end{table}
\begin{table}[htbp]\centering\small
\caption{Competing risks: ATLAS-GRU versus the strongest fitted direct comparator. S is survival; A and B are the separate CIFs. Differences are ATLAS minus comparator; negative favors ATLAS. Pointwise 95\% paired patient-bootstrap intervals condition on the saved fits; comparison pairs are held fixed.}\label{tab:patientboot_cr}
\resizebox{\textwidth}{!}{%

\begin{tabular}{llllll}
\hline
L & n & Target & Comparator & Difference & 95\% interval\\\hline
0 & 300 & S & Weibull-MLP & -0.007608 & [-0.008439, -0.006784]\\
0 & 300 & A & Dynamic-DeepHit (optimized) & -0.005667 & [-0.006209, -0.005160]\\
0 & 300 & B & Weibull-MLP & -0.001602 & [-0.002041, -0.001190]\\
10 & 251 & S & Dynamic-DeepHit (optimized) & -0.014236 & [-0.015992, -0.012561]\\
10 & 251 & A & Dynamic-DeepHit (optimized) & -0.010763 & [-0.012401, -0.009266]\\
10 & 251 & B & Dynamic-DeepHit (optimized) & -0.008086 & [-0.009554, -0.006745]\\
20 & 179 & S & Dynamic-DeepHit (optimized) & -0.016049 & [-0.018552, -0.013629]\\
20 & 179 & A & Dynamic-DeepHit (optimized) & -0.011393 & [-0.013524, -0.009444]\\
20 & 179 & B & Dynamic-DeepHit (optimized) & -0.007355 & [-0.009055, -0.005812]\\
\hline\end{tabular}}
\end{table}
\subsection{Scope and limitations}
\paragraph{Patient and training variability answer different questions} The results show that the later-landmark differences persist when the composition of the evaluation sample is resampled. This complements between-training-run variation, while leaving the training data and fitted weights fixed. An interval for the average error across saved fits is not a prediction interval for the performance of a newly trained individual model, and does not imply that every training run wins.

\paragraph{Landmark comparisons} Later landmarks contain fewer eligible patients and different prediction windows. The bootstrap retains these changes rather than treating the same 300 patients as eligible at every landmark. The resulting intervals describe each landmark-specific comparison; differences between landmarks do not isolate the contribution of additional history.

\paragraph{Benchmark scope} The reference curves are known synthetic functionals, so these intervals concern errors in approximating those functionals. They are not confidence bands for individual survival probabilities, reductions in death risk, or evidence of clinical calibration. Generalization to new cohorts or clinical settings requires additional evaluation. The analysis concerns the eight fitted main-body direct comparators.

\clearpage
\begingroup
\newgeometry{layoutwidth=11in,layoutheight=8.5in,margin=0.5in}
\paperwidth=11in\paperheight=8.5in
\ifdefined\pdfpagewidth\pdfpagewidth=\paperwidth\pdfpageheight=\paperheight
\else\ifdefined\pagewidth\pagewidth=\paperwidth\pageheight=\paperheight
\else\fi\fi
\captionsetup{font=small}
\centering
\noindent\includegraphics[width=0.95\linewidth,height=0.80\textheight,keepaspectratio]{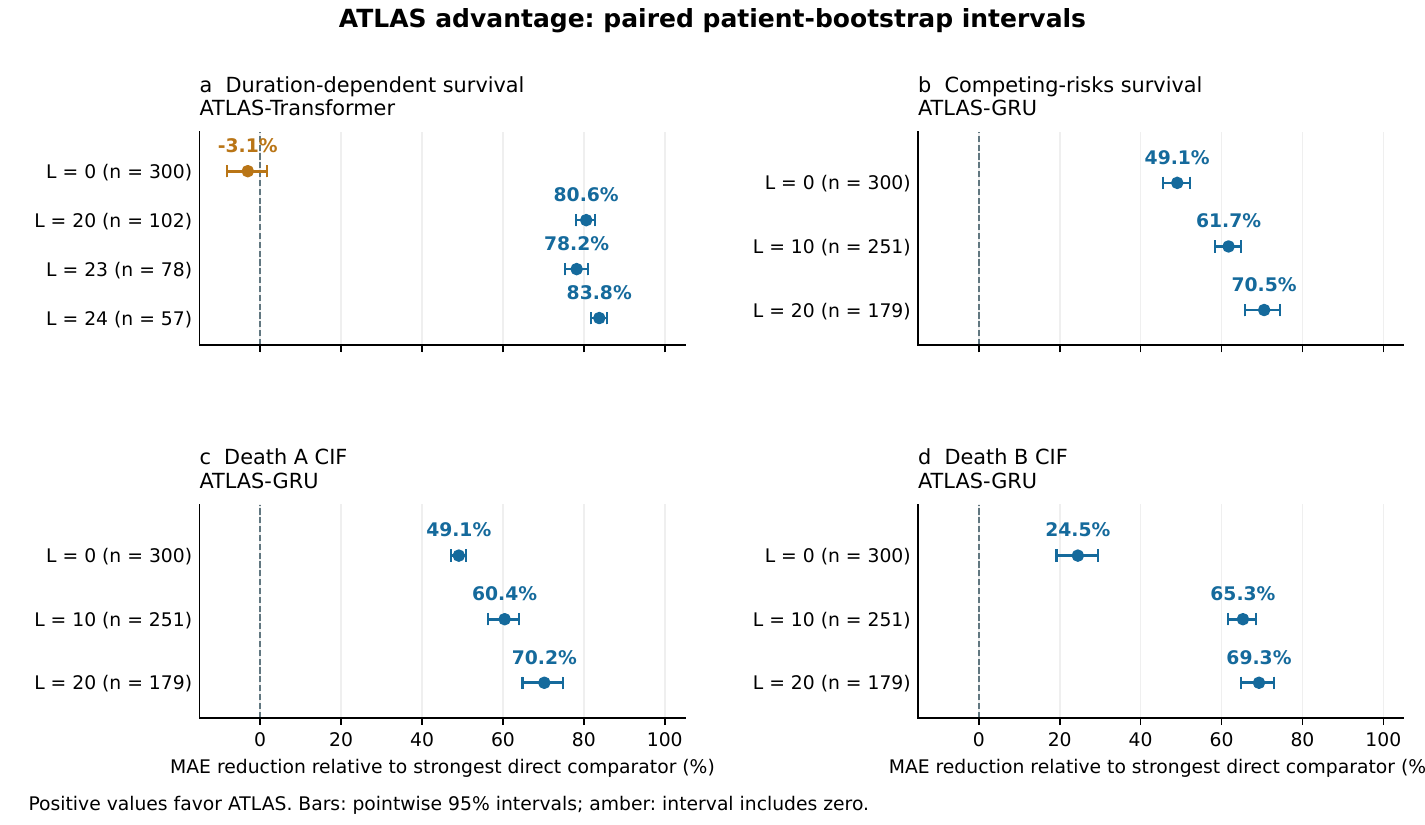}\par
\captionof{figure}{ATLAS-Transformer in the duration-dependent experiment and ATLAS-GRU in the competing-risks experiment are compared with the strongest fitted direct comparator at each landmark and target. Points show relative MAE reductions, 100 times (comparator MAE minus ATLAS MAE) divided by comparator MAE; positive values favor ATLAS. Bars are pointwise 95\% percentile intervals from 5,000 paired patient resamples. The duration-dependent comparators are DynamicDeepHit-Lite at baseline, SurvTransformer-Discrete at landmarks 20 and 23, and Dynamic-DeepHit (optimized) at landmark 24. In competing risks, Weibull-MLP is used for baseline survival and Death B incidence, and Dynamic-DeepHit (optimized) for baseline Death A incidence and all post-baseline targets. Amber marks an interval including zero. ATLAS uses the common Sobol sampler. Comparisons are fixed by the unresampled updated results; fits are held fixed and intervals do not account for model selection or multiplicity.}\label{fig:patientboot_1}\par
\clearpage
\noindent\includegraphics[width=0.95\linewidth,height=0.80\textheight,keepaspectratio]{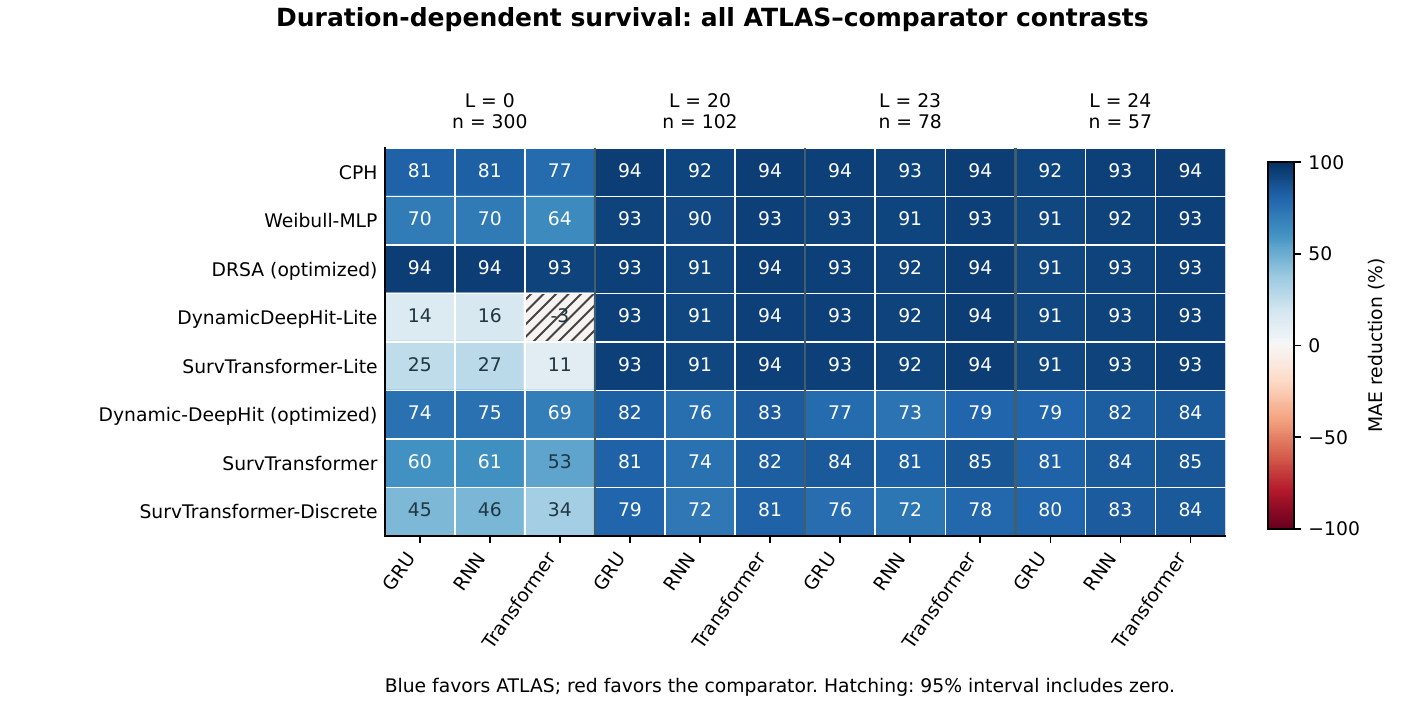}\par
\captionof{figure}{Cells show percentage reductions in mean survival-curve MAE for ATLAS-GRU, ATLAS-RNN and ATLAS-Transformer relative to each of the eight fitted direct comparators. Blue favors ATLAS and red favors the comparator; a common scale from -100\% to 100\% is centered at zero. Hatching marks a pointwise 95\% paired patient-bootstrap interval for the MAE difference that includes zero. This occurs only for baseline Transformer versus DynamicDeepHit-Lite. ATLAS means average common-Sobol errors over nine saved training fits. Cell values are rounded to whole percentages. Different eligible patient populations and prediction windows apply across landmarks; intervals are not simultaneous.}\label{fig:patientboot_2}\par
\clearpage
\noindent\includegraphics[width=0.95\linewidth,height=0.80\textheight,keepaspectratio]{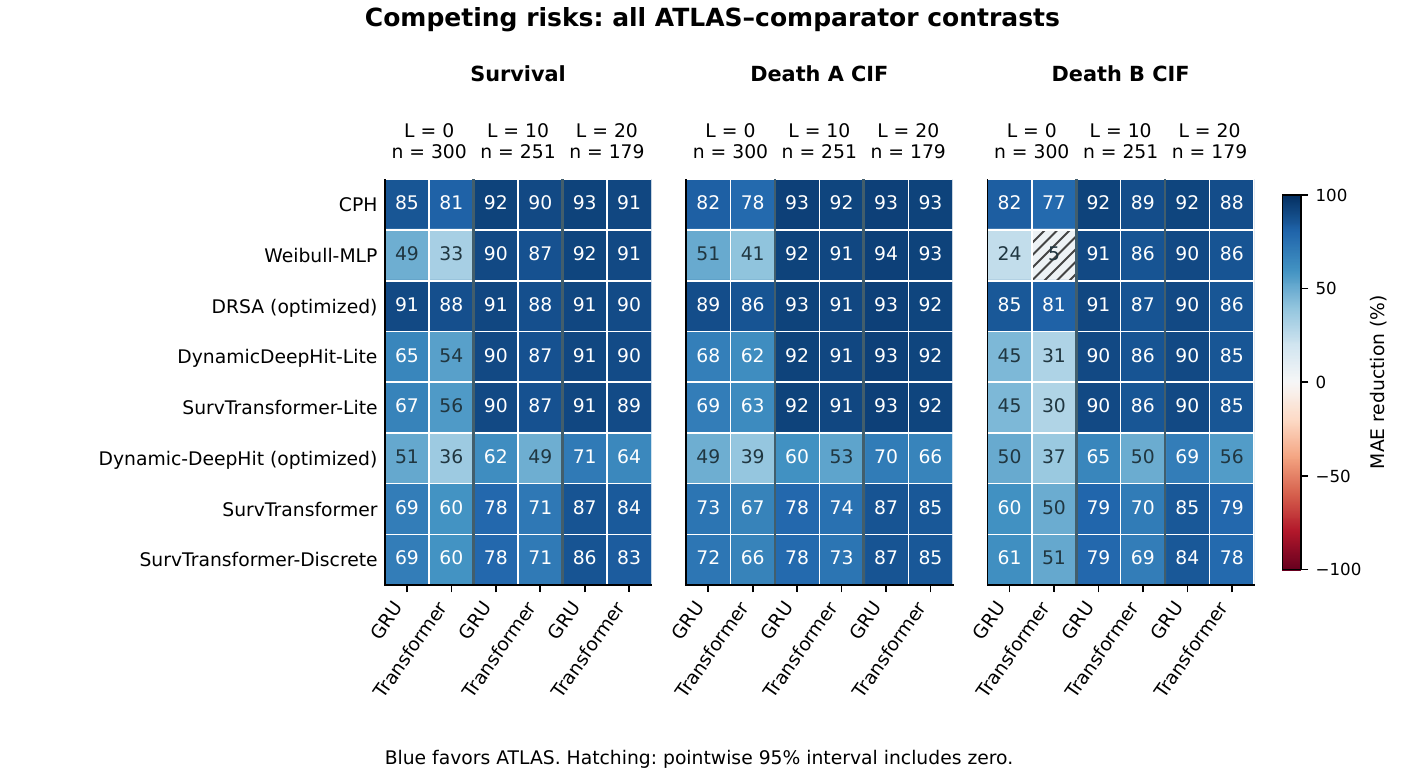}\par
\captionof{figure}{The three panels show survival, Death A CIF and Death B CIF comparisons for ATLAS-GRU and ATLAS-Transformer against the eight fitted direct comparators. Cells show relative reductions in mean MAE with the same color scale and hatching convention as Figure~\ref{fig:patientboot_2}. All point estimates favor ATLAS, but the baseline Transformer versus Weibull-MLP Death B CIF interval includes zero. ATLAS means use the common Sobol sampler and five saved training fits. Errors and bootstrap resamples are paired by patient across models, landmarks and targets. These are pointwise intervals conditional on saved fits, without adjustment for multiple comparisons; percentage error reductions are not reductions in clinical event risk.}\label{fig:patientboot_3}\par
\clearpage
\restoregeometry
\paperwidth=8.5in\paperheight=11in
\ifdefined\pdfpagewidth\pdfpagewidth=\paperwidth\pdfpageheight=\paperheight
\else\ifdefined\pagewidth\pagewidth=\paperwidth\pageheight=\paperheight
\else\fi\fi
\endgroup

\clearpage
\subsection{Complete numerical comparisons}
Table~\ref{tab:patientboot_all} reports all 240 comparisons. Architecture names identify ATLAS models using the common Sobol sampler. Targets S, A and B denote survival, Death A CIF and Death B CIF. Differences are ATLAS minus comparator MAE, with negative values favoring ATLAS; positive relative reductions favor ATLAS. Values in brackets are pointwise 95\% paired patient-bootstrap intervals. Comparator names identify the fitted models; Weibull denotes Weibull-MLP and Dynamic-DeepHit (optimized) denotes the validation-selected adaptation.
\begingroup\scriptsize
\setlength{\tabcolsep}{3pt}
\begin{longtable}{lll p{3.0cm} rr}
\caption{Complete paired patient-bootstrap comparisons. MAE differences retain the probability scale; relative reductions are percentages.}\label{tab:patientboot_all}\\
\toprule
ATLAS & L & Target & Comparator & Difference [95\% interval] & Reduction [95\% interval]\\\midrule
\endfirsthead
\multicolumn{6}{l}{\tablename~\thetable\ (continued)}\\\toprule
ATLAS & L & Target & Comparator & Difference [95\% interval] & Reduction [95\% interval]\\\midrule
\endhead
\bottomrule\endfoot
\multicolumn{6}{l}{\textbf{Duration-dependent experiment}}\\\midrule
GRU & 0 & S & CPH & -0.031475 [-0.034783, -0.028406] & 81.0 [79.4, 82.5]\\
GRU & 0 & S & Weibull-MLP & -0.016882 [-0.017068, -0.016704] & 69.6 [69.3, 69.9]\\
GRU & 0 & S & DRSA (optimized) & -0.111191 [-0.112694, -0.109723] & 93.8 [93.7, 93.9]\\
GRU & 0 & S & DynamicDeepHit-Lite & -0.001186 [-0.001528, -0.000829] & 13.9 [10.0, 17.3]\\
GRU & 0 & S & SurvTransformer-Lite & -0.002506 [-0.002822, -0.002186] & 25.4 [22.8, 27.8]\\
GRU & 0 & S & Dynamic-DeepHit (optimized) & -0.021235 [-0.022451, -0.020027] & 74.2 [72.9, 75.4]\\
GRU & 0 & S & SurvTransformer & -0.011232 [-0.011862, -0.010644] & 60.4 [59.2, 61.6]\\
GRU & 0 & S & SurvTransformer-Discrete & -0.005925 [-0.006530, -0.005347] & 44.6 [42.1, 47.0]\\
RNN & 0 & S & CPH & -0.031650 [-0.034968, -0.028555] & 81.5 [79.9, 82.9]\\
RNN & 0 & S & Weibull-MLP & -0.017057 [-0.017303, -0.016812] & 70.3 [69.8, 70.9]\\
RNN & 0 & S & DRSA (optimized) & -0.111365 [-0.112812, -0.109960] & 93.9 [93.8, 94.0]\\
RNN & 0 & S & DynamicDeepHit-Lite & -0.001360 [-0.001724, -0.000986] & 15.9 [11.9, 19.5]\\
RNN & 0 & S & SurvTransformer-Lite & -0.002681 [-0.002997, -0.002355] & 27.1 [24.5, 29.6]\\
RNN & 0 & S & Dynamic-DeepHit (optimized) & -0.021409 [-0.022555, -0.020263] & 74.8 [73.8, 75.8]\\
RNN & 0 & S & SurvTransformer & -0.011407 [-0.012066, -0.010790] & 61.3 [60.0, 62.7]\\
RNN & 0 & S & SurvTransformer-Discrete & -0.006100 [-0.006721, -0.005504] & 45.9 [43.3, 48.3]\\
Transformer & 0 & S & CPH & -0.030027 [-0.033193, -0.027056] & 77.3 [75.7, 78.7]\\
Transformer & 0 & S & Weibull-MLP & -0.015435 [-0.015663, -0.015212] & 63.6 [62.8, 64.4]\\
Transformer & 0 & S & DRSA (optimized) & -0.109743 [-0.111273, -0.108232] & 92.6 [92.3, 92.8]\\
Transformer & 0 & S & DynamicDeepHit-Lite & 0.000262 [-0.000146, 0.000685] & -3.1 [-8.3, 1.6]\\
Transformer & 0 & S & SurvTransformer-Lite & -0.001059 [-0.001395, -0.000715] & 10.7 [7.4, 13.8]\\
Transformer & 0 & S & Dynamic-DeepHit (optimized) & -0.019787 [-0.021036, -0.018567] & 69.2 [67.5, 70.7]\\
Transformer & 0 & S & SurvTransformer & -0.009785 [-0.010344, -0.009256] & 52.6 [51.3, 53.9]\\
Transformer & 0 & S & SurvTransformer-Discrete & -0.004478 [-0.005025, -0.003955] & 33.7 [30.9, 36.3]\\
GRU & 20 & S & CPH & -0.148154 [-0.168786, -0.129227] & 93.8 [92.8, 94.7]\\
GRU & 20 & S & Weibull-MLP & -0.127261 [-0.148976, -0.107806] & 92.9 [91.5, 94.1]\\
GRU & 20 & S & DRSA (optimized) & -0.136800 [-0.153926, -0.121912] & 93.3 [92.3, 94.3]\\
GRU & 20 & S & DynamicDeepHit-Lite & -0.132828 [-0.150728, -0.117009] & 93.2 [92.1, 94.2]\\
GRU & 20 & S & SurvTransformer-Lite & -0.132497 [-0.150878, -0.116336] & 93.1 [92.0, 94.2]\\
GRU & 20 & S & Dynamic-DeepHit (optimized) & -0.043980 [-0.051081, -0.037115] & 81.8 [79.0, 84.1]\\
GRU & 20 & S & SurvTransformer & -0.041137 [-0.046600, -0.036182] & 80.8 [78.2, 83.2]\\
GRU & 20 & S & SurvTransformer-Discrete & -0.036778 [-0.041082, -0.032795] & 79.0 [76.6, 81.1]\\
RNN & 20 & S & CPH & -0.144804 [-0.165515, -0.125996] & 91.7 [90.3, 92.9]\\
RNN & 20 & S & Weibull-MLP & -0.123911 [-0.145487, -0.104406] & 90.4 [88.6, 92.0]\\
RNN & 20 & S & DRSA (optimized) & -0.133449 [-0.150533, -0.118583] & 91.1 [89.8, 92.2]\\
RNN & 20 & S & DynamicDeepHit-Lite & -0.129478 [-0.147370, -0.113640] & 90.8 [89.5, 92.1]\\
RNN & 20 & S & SurvTransformer-Lite & -0.129147 [-0.147521, -0.113013] & 90.8 [89.4, 92.1]\\
RNN & 20 & S & Dynamic-DeepHit (optimized) & -0.040630 [-0.047686, -0.033848] & 75.6 [71.8, 78.5]\\
RNN & 20 & S & SurvTransformer & -0.037786 [-0.043316, -0.032730] & 74.2 [70.9, 77.3]\\
RNN & 20 & S & SurvTransformer-Discrete & -0.033428 [-0.037597, -0.029587] & 71.8 [69.0, 74.3]\\
Transformer & 20 & S & CPH & -0.148876 [-0.169278, -0.130193] & 94.3 [93.4, 95.1]\\
Transformer & 20 & S & Weibull-MLP & -0.127984 [-0.149552, -0.108670] & 93.4 [92.1, 94.5]\\
Transformer & 20 & S & DRSA (optimized) & -0.137522 [-0.154279, -0.122712] & 93.8 [92.9, 94.7]\\
Transformer & 20 & S & DynamicDeepHit-Lite & -0.133550 [-0.151199, -0.117659] & 93.7 [92.6, 94.6]\\
Transformer & 20 & S & SurvTransformer-Lite & -0.133220 [-0.151385, -0.117229] & 93.6 [92.6, 94.6]\\
Transformer & 20 & S & Dynamic-DeepHit (optimized) & -0.044702 [-0.051662, -0.037999] & 83.2 [80.7, 85.2]\\
Transformer & 20 & S & SurvTransformer & -0.041859 [-0.047185, -0.036969] & 82.2 [79.7, 84.5]\\
Transformer & 20 & S & SurvTransformer-Discrete & -0.037501 [-0.041791, -0.033499] & 80.6 [78.1, 82.8]\\
GRU & 23 & S & CPH & -0.198253 [-0.220212, -0.177872] & 93.6 [92.7, 94.4]\\
GRU & 23 & S & Weibull-MLP & -0.172119 [-0.206457, -0.138461] & 92.7 [91.1, 93.9]\\
GRU & 23 & S & DRSA (optimized) & -0.194563 [-0.213263, -0.177625] & 93.4 [92.6, 94.2]\\
GRU & 23 & S & DynamicDeepHit-Lite & -0.191104 [-0.210697, -0.173406] & 93.3 [92.5, 94.1]\\
GRU & 23 & S & SurvTransformer-Lite & -0.190595 [-0.210256, -0.172648] & 93.3 [92.5, 94.1]\\
GRU & 23 & S & Dynamic-DeepHit (optimized) & -0.046195 [-0.052315, -0.040429] & 77.2 [75.1, 79.1]\\
GRU & 23 & S & SurvTransformer & -0.071642 [-0.083159, -0.061043] & 84.0 [81.6, 86.2]\\
GRU & 23 & S & SurvTransformer-Discrete & -0.043848 [-0.048405, -0.039422] & 76.3 [73.3, 78.9]\\
RNN & 23 & S & CPH & -0.196011 [-0.217650, -0.175723] & 92.5 [91.4, 93.5]\\
RNN & 23 & S & Weibull-MLP & -0.169877 [-0.203812, -0.136718] & 91.4 [89.8, 92.7]\\
RNN & 23 & S & DRSA (optimized) & -0.192321 [-0.210842, -0.175318] & 92.4 [91.3, 93.4]\\
RNN & 23 & S & DynamicDeepHit-Lite & -0.188862 [-0.208488, -0.170938] & 92.2 [91.1, 93.3]\\
RNN & 23 & S & SurvTransformer-Lite & -0.188354 [-0.207942, -0.170365] & 92.2 [91.1, 93.3]\\
RNN & 23 & S & Dynamic-DeepHit (optimized) & -0.043953 [-0.049672, -0.038543] & 73.4 [70.8, 75.8]\\
RNN & 23 & S & SurvTransformer & -0.069400 [-0.080960, -0.058928] & 81.4 [78.4, 84.1]\\
RNN & 23 & S & SurvTransformer-Discrete & -0.041606 [-0.046294, -0.037043] & 72.4 [68.4, 75.9]\\
Transformer & 23 & S & CPH & -0.199375 [-0.221213, -0.179032] & 94.1 [93.2, 94.9]\\
Transformer & 23 & S & Weibull-MLP & -0.173241 [-0.207110, -0.140071] & 93.3 [92.0, 94.2]\\
Transformer & 23 & S & DRSA (optimized) & -0.195685 [-0.213985, -0.178909] & 94.0 [93.1, 94.7]\\
Transformer & 23 & S & DynamicDeepHit-Lite & -0.192226 [-0.211599, -0.174476] & 93.9 [93.0, 94.6]\\
Transformer & 23 & S & SurvTransformer-Lite & -0.191717 [-0.211078, -0.173967] & 93.9 [93.0, 94.6]\\
Transformer & 23 & S & Dynamic-DeepHit (optimized) & -0.047317 [-0.053246, -0.041569] & 79.1 [76.9, 81.0]\\
Transformer & 23 & S & SurvTransformer & -0.072763 [-0.083890, -0.062557] & 85.3 [83.3, 87.2]\\
Transformer & 23 & S & SurvTransformer-Discrete & -0.044970 [-0.049395, -0.040651] & 78.2 [75.4, 81.0]\\
GRU & 24 & S & CPH & -0.158125 [-0.186520, -0.132199] & 92.1 [90.6, 93.3]\\
GRU & 24 & S & Weibull-MLP & -0.137082 [-0.168547, -0.108566] & 91.0 [89.0, 92.5]\\
GRU & 24 & S & DRSA (optimized) & -0.144263 [-0.171189, -0.119082] & 91.4 [89.7, 92.7]\\
GRU & 24 & S & DynamicDeepHit-Lite & -0.141634 [-0.168859, -0.116279] & 91.2 [89.5, 92.6]\\
GRU & 24 & S & SurvTransformer-Lite & -0.142583 [-0.170008, -0.117227] & 91.3 [89.6, 92.6]\\
GRU & 24 & S & Dynamic-DeepHit (optimized) & -0.051175 [-0.061379, -0.041556] & 79.0 [76.2, 81.4]\\
GRU & 24 & S & SurvTransformer & -0.057890 [-0.066434, -0.050214] & 81.0 [78.7, 83.1]\\
GRU & 24 & S & SurvTransformer-Discrete & -0.053342 [-0.059781, -0.047407] & 79.7 [77.1, 82.2]\\
RNN & 24 & S & CPH & -0.160249 [-0.189083, -0.133990] & 93.3 [91.8, 94.5]\\
RNN & 24 & S & Weibull-MLP & -0.139207 [-0.170737, -0.110504] & 92.4 [90.5, 93.8]\\
RNN & 24 & S & DRSA (optimized) & -0.146388 [-0.173583, -0.120924] & 92.7 [91.2, 94.0]\\
RNN & 24 & S & DynamicDeepHit-Lite & -0.143759 [-0.171329, -0.118078] & 92.6 [91.0, 93.9]\\
RNN & 24 & S & SurvTransformer-Lite & -0.144708 [-0.171965, -0.119044] & 92.6 [91.1, 93.9]\\
RNN & 24 & S & Dynamic-DeepHit (optimized) & -0.053300 [-0.064125, -0.043043] & 82.3 [78.9, 85.0]\\
RNN & 24 & S & SurvTransformer & -0.060015 [-0.068871, -0.051920] & 83.9 [81.3, 86.1]\\
RNN & 24 & S & SurvTransformer-Discrete & -0.055467 [-0.061896, -0.049464] & 82.8 [80.3, 85.0]\\
Transformer & 24 & S & CPH & -0.161247 [-0.190084, -0.134776] & 93.9 [92.4, 95.0]\\
Transformer & 24 & S & Weibull-MLP & -0.140204 [-0.171871, -0.111647] & 93.0 [91.1, 94.4]\\
Transformer & 24 & S & DRSA (optimized) & -0.147386 [-0.174463, -0.121816] & 93.4 [91.6, 94.6]\\
Transformer & 24 & S & DynamicDeepHit-Lite & -0.144757 [-0.172341, -0.119006] & 93.2 [91.5, 94.5]\\
Transformer & 24 & S & SurvTransformer-Lite & -0.145706 [-0.172996, -0.119954] & 93.3 [91.5, 94.5]\\
Transformer & 24 & S & Dynamic-DeepHit (optimized) & -0.054298 [-0.064475, -0.044688] & 83.8 [81.7, 85.7]\\
Transformer & 24 & S & SurvTransformer & -0.061013 [-0.069266, -0.053500] & 85.3 [83.2, 87.1]\\
Transformer & 24 & S & SurvTransformer-Discrete & -0.056465 [-0.062487, -0.050947] & 84.3 [82.1, 86.4]\\
\multicolumn{6}{l}{\textbf{Competing-risks experiment}}\\\midrule
GRU & 0 & S & CPH & -0.046281 [-0.049271, -0.043216] & 85.4 [84.4, 86.4]\\
GRU & 0 & S & Weibull-MLP & -0.007608 [-0.008439, -0.006784] & 49.1 [45.6, 52.3]\\
GRU & 0 & S & DRSA (optimized) & -0.075408 [-0.078923, -0.071812] & 90.5 [89.9, 91.1]\\
GRU & 0 & S & DynamicDeepHit-Lite & -0.014804 [-0.016700, -0.012929] & 65.2 [61.9, 68.1]\\
GRU & 0 & S & SurvTransformer-Lite & -0.015952 [-0.017955, -0.014049] & 66.9 [63.7, 69.6]\\
GRU & 0 & S & Dynamic-DeepHit (optimized) & -0.008340 [-0.009127, -0.007591] & 51.3 [48.8, 53.8]\\
GRU & 0 & S & SurvTransformer & -0.017784 [-0.018860, -0.016750] & 69.2 [67.3, 71.1]\\
GRU & 0 & S & SurvTransformer-Discrete & -0.017876 [-0.018875, -0.016890] & 69.3 [67.7, 71.0]\\
Transformer & 0 & S & CPH & -0.043805 [-0.047038, -0.040474] & 80.8 [79.2, 82.4]\\
Transformer & 0 & S & Weibull-MLP & -0.005132 [-0.006075, -0.004179] & 33.1 [28.0, 37.5]\\
Transformer & 0 & S & DRSA (optimized) & -0.072932 [-0.076316, -0.069580] & 87.5 [86.9, 88.1]\\
Transformer & 0 & S & DynamicDeepHit-Lite & -0.012327 [-0.014194, -0.010420] & 54.3 [50.1, 57.9]\\
Transformer & 0 & S & SurvTransformer-Lite & -0.013475 [-0.015459, -0.011500] & 56.5 [52.4, 59.9]\\
Transformer & 0 & S & Dynamic-DeepHit (optimized) & -0.005864 [-0.006742, -0.005042] & 36.1 [32.5, 39.6]\\
Transformer & 0 & S & SurvTransformer & -0.015308 [-0.016384, -0.014332] & 59.6 [57.5, 61.7]\\
Transformer & 0 & S & SurvTransformer-Discrete & -0.015399 [-0.016492, -0.014357] & 59.7 [57.6, 62.0]\\
GRU & 0 & A & CPH & -0.026439 [-0.028898, -0.024139] & 81.8 [80.4, 83.0]\\
GRU & 0 & A & Weibull-MLP & -0.006063 [-0.006799, -0.005340] & 50.8 [47.4, 53.8]\\
GRU & 0 & A & DRSA (optimized) & -0.045418 [-0.047886, -0.043011] & 88.5 [87.7, 89.4]\\
GRU & 0 & A & DynamicDeepHit-Lite & -0.012760 [-0.014701, -0.010832] & 68.5 [65.1, 71.2]\\
GRU & 0 & A & SurvTransformer-Lite & -0.013299 [-0.015202, -0.011407] & 69.3 [66.1, 72.1]\\
GRU & 0 & A & Dynamic-DeepHit (optimized) & -0.005667 [-0.006209, -0.005160] & 49.1 [47.1, 50.9]\\
GRU & 0 & A & SurvTransformer & -0.015697 [-0.016602, -0.014829] & 72.8 [71.1, 74.4]\\
GRU & 0 & A & SurvTransformer-Discrete & -0.015273 [-0.016281, -0.014272] & 72.2 [70.8, 73.6]\\
Transformer & 0 & A & CPH & -0.025227 [-0.027972, -0.022708] & 78.1 [75.8, 80.1]\\
Transformer & 0 & A & Weibull-MLP & -0.004850 [-0.005728, -0.004023] & 40.6 [36.1, 44.9]\\
Transformer & 0 & A & DRSA (optimized) & -0.044206 [-0.046615, -0.041882] & 86.2 [85.6, 86.8]\\
Transformer & 0 & A & DynamicDeepHit-Lite & -0.011547 [-0.013592, -0.009535] & 62.0 [57.5, 65.7]\\
Transformer & 0 & A & SurvTransformer-Lite & -0.012087 [-0.014103, -0.010111] & 63.0 [58.9, 66.6]\\
Transformer & 0 & A & Dynamic-DeepHit (optimized) & -0.004455 [-0.005229, -0.003748] & 38.6 [34.5, 42.5]\\
Transformer & 0 & A & SurvTransformer & -0.014485 [-0.015432, -0.013579] & 67.1 [65.5, 68.7]\\
Transformer & 0 & A & SurvTransformer-Discrete & -0.014061 [-0.015224, -0.012960] & 66.5 [64.5, 68.4]\\
GRU & 0 & B & CPH & -0.022580 [-0.024310, -0.020824] & 82.0 [80.7, 83.3]\\
GRU & 0 & B & Weibull-MLP & -0.001602 [-0.002041, -0.001190] & 24.5 [19.2, 29.5]\\
GRU & 0 & B & DRSA (optimized) & -0.027619 [-0.029498, -0.025735] & 84.8 [84.0, 85.6]\\
GRU & 0 & B & DynamicDeepHit-Lite & -0.004053 [-0.005049, -0.003121] & 45.1 [38.9, 50.1]\\
GRU & 0 & B & SurvTransformer-Lite & -0.003989 [-0.004757, -0.003232] & 44.7 [39.8, 48.8]\\
GRU & 0 & B & Dynamic-DeepHit (optimized) & -0.004971 [-0.005427, -0.004532] & 50.2 [47.8, 52.5]\\
GRU & 0 & B & SurvTransformer & -0.007481 [-0.008187, -0.006792] & 60.2 [57.9, 62.4]\\
GRU & 0 & B & SurvTransformer-Discrete & -0.007713 [-0.008330, -0.007102] & 61.0 [58.9, 62.9]\\
Transformer & 0 & B & CPH & -0.021309 [-0.022960, -0.019583] & 77.4 [75.9, 78.9]\\
Transformer & 0 & B & Weibull-MLP & -0.000331 [-0.000773, 0.000104] & 5.1 [-1.6, 11.2]\\
Transformer & 0 & B & DRSA (optimized) & -0.026348 [-0.028169, -0.024541] & 80.9 [79.9, 81.8]\\
Transformer & 0 & B & DynamicDeepHit-Lite & -0.002782 [-0.003742, -0.001894] & 30.9 [23.7, 37.2]\\
Transformer & 0 & B & SurvTransformer-Lite & -0.002718 [-0.003462, -0.001987] & 30.4 [24.3, 35.6]\\
Transformer & 0 & B & Dynamic-DeepHit (optimized) & -0.003701 [-0.004145, -0.003275] & 37.3 [34.4, 40.2]\\
Transformer & 0 & B & SurvTransformer & -0.006210 [-0.006832, -0.005609] & 50.0 [47.7, 52.2]\\
Transformer & 0 & B & SurvTransformer-Discrete & -0.006442 [-0.007011, -0.005901] & 50.9 [48.8, 53.0]\\
GRU & 10 & S & CPH & -0.106974 [-0.117243, -0.097523] & 92.4 [91.5, 93.2]\\
GRU & 10 & S & Weibull-MLP & -0.081152 [-0.091051, -0.071408] & 90.2 [89.0, 91.2]\\
GRU & 10 & S & DRSA (optimized) & -0.086826 [-0.096177, -0.077544] & 90.8 [89.6, 91.8]\\
GRU & 10 & S & DynamicDeepHit-Lite & -0.082591 [-0.092225, -0.072883] & 90.3 [89.1, 91.4]\\
GRU & 10 & S & SurvTransformer-Lite & -0.081760 [-0.091450, -0.071897] & 90.3 [89.0, 91.3]\\
GRU & 10 & S & Dynamic-DeepHit (optimized) & -0.014236 [-0.015992, -0.012561] & 61.7 [58.3, 64.8]\\
GRU & 10 & S & SurvTransformer & -0.031780 [-0.035839, -0.028036] & 78.3 [75.9, 80.5]\\
GRU & 10 & S & SurvTransformer-Discrete & -0.031639 [-0.035310, -0.028146] & 78.2 [75.9, 80.2]\\
Transformer & 10 & S & CPH & -0.104072 [-0.114470, -0.094505] & 89.9 [88.7, 90.9]\\
Transformer & 10 & S & Weibull-MLP & -0.078249 [-0.088064, -0.068535] & 87.0 [85.4, 88.3]\\
Transformer & 10 & S & DRSA (optimized) & -0.083924 [-0.093296, -0.074658] & 87.7 [86.3, 89.0]\\
Transformer & 10 & S & DynamicDeepHit-Lite & -0.079688 [-0.089372, -0.069991] & 87.2 [85.6, 88.5]\\
Transformer & 10 & S & SurvTransformer-Lite & -0.078857 [-0.088568, -0.069135] & 87.1 [85.5, 88.4]\\
Transformer & 10 & S & Dynamic-DeepHit (optimized) & -0.011334 [-0.013054, -0.009696] & 49.1 [45.0, 52.9]\\
Transformer & 10 & S & SurvTransformer & -0.028877 [-0.032871, -0.025198] & 71.1 [68.2, 73.8]\\
Transformer & 10 & S & SurvTransformer-Discrete & -0.028737 [-0.032337, -0.025329] & 71.0 [68.4, 73.5]\\
GRU & 10 & A & CPH & -0.094703 [-0.104863, -0.085399] & 93.1 [92.3, 93.8]\\
GRU & 10 & A & Weibull-MLP & -0.084648 [-0.094257, -0.075459] & 92.3 [91.5, 93.0]\\
GRU & 10 & A & DRSA (optimized) & -0.088312 [-0.097852, -0.079047] & 92.6 [91.8, 93.3]\\
GRU & 10 & A & DynamicDeepHit-Lite & -0.085922 [-0.095742, -0.076496] & 92.4 [91.6, 93.1]\\
GRU & 10 & A & SurvTransformer-Lite & -0.085240 [-0.095116, -0.075747] & 92.4 [91.5, 93.1]\\
GRU & 10 & A & Dynamic-DeepHit (optimized) & -0.010763 [-0.012401, -0.009266] & 60.4 [56.4, 64.0]\\
GRU & 10 & A & SurvTransformer & -0.025738 [-0.029212, -0.022507] & 78.5 [76.1, 80.7]\\
GRU & 10 & A & SurvTransformer-Discrete & -0.024586 [-0.028062, -0.021305] & 77.7 [75.1, 80.0]\\
Transformer & 10 & A & CPH & -0.093336 [-0.103486, -0.084002] & 91.7 [90.9, 92.5]\\
Transformer & 10 & A & Weibull-MLP & -0.083281 [-0.092922, -0.074053] & 90.8 [89.9, 91.6]\\
Transformer & 10 & A & DRSA (optimized) & -0.086945 [-0.096524, -0.077774] & 91.2 [90.3, 92.0]\\
Transformer & 10 & A & DynamicDeepHit-Lite & -0.084555 [-0.094356, -0.075075] & 90.9 [90.0, 91.8]\\
Transformer & 10 & A & SurvTransformer-Lite & -0.083873 [-0.093721, -0.074379] & 90.9 [90.0, 91.7]\\
Transformer & 10 & A & Dynamic-DeepHit (optimized) & -0.009396 [-0.011015, -0.007935] & 52.7 [48.7, 56.5]\\
Transformer & 10 & A & SurvTransformer & -0.024371 [-0.027789, -0.021151] & 74.3 [71.8, 76.7]\\
Transformer & 10 & A & SurvTransformer-Discrete & -0.023219 [-0.026698, -0.019969] & 73.4 [70.7, 76.0]\\
GRU & 10 & B & CPH & -0.051976 [-0.058860, -0.045857] & 92.4 [91.4, 93.2]\\
GRU & 10 & B & Weibull-MLP & -0.041295 [-0.047441, -0.035586] & 90.6 [89.3, 91.7]\\
GRU & 10 & B & DRSA (optimized) & -0.042345 [-0.048451, -0.036633] & 90.8 [89.5, 91.9]\\
GRU & 10 & B & DynamicDeepHit-Lite & -0.039958 [-0.046107, -0.034170] & 90.3 [89.0, 91.5]\\
GRU & 10 & B & SurvTransformer-Lite & -0.039722 [-0.045823, -0.033936] & 90.2 [88.9, 91.4]\\
GRU & 10 & B & Dynamic-DeepHit (optimized) & -0.008086 [-0.009554, -0.006745] & 65.3 [61.6, 68.5]\\
GRU & 10 & B & SurvTransformer & -0.016495 [-0.019411, -0.013872] & 79.3 [76.6, 81.7]\\
GRU & 10 & B & SurvTransformer-Discrete & -0.016068 [-0.018809, -0.013564] & 78.9 [76.1, 81.3]\\
Transformer & 10 & B & CPH & -0.050057 [-0.056652, -0.044072] & 89.0 [87.7, 90.1]\\
Transformer & 10 & B & Weibull-MLP & -0.039376 [-0.045385, -0.033831] & 86.4 [84.7, 87.9]\\
Transformer & 10 & B & DRSA (optimized) & -0.040426 [-0.046349, -0.034909] & 86.7 [84.9, 88.2]\\
Transformer & 10 & B & DynamicDeepHit-Lite & -0.038039 [-0.044071, -0.032450] & 86.0 [84.1, 87.6]\\
Transformer & 10 & B & SurvTransformer-Lite & -0.037803 [-0.043791, -0.032138] & 85.9 [84.1, 87.5]\\
Transformer & 10 & B & Dynamic-DeepHit (optimized) & -0.006167 [-0.007486, -0.004990] & 49.8 [45.2, 54.0]\\
Transformer & 10 & B & SurvTransformer & -0.014576 [-0.017294, -0.012092] & 70.1 [66.7, 73.1]\\
Transformer & 10 & B & SurvTransformer-Discrete & -0.014149 [-0.016772, -0.011823] & 69.5 [66.0, 72.4]\\
GRU & 20 & S & CPH & -0.086375 [-0.098142, -0.075017] & 92.8 [91.5, 93.9]\\
GRU & 20 & S & Weibull-MLP & -0.078665 [-0.090022, -0.068056] & 92.1 [90.6, 93.4]\\
GRU & 20 & S & DRSA (optimized) & -0.070744 [-0.082514, -0.059112] & 91.3 [89.8, 92.6]\\
GRU & 20 & S & DynamicDeepHit-Lite & -0.070665 [-0.082192, -0.059401] & 91.3 [89.8, 92.6]\\
GRU & 20 & S & SurvTransformer-Lite & -0.069957 [-0.081749, -0.058434] & 91.3 [89.7, 92.5]\\
GRU & 20 & S & Dynamic-DeepHit (optimized) & -0.016049 [-0.018552, -0.013629] & 70.5 [65.7, 74.5]\\
GRU & 20 & S & SurvTransformer & -0.043766 [-0.051553, -0.036586] & 86.7 [84.2, 88.8]\\
GRU & 20 & S & SurvTransformer-Discrete & -0.042237 [-0.050136, -0.035123] & 86.3 [83.5, 88.6]\\
Transformer & 20 & S & CPH & -0.084985 [-0.096794, -0.073741] & 91.3 [89.9, 92.5]\\
Transformer & 20 & S & Weibull-MLP & -0.077274 [-0.088470, -0.066578] & 90.5 [88.8, 92.0]\\
Transformer & 20 & S & DRSA (optimized) & -0.069354 [-0.081138, -0.057684] & 89.5 [87.8, 91.0]\\
Transformer & 20 & S & DynamicDeepHit-Lite & -0.069275 [-0.080863, -0.057945] & 89.5 [87.7, 91.0]\\
Transformer & 20 & S & SurvTransformer-Lite & -0.068567 [-0.080301, -0.057098] & 89.4 [87.6, 91.0]\\
Transformer & 20 & S & Dynamic-DeepHit (optimized) & -0.014658 [-0.017134, -0.012296] & 64.4 [59.2, 68.9]\\
Transformer & 20 & S & SurvTransformer & -0.042375 [-0.050017, -0.035214] & 84.0 [81.1, 86.2]\\
Transformer & 20 & S & SurvTransformer-Discrete & -0.040847 [-0.048600, -0.033844] & 83.5 [80.4, 85.9]\\
GRU & 20 & A & CPH & -0.069452 [-0.079766, -0.059698] & 93.5 [92.3, 94.5]\\
GRU & 20 & A & Weibull-MLP & -0.070199 [-0.080944, -0.059734] & 93.6 [92.2, 94.7]\\
GRU & 20 & A & DRSA (optimized) & -0.063901 [-0.074607, -0.053457] & 93.0 [91.7, 94.1]\\
GRU & 20 & A & DynamicDeepHit-Lite & -0.064855 [-0.075331, -0.054579] & 93.1 [91.7, 94.2]\\
GRU & 20 & A & SurvTransformer-Lite & -0.064568 [-0.075066, -0.054113] & 93.0 [91.7, 94.2]\\
GRU & 20 & A & Dynamic-DeepHit (optimized) & -0.011393 [-0.013524, -0.009444] & 70.2 [64.8, 74.8]\\
GRU & 20 & A & SurvTransformer & -0.031340 [-0.037605, -0.025398] & 86.6 [83.7, 89.0]\\
GRU & 20 & A & SurvTransformer-Discrete & -0.031273 [-0.037772, -0.025199] & 86.6 [83.3, 89.1]\\
Transformer & 20 & A & CPH & -0.068828 [-0.079022, -0.059236] & 92.7 [91.5, 93.7]\\
Transformer & 20 & A & Weibull-MLP & -0.069575 [-0.080199, -0.059213] & 92.7 [91.4, 93.9]\\
Transformer & 20 & A & DRSA (optimized) & -0.063277 [-0.073731, -0.052925] & 92.1 [90.9, 93.2]\\
Transformer & 20 & A & DynamicDeepHit-Lite & -0.064231 [-0.074619, -0.054103] & 92.2 [90.9, 93.3]\\
Transformer & 20 & A & SurvTransformer-Lite & -0.063944 [-0.074294, -0.053704] & 92.1 [90.9, 93.3]\\
Transformer & 20 & A & Dynamic-DeepHit (optimized) & -0.010769 [-0.012819, -0.008867] & 66.4 [61.1, 70.9]\\
Transformer & 20 & A & SurvTransformer & -0.030717 [-0.036848, -0.024918] & 84.9 [82.0, 87.2]\\
Transformer & 20 & A & SurvTransformer-Discrete & -0.030649 [-0.037147, -0.024671] & 84.9 [81.7, 87.4]\\
GRU & 20 & B & CPH & -0.037444 [-0.044744, -0.030860] & 92.0 [90.6, 93.2]\\
GRU & 20 & B & Weibull-MLP & -0.030188 [-0.036176, -0.024615] & 90.2 [88.3, 91.9]\\
GRU & 20 & B & DRSA (optimized) & -0.029284 [-0.035870, -0.023300] & 90.0 [88.0, 91.6]\\
GRU & 20 & B & DynamicDeepHit-Lite & -0.029019 [-0.035491, -0.023121] & 89.9 [87.9, 91.6]\\
GRU & 20 & B & SurvTransformer-Lite & -0.029019 [-0.035555, -0.023062] & 89.9 [87.9, 91.6]\\
GRU & 20 & B & Dynamic-DeepHit (optimized) & -0.007355 [-0.009055, -0.005812] & 69.3 [64.9, 73.1]\\
GRU & 20 & B & SurvTransformer & -0.018626 [-0.022495, -0.014865] & 85.1 [82.5, 87.2]\\
GRU & 20 & B & SurvTransformer-Discrete & -0.017705 [-0.021483, -0.014084] & 84.4 [81.9, 86.5]\\
Transformer & 20 & B & CPH & -0.036008 [-0.043087, -0.029764] & 88.5 [86.6, 90.2]\\
Transformer & 20 & B & Weibull-MLP & -0.028753 [-0.034368, -0.023490] & 86.0 [83.4, 88.2]\\
Transformer & 20 & B & DRSA (optimized) & -0.027849 [-0.033999, -0.022169] & 85.6 [83.1, 87.8]\\
Transformer & 20 & B & DynamicDeepHit-Lite & -0.027584 [-0.033579, -0.021914] & 85.4 [82.9, 87.7]\\
Transformer & 20 & B & SurvTransformer-Lite & -0.027584 [-0.033683, -0.021881] & 85.4 [82.9, 87.7]\\
Transformer & 20 & B & Dynamic-DeepHit (optimized) & -0.005920 [-0.007505, -0.004489] & 55.7 [48.6, 61.9]\\
Transformer & 20 & B & SurvTransformer & -0.017191 [-0.020867, -0.013725] & 78.5 [74.8, 81.5]\\
Transformer & 20 & B & SurvTransformer-Discrete & -0.016270 [-0.019638, -0.012952] & 77.6 [74.3, 80.4]\\
\end{longtable}
\endgroup

\clearpage
\setcounter{section}{8}
\includepdf[pages=-,pagecommand={},addtotoc={1,section,1,Exploratory MMLR and HAMMLR comparators,app:mlr}]{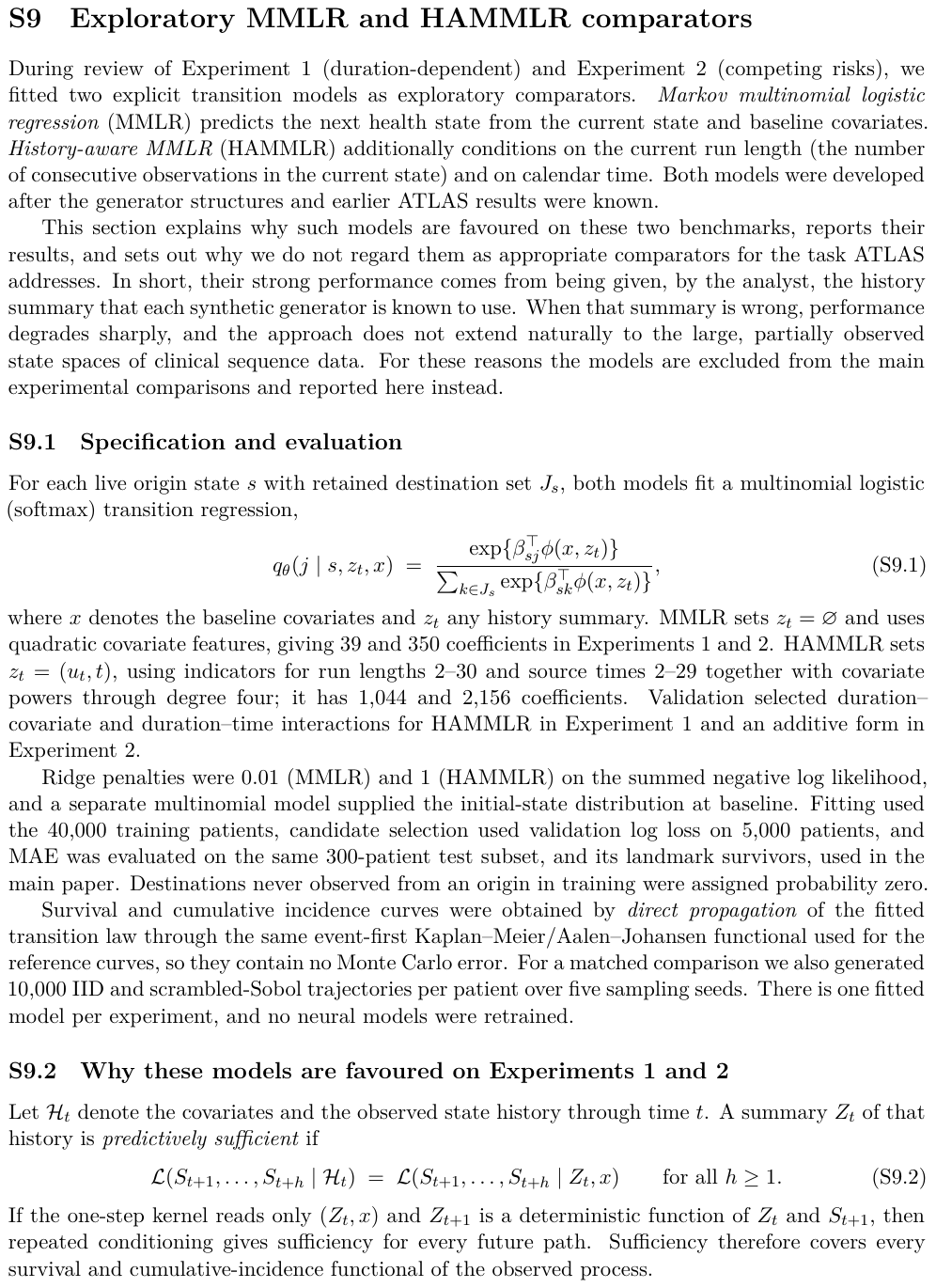}
\setcounter{table}{13}

\clearpage
\section{Original history-aware Dynamic-DeepHit results}\label{app:originalddh}
In the initial comparisons, the history-aware Dynamic-DeepHit adaptation had substantially larger curve errors than several other comparators. To determine whether its experimental specification contributed to this difference, we reviewed the implementation against the published formulation and conducted a staged search over architecture, optimization, regularization, and loss settings. Configurations were selected using observed validation outcomes, without using analytical reference curves or test-set MAE. The selected configurations were then evaluated using the original training seeds (nine and five, respectively) and the same test patients. These optimized configurations replace the initial adaptation in the main comparisons, and the original results are retained here for transparency. This follow-up was motivated by the initial test findings; it evaluates a revised fitted procedure rather than isolating the effect of checkpoint selection alone.

The tables below retain the original history-aware adaptation, rather than DynamicDeepHit-Lite. MAE and RMSE are reported on the probability scale. Between-run standard deviations describe the nine duration-dependent or five competing-risks fits. The 95\% intervals for mean MAE use the paired patient-bootstrap procedure in Section~S8, applied to the original saved fits. Full-precision metrics, the original ATLAS--DDH bootstrap comparisons, checkpoints, training logs and all 51 saved prediction files are preserved in \texttt{supplementary/original\_DDH\_results\_2026-11-02/} in the accompanying data and code package.

\begin{table}[htbp]\centering\small
\caption{Original history-aware Dynamic-DeepHit mean curve errors. Targets S, A and B denote survival, Death A CIF and Death B CIF. SD is the sample standard deviation across training runs.}\label{tab:originalddh_summary}
\begin{tabular}{llrrrrl}\toprule
Experiment / L & Target & Fits & Mean MAE & SD & Mean RMSE & 95\% MAE interval\\\midrule
Duration / 0 & S & 9 & 0.162647 & 0.001402 & 0.185249 & [0.157734, 0.167620]\\
Duration / 20 & S & 9 & 0.157081 & 0.004023 & 0.206254 & [0.134957, 0.180032]\\
Duration / 23 & S & 9 & 0.184387 & 0.004170 & 0.244296 & [0.152025, 0.217526]\\
Duration / 24 & S & 9 & 0.147417 & 0.004957 & 0.191851 & [0.120341, 0.178009]\\
Competing risks / 0 & A & 5 & 0.077074 & 0.007877 & 0.095116 & [0.074343, 0.079912]\\
Competing risks / 0 & B & 5 & 0.049179 & 0.002733 & 0.060681 & [0.046603, 0.051829]\\
Competing risks / 0 & S & 5 & 0.103898 & 0.005358 & 0.122215 & [0.100926, 0.106917]\\
Competing risks / 10 & A & 5 & 0.091127 & 0.004701 & 0.127368 & [0.082608, 0.100372]\\
Competing risks / 10 & B & 5 & 0.049876 & 0.004455 & 0.071509 & [0.044983, 0.055073]\\
Competing risks / 10 & S & 5 & 0.095754 & 0.005709 & 0.119784 & [0.089541, 0.102349]\\
Competing risks / 20 & A & 5 & 0.050539 & 0.003007 & 0.071648 & [0.044714, 0.056673]\\
Competing risks / 20 & B & 5 & 0.036874 & 0.004719 & 0.055889 & [0.032114, 0.041926]\\
Competing risks / 20 & S & 5 & 0.068788 & 0.004171 & 0.091942 & [0.062021, 0.075975]\\
\bottomrule\end{tabular}
\end{table}

\begingroup\small
\setlength{\tabcolsep}{5pt}
\begin{longtable}{rrrrr}
\caption{Original history-aware Dynamic-DeepHit individual-run errors: duration-dependent experiment. Post-rule MAE restricts scoring to $t\ge24$.}\label{tab:originalddh_duration}\\
\toprule
Seed & L & MAE & RMSE & Post-rule MAE\\\midrule
\endfirsthead
\multicolumn{5}{l}{\tablename~\thetable\ (continued)}\\\toprule
Seed & L & MAE & RMSE & Post-rule MAE\\\midrule
\endhead
\bottomrule\endfoot
42 & 0 & 0.164452 & 0.187311 & 0.078990\\
43 & 0 & 0.162671 & 0.185291 & 0.078078\\
44 & 0 & 0.163136 & 0.185768 & 0.077443\\
45 & 0 & 0.159550 & 0.182313 & 0.078477\\
50 & 0 & 0.163793 & 0.186773 & 0.079459\\
51 & 0 & 0.162768 & 0.184808 & 0.082150\\
52 & 0 & 0.161711 & 0.184331 & 0.078435\\
62 & 0 & 0.162468 & 0.184747 & 0.078215\\
63 & 0 & 0.163277 & 0.185899 & 0.079047\\
42 & 20 & 0.158612 & 0.208085 & 0.168055\\
43 & 20 & 0.159394 & 0.206690 & 0.165822\\
44 & 20 & 0.153823 & 0.203778 & 0.167551\\
45 & 20 & 0.152442 & 0.204100 & 0.163127\\
50 & 20 & 0.156150 & 0.209314 & 0.167820\\
51 & 20 & 0.164053 & 0.215745 & 0.174180\\
52 & 20 & 0.156281 & 0.204805 & 0.165871\\
62 & 20 & 0.152081 & 0.194163 & 0.157139\\
63 & 20 & 0.160895 & 0.209607 & 0.169500\\
42 & 23 & 0.184283 & 0.244225 & 0.184283\\
43 & 23 & 0.186854 & 0.246912 & 0.186854\\
44 & 23 & 0.186335 & 0.244295 & 0.186335\\
45 & 23 & 0.184564 & 0.248986 & 0.184564\\
50 & 23 & 0.179673 & 0.240884 & 0.179673\\
51 & 23 & 0.189392 & 0.252234 & 0.189392\\
52 & 23 & 0.184688 & 0.239239 & 0.184688\\
62 & 23 & 0.175975 & 0.232665 & 0.175975\\
63 & 23 & 0.187715 & 0.249223 & 0.187715\\
42 & 24 & 0.151343 & 0.190655 & 0.151343\\
43 & 24 & 0.151270 & 0.195729 & 0.151270\\
44 & 24 & 0.149347 & 0.196119 & 0.149347\\
45 & 24 & 0.144734 & 0.190137 & 0.144734\\
50 & 24 & 0.139537 & 0.189122 & 0.139537\\
51 & 24 & 0.154024 & 0.204463 & 0.154024\\
52 & 24 & 0.142678 & 0.183337 & 0.142678\\
62 & 24 & 0.150562 & 0.192880 & 0.150562\\
63 & 24 & 0.143256 & 0.184220 & 0.143256\\
\end{longtable}
\endgroup

\begingroup\small
\setlength{\tabcolsep}{5pt}
\begin{longtable}{rrlrr}
\caption{Original history-aware Dynamic-DeepHit individual-run errors: competing-risks experiment. Targets S, A and B denote survival and the separate cause-specific CIFs.}\label{tab:originalddh_competing}\\
\toprule
Seed & L & Target & MAE & RMSE\\\midrule
\endfirsthead
\multicolumn{5}{l}{\tablename~\thetable\ (continued)}\\\toprule
Seed & L & Target & MAE & RMSE\\\midrule
\endhead
\bottomrule\endfoot
42 & 0 & A & 0.083774 & 0.100567\\
43 & 0 & A & 0.076267 & 0.093468\\
45 & 0 & A & 0.078567 & 0.098774\\
46 & 0 & A & 0.082686 & 0.100146\\
47 & 0 & A & 0.064078 & 0.082624\\
42 & 0 & B & 0.049692 & 0.061803\\
43 & 0 & B & 0.045496 & 0.057829\\
45 & 0 & B & 0.051508 & 0.062899\\
46 & 0 & B & 0.047333 & 0.059247\\
47 & 0 & B & 0.051866 & 0.061627\\
42 & 0 & S & 0.111043 & 0.129870\\
43 & 0 & S & 0.099218 & 0.117340\\
45 & 0 & S & 0.105044 & 0.123854\\
46 & 0 & S & 0.106218 & 0.123336\\
47 & 0 & S & 0.097966 & 0.116675\\
42 & 10 & A & 0.097497 & 0.135453\\
43 & 10 & A & 0.088757 & 0.124313\\
45 & 10 & A & 0.094683 & 0.135999\\
46 & 10 & A & 0.088053 & 0.123765\\
47 & 10 & A & 0.086645 & 0.117311\\
42 & 10 & B & 0.053182 & 0.075491\\
43 & 10 & B & 0.055164 & 0.077268\\
45 & 10 & B & 0.049974 & 0.073320\\
46 & 10 & B & 0.046702 & 0.063442\\
47 & 10 & B & 0.044360 & 0.068027\\
42 & 10 & S & 0.103083 & 0.128884\\
43 & 10 & S & 0.096846 & 0.120588\\
45 & 10 & S & 0.098735 & 0.126244\\
46 & 10 & S & 0.090949 & 0.112579\\
47 & 10 & S & 0.089159 & 0.110625\\
42 & 20 & A & 0.048831 & 0.068010\\
43 & 20 & A & 0.047147 & 0.069628\\
45 & 20 & A & 0.054703 & 0.076626\\
46 & 20 & A & 0.052417 & 0.071330\\
47 & 20 & A & 0.049597 & 0.072644\\
42 & 20 & B & 0.033146 & 0.049366\\
43 & 20 & B & 0.044727 & 0.070708\\
45 & 20 & B & 0.035960 & 0.056325\\
46 & 20 & B & 0.037210 & 0.052391\\
47 & 20 & B & 0.033327 & 0.050658\\
42 & 20 & S & 0.064170 & 0.086260\\
43 & 20 & S & 0.073646 & 0.104626\\
45 & 20 & S & 0.070822 & 0.091158\\
46 & 20 & S & 0.070645 & 0.090653\\
47 & 20 & S & 0.064657 & 0.087012\\
\end{longtable}
\endgroup

\fi


\clearpage
\begin{thebibliography}{99}
\small

\bibitem{aalen1978}
O.~O. Aalen and S.~Johansen.
An empirical transition matrix for non-homogeneous Markov chains based on
censored observations.
\emph{Scandinavian Journal of Statistics}, 5(3):141--150, 1978.

\bibitem{cho2014}
K.~Cho, B.~van Merri\"{e}nboer, D.~Bahdanau, and Y.~Bengio.
On the properties of neural machine translation: encoder--decoder approaches.
\emph{arXiv:1409.1259}, 2014.

\bibitem{cho2023survival}
H.~Cho, S.~T. Holloway, D.~J. Couper, and M.~R. Kosorok.
Multi-stage optimal dynamic treatment regimes for survival outcomes with dependent censoring.
\emph{Biometrika}, 110(2):395--410, 2023.
\url{https://doi.org/10.1093/biomet/asac047}.

\bibitem{glasserman2004}
P.~Glasserman.
\emph{Monte Carlo Methods in Financial Engineering}.
Springer, New York, 2004.

\bibitem{glazier2019}
S.~W. Glazier.
\emph{Sequential Survival Analysis with Deep Learning}.
Master's thesis, Brigham Young University, 2019.

\bibitem{goldberg2012}
Y.~Goldberg and M.~R. Kosorok.
Q-learning with censored data.
\emph{Annals of Statistics}, 40(1):529--560, 2012.
\url{https://doi.org/10.1214/12-AOS968}.

\bibitem{graf1999}
E.~Graf, C.~Schmoor, W.~Sauerbrei, and M.~Schumacher.
Assessment and comparison of prognostic classification schemes for survival
data.
\emph{Statistics in Medicine}, 18(17--18):2529--2545, 1999.

\bibitem{hochreiter1997}
S.~Hochreiter and J.~Schmidhuber.
Long short-term memory.
\emph{Neural Computation}, 9(8):1735--1780, 1997.

\bibitem{kaplan1958}
E.~L. Kaplan and P.~Meier.
Nonparametric estimation from incomplete observations.
\emph{Journal of the American Statistical Association}, 53(282):457--481, 1958.

\bibitem{lee2018deephit}
C.~Lee, W.~Zame, J.~Yoon, and M.~van~der Schaar.
DeepHit: a deep learning approach to survival analysis with competing risks.
In \emph{Proceedings of the AAAI Conference on Artificial Intelligence},
volume~32, pages 2314--2321, 2018.

\bibitem{lee2019dynamic}
C.~Lee, J.~Yoon, and M.~van~der Schaar.
Dynamic-DeepHit: a deep learning approach for dynamic survival analysis with
competing risks based on longitudinal data.
\emph{IEEE Transactions on Biomedical Engineering}, 67(1):122--133, 2020.

\bibitem{martinsson2017}
E.~Martinsson.
\emph{WTTE-RNN: Weibull Time To Event Recurrent Neural Network}.
Master's thesis, Chalmers University of Technology, 2017.

\bibitem{owen1998}
A.~B. Owen.
Scrambling Sobol' and Niederreiter--Xing points.
\emph{Journal of Complexity}, 14(4):466--489, 1998.

\bibitem{owen2013}
A.~B. Owen.
\emph{Monte Carlo Theory, Methods and Examples}.
2013. \url{https://artowen.su.domains/mc/}.

\bibitem{putter2007}
H.~Putter, M.~Fiocco, and R.~B. Geskus.
Tutorial in biostatistics: competing risks and multi-state models.
\emph{Statistics in Medicine}, 26(11):2389--2430, 2007.

\bibitem{ren2019}
K.~Ren, J.~Qin, L.~Zheng, Z.~Yang, W.~Zhang, L.~Qiu, and Y.~Yu.
Deep recurrent survival analysis.
In \emph{Proceedings of the AAAI Conference on Artificial Intelligence},
volume~33, pages 4798--4805, 2019.

\bibitem{renc2024}
P.~Renc, Y.~Jia, A.~E. Samir, J.~Was, Q.~Li, D.~W. Bates, and A.~Sitek.
Zero shot health trajectory prediction using transformer.
\emph{npj Digital Medicine}, 7:256, 2024.

\bibitem{rizopoulos2012}
D.~Rizopoulos.
\emph{Joint Models for Longitudinal and Time-to-Event Data: With Applications in R}.
Chapman and Hall/CRC, Boca Raton, 2012.

\bibitem{shmatko2025}
A.~Shmatko, A.~W. Jung, K.~Gaurav, S.~Brunak, L.~H. Mortensen, E.~Birney, T.~Fitzgerald, and M.~Gerstung.
Learning the natural history of human disease with generative transformers.
\emph{Nature}, 647:248--256, 2025. \url{https://doi.org/10.1038/s41586-025-09529-3}.

\bibitem{siebert2012}
U.~Siebert, O.~Alagoz, A.~M. Bayoumi, B.~Jahn, D.~K. Owens, D.~J. Cohen, and K.~M. Kuntz.
State-transition modeling: a report of the ISPOR-SMDM Modeling Good Research Practices Task Force-3.
\emph{Medical Decision Making}, 32(5):690--700, 2012.

\bibitem{survtracecode}
Z.~Wang and J.~Sun. SurvTRACE author implementation: training utilities.
\url{https://github.com/RyanWangZf/SurvTRACE/blob/main/survtrace/train_utils.py}.
Accessed September 2026.

\bibitem{vanhouwelingen2007}
H.~C. van Houwelingen.
Dynamic prediction by landmarking in event history analysis.
\emph{Scandinavian Journal of Statistics}, 34(1):70--85, 2007.

\bibitem{vaswani2017}
A.~Vaswani, N.~Shazeer, N.~Parmar, J.~Uszkoreit, L.~Jones, A.~N. Gomez,
L.~Kaiser, and I.~Polosukhin.
Attention is all you need.
In \emph{Advances in Neural Information Processing Systems}, pages 5998--6008,
2017.

\bibitem{wang2022survtrace}
Z.~Wang and J.~Sun.
SurvTRACE: Transformers for survival analysis with competing events.
In \emph{ACM Conference on Bioinformatics, Computational Biology, and Health Informatics}, 2022.
\url{https://doi.org/10.1145/3535508.3545521}.
\end{thebibliography}
\end{document}